\documentclass[smallcondensed,nospthms]{svjour3}
 
\usepackage{fullpage}
\usepackage{amsmath}
\usepackage{wasysym}
\usepackage{mathrsfs}	
\usepackage{graphicx}	
\usepackage{float}
\usepackage{ulem}		
\usepackage{epstopdf}
\usepackage{color}
\usepackage{gensymb}
\usepackage{amsthm}

\newtheorem{theorem}{Theorem}[section]

\newtheorem{corollary}{Corollary}[section]

\newtheorem{remark}{Remark}[section]

\newtheoremstyle{example}
     {10pt}
     {10pt}
     {\sffamily}
     {}
     {\scshape}
     {:}
     {0.5em}
     {}
\theoremstyle{example}

\DeclareGraphicsExtensions{.png,.pdf,.jpg}

\begin{document}

\normalem	

\title{Dynamic Limits on Planar Libration-Orbit Coupling Around an Oblate Primary}

\author{Jay W. McMahon \and Daniel J. Scheeres}

\institute{Jay W. McMahon \at Research Associate, Aerospace Engineering Sciences, University of Colorado - Boulder, 431 UCB, University of Colorado, Boulder, CO 80309-0431, \email{mcmahojw@colorado.edu}
\and Daniel J. Scheeres \at A. Richard Seebass Endowed Professor, Aerospace Engineering Sciences, University of Colorado at Boulder, 429 UCB, University of Colorado, Boulder, CO 80309-0429.}

\date{Received: date / Accepted: date}

\maketitle

\begin{abstract}

This paper explores the dynamic properties of the planar system of an ellipsoidal satellite in an equatorial orbit about an oblate primary. In particular, we investigate the conditions for which the satellite is bound in librational motion or when the satellite will circulate with respect to the primary.  
We find the existence of stable equilibrium points about which the satellite can librate, and explore both the linearized and non-linear dynamics around these points. Absolute bounds are placed on the phase space of the libration-orbit coupling through the use of zero-velocity curves that exist in the system. These zero-velocity curves are used to derive a sufficient condition for when the satellite's libration is bound to less than 90$\degree$. When this condition is not satisfied so that circulation of the satellite is possible, the initial conditions at zero libration angle are determined which lead to circulation of the satellite. Exact analytical conditions for circulation and the maximum libration angle are derived for the case of a small satellite in orbits of any eccentricity.
\end{abstract}

\section{Introduction}







The investigation of the translational-rotational coupling for a finite orbiting body, referred to in the literature as the Full Two Body Problem, has received renewed attention in recent years \cite{hamiltonian_rigid,maciejewski,koon,cendra,scheeres}. Much of this work has been motivated by interest in the dynamic evolution of binary asteroid systems, which comprise 16\% of Near-Earth asteroids \cite{margot}. Scheeres \cite{scheeres_rot_fiss} studied the stability of bodies resting on one-another which can lead to the formation of a binary asteroid system through rotational fission. Bellerose \cite{bellerose} looked at the dynamics and stability in the planar Full Two Body Problem from an energetic standpoint. Fahenstock \cite{gene1} studied the problem by modeling the gravitational interaction with polyhedral models, which were applied to simulate the dynamics of the near-Earth binary asteroid 1999 KW4 \cite{gene1999kw4}.

Over time, many simplifications have been made to the Full Two Body Problem in order to make analytical progress in the study of specific applications. The most common simplification is to treat the finite bodies to second-order in their mass properties, which allows for the use of inertia matrices for modeling the attitude dynamics \cite{hughes,beletskii,scheeres_planarF2BP_stability,bellerose}. Analytical expressions have been constructed to fourth order \cite{hughes_high_order}, however most analytical studies stop at second order. A second simplification comes by reducing the problem to the planar problem. If the second order simplification is made on mass properties, then the in orbit plane libration angle (commonly referred to as the pitch angle) dynamics are decoupled from the out-of-plane (roll/yaw) dynamics if they are initially quiescent \cite{hughes}. This reduces the dimensionality of the problem from six to three degrees-of-freedom. The third major simplification that is often seen, particularly in the spacecraft community, is decoupling the translational and rotational dynamics \cite{hughes,beletskii}. In this case, it is assumed that the mass of the orbiting body is insignificant compared to the primary, so that the perturbation to the orbital dynamics is negligible. The orbital dynamics are known from the given Keplerian motion, and these are used as inputs to the attitude dynamics. A notable work which explores the effects of the coupling for the case of spacecraft sized objects was carried out by Mohan \cite{breakwell}. 

In this paper, we analyze the problem of a triaxial body in orbit about a spherical primary, or in an equatorial orbit about an oblate primary. These bodies are represented to second order with their moments of inertia, but all coupling between the translation and attitude motions for this case are preserved. This model can be used as a first order representation of common dynamic problems including binary asteroids (secondaries tend to be in near equatorial orbits of oblate primaries \cite{ostro}), spacecraft orbiters, and planet-moon systems. This paper is closely related to those by Scheeres \cite{scheeres_planarF2BP_stability} and Bellerose \cite{bellerose}, however we extend the foundation laid by those works.



The main contribution of this paper is the study of the limits of librational motion for a given system. There is a small amount of literature dedicated to bounding librational motion. Auelmann \cite{auelmann} laid out the bounding conditions for an axially symmetric spacecraft in circular orbit, using the uncoupled attitude dynamics with a zero spin rate about the axis of symmetry. Pringle \cite{pringle} extended this idea, analyzing the equilibrium cases and librational bounds for all spin states. 

This paper only looks at the planar libration, however we account for coupling, triaxial shapes, and are not limited to circular orbits, or indeed even Keplerian orbits due to the perturbations from the attitude coupling. The methodology used to bound the libration is initially similar to \cite{auelmann,pringle}, in that we use the energy to limit the amount of libration that is dynamically possible by computing zero-velocity curves for the system. We extend this, however, to cases when the energy is high enough so that circulation is possible. In this case, we show that although the energy zero-velocity curve does not bound the libration angle, certain initial conditions can lead to trajectories which have bounded maximum libration angles. In particular, for the case where the ellipsoid is very small compared to the other body (such as a spacecraft in Earth orbit), we can derive analytical expressions which predict the maximum libration amplitude for an orbit of any eccentricity. For cases with significant coupling, we present a method of analysis which determines the structure of bound and unbound trajectories.

The paper is organized as follows. Section \ref{sec:system} reviews the mathematical description of the system in question as originally derived by Scheeres \cite{scheeres_planarF2BP_stability}. Section \ref{sec:equilibrium} discusses the location and properties of equilibria in the system. Section \ref{sec:suff_bounded} presents a sufficient condition for bounded librational motion. The fact that this is only a sufficient condition is shown in Section \ref{sec:transient}. Finally, Section \ref{sec:suff_unbounded} presents the main results of the paper which describe conditions for unbound libration.

\section{Physical System Description}\label{sec:system}

\begin{figure}[htb]
\begin{center}
	\includegraphics[width=0.4\textwidth]{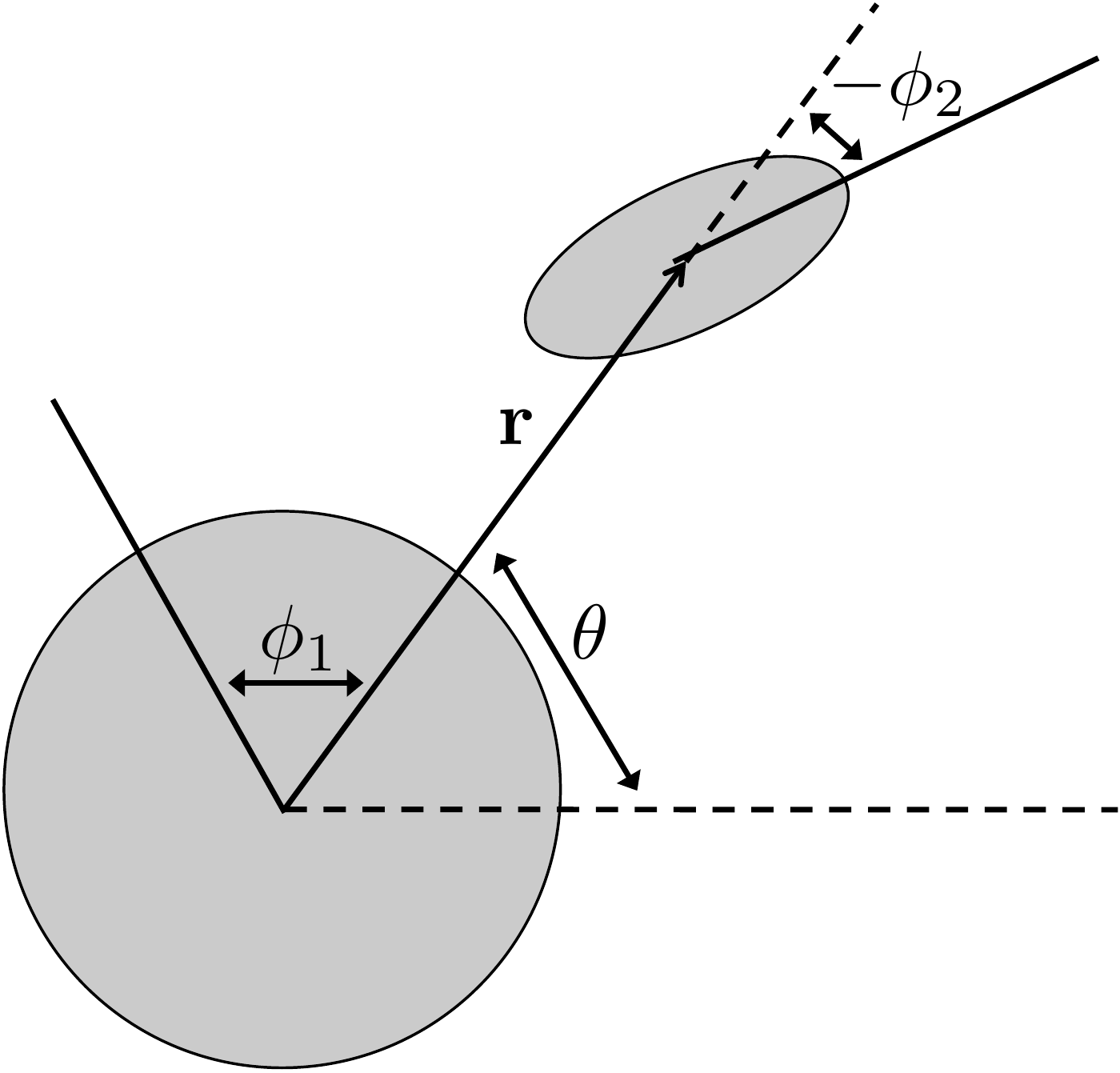}
	\caption{Definition of the system angles and relative radial vector.}
	\label{fig:system}
\end{center}
\end{figure}

In general, we begin with the planar full two-body problem where both bodies are finite, but constrained to have their individual rotation poles perpendicular to the mutual orbit plane. This situation is illustrated in Fig. \ref{fig:system} with the angles of interest and radius defined. 

In this section we present the key mathematical relationships used to describe this system, which are simplifications of the general relationships derived in \cite{scheeres_planarF2BP_stability}. Further details of the derivations are available in the Appendix.

\subsection{Equations of Motion}\label{sec:EoM}

Scheeres \cite{scheeres_planarF2BP_stability} showed that in this case, the potential energy using a second order expansion in the moments of inertia is,
\begin{equation}
	\begin{split}
	V(r,\phi_1,\phi_2) = &-\frac{G M_1 M_2}{r} \bigg\{ 1 + \frac{1}{2r^2}\bigg[ \text{Tr}\left(\mathbf{I}'_1\right) + \text{Tr}\left(\mathbf{I}'_2\right) - \frac{3}{2}\Big(I'_{1_x} +I'_{1_y} \\
	&-  \left(I'_{1_y} -I'_{1_x}\right) \cos 2\phi_1 +I'_{2_x} +I'_{2_y} - \left(I'_{2_y} -I'_{2_x}\right) \cos 2 \phi_2  \Big) \bigg] \bigg\}
	\end{split}
\end{equation}
where $\mathbf{I}'_i$ is the mass normalized (signified by the prime) inertia dyad of body $i$, and the subscripts on the non-bold versions indicate a principle moment of inertia. $G$ is the gravitational constant, $M_i$ the mass of body $i$, Tr() indicates the trace of the dyad.

In this paper, we will normalize the lengths with respect to the maximum ellipsoid semi-axis, $\alpha$; the time will be normalized by the mean motion of the system at this distance, $n = \sqrt{G(M_1 + M_2)/\alpha^3}$ (units of 1/s); and the mass will be normalized by the ellipsoid mass, $M_2$. Note that using these normalization factors, the normalized value of $\mu = G(M_1 + M_2)$ is 1. All equations from this point on will be written in normalized units.

First, let us define the reduced mass as,
\begin{equation}
	m = \frac{M_1 M_2}{M_1 + M_2}
\end{equation}
the mass fraction is then,
\begin{equation}\label{eq:mass_fraction}
	\nu = \frac{M_1}{M_1+M_2}
\end{equation}
which is just the reduced mass in normalized units. Next, we define the general moment of inertia as,
\begin{equation}\label{eq:Iz}
	I_z = \overline{I}_{2_z} + \nu r^2
\end{equation}
Note that the general moment of inertia is a function of $r$. The normalized moment of intertias are computed as
\begin{equation}
	\overline{I}_{2_z} = \frac{I_{2_z}}{M_2 \alpha^2} =  \frac{I'_{2_z}}{\alpha^2}
\end{equation}
\begin{equation}
	\overline{I}_{1_z} = \frac{I_{1_z}}{M_1 \alpha^2} = \frac{I'_{1_z}}{\alpha^2}
\end{equation}

For a system with an oblate primary, we have that $\overline{I}_{1_x} = \overline{I}_{1_y} = \overline{I}_s$, where $\overline{I}_s < \overline{I}_{1_z}$, so the potential becomes in normalized units,
\begin{equation}\label{eq:potential_oblate}
	V(r,\phi_2) = -\frac{\nu}{r} \bigg\{ 1 + \frac{1}{2r^2}\bigg[\left(\overline{I}_{1_z} - \overline{I}_s\right) -\frac{1}{2}\overline{I}_{2_x} -\frac{1}{2} \overline{I}_{2_y} + \overline{I}_{2_z} + \frac{3}{2} \left(\overline{I}_{2_y} - \overline{I}_{2_x}\right) \cos 2 \phi_2 \bigg] \bigg\}
\end{equation}

In the case of a spherical primary, which means that $\overline{I}_{1_x} = \overline{I}_{1_y} = \overline{I}_{1_z}$, and therefore the potential simplifies to,
\begin{equation}\label{eq:potential_sphere}
	V(r,\phi_2) = -\frac{\nu}{r} \bigg\{ 1 + \frac{1}{2r^2}\bigg[ -\frac{1}{2}\overline{I}_{2_x} -\frac{1}{2} \overline{I}_{2_y} + \overline{I}_{2_z} + \frac{3}{2} \left(\overline{I}_{2_y} - \overline{I}_{2_x}\right) \cos 2 \phi_2  \bigg] \bigg\}
\end{equation}

Note that in both the spherical and oblate primary case, the spherical symmetry causes the potential energy to no longer be dependent on the primary's orientation, as represented in the planar problem by $\phi_1$. 

Starting from Scheeres' \cite{scheeres_planarF2BP_stability} expression, in this case the kinetic energy is,
\begin{equation}\label{eq:kinetic_total}
	T = \frac{1}{2}\overline{I}_{1_z}\frac{M_1}{M_2}\dot{\theta}_1^2 + \frac{1}{2}\overline{I}_{2_z}\dot{\phi}_2^2 + \frac{1}{2}\nu\dot{r}^2 + \frac{1}{2}\left( \overline{I}_{2_z} + \nu r^2 \right)  \dot{\theta}^2 +  \overline{I}_{2_z}\dot{\phi}_2 \dot{\theta}
\end{equation}
where $\dot{\theta}_1 = \dot{\phi}_1 + \dot{\theta}$. Note that the dot indicates derivatives with respect to unit-less time.

\subsection{Integrals of Motion for the System}\label{sec:integrals}

The system described in Section \ref{sec:EoM} has three integrals of motion that can be used to simplify the problem: the Jacobi constant (or total energy), the total angular momentum ($K_{tot}$), and the inertial angular velocity of the primary ($\dot{\theta}_1$). That these quantities are integrals is shown in the Appendix. 

Using the integrals, the kinetic energy from Eq. \eqref{eq:kinetic_total} can be rewritten as
\begin{equation}\label{eq:kinetic}
	T = T_1 + \frac{1}{2}\overline{I}_{2_z}\dot{\phi}_2^2 + \frac{1}{2}\nu\dot{r}^2 + \frac{1}{2} I_z \dot{\theta}^2+  \overline{I}_{2_z}\dot{\phi}_2 \dot{\theta}
\end{equation}
where $T_1 = (M_1/(2M_2))\overline{I}_{1_z}\dot{\theta}_1^2$ is the kinetic energy of the primary, which is constant. Likewise, we can define the free angular momentum from Eq. \eqref{eq:angmom} to be,
\begin{equation}\label{eq:angmom}
	K = K_{tot} - K_1 = I_z \dot{\theta} + \overline{I}_{2_z}\dot{\phi}_2
\end{equation}
where $K_1 = (M_1/M_2)\overline{I}_{1_z}\dot{\theta}_1$ is the angular momentum of the primary, which is constant. This relationship between the free angular momentum and the orbit angular velocity will allow us to eliminate $\dot{\theta}$ from the system. Solving Eq. \eqref{eq:angmom} for the angular velocity gives,
\begin{equation}\label{eq:thetadot}
	\dot{\theta} = \frac{K - \overline{I}_{2_z}\dot{\phi}_2}{I_z}
\end{equation}
Substituting Eq. \eqref{eq:thetadot} into Eq. \eqref{eq:kinetic} allows us to write the kinetic energy as,
\begin{equation}
	T = T_1 + \frac{1}{2}\frac{K^2}{I_z} + \frac{1}{2}\nu\dot{r}^2 + \frac{1}{2}\frac{\overline{I}_{2_z}\nu r^2\dot{\phi}_2^2}{I_z}
\end{equation}
so that the total energy of the system can be written as,
\begin{equation}
	E_{tot} = T_1 + \frac{1}{2}\frac{K^2}{I_z} + \frac{1}{2}\nu \dot{r}^2 + \frac{1}{2}\frac{\overline{I}_{2_z}\nu r^2\dot{\phi}_2^2}{I_z} + V(r,\phi_2)
\end{equation}
or alternatively the free energy of the system can be written as,
\begin{equation}\label{eq:free_energy}
	E= E_{tot} - T_1 = \frac{1}{2}\frac{K^2}{I_z} + \frac{1}{2}\nu \dot{r}^2 + \frac{1}{2}\frac{\overline{I}_{2_z}\nu r^2\dot{\phi}_2^2}{I_z} + V(r,\phi_2)
\end{equation}

It is interesting to note that the description of the free energy of the system does not require any knowledge about the primary spin state as neither $\phi_1$ or $\dot{\phi}_1$ appear anywhere in Eq. \eqref{eq:free_energy}.

\subsection{Dynamic System Analysis}\label{sec:dyn}


In order to look at the stability of any relative states for the planar system, we derive the dynamic matrix corresponding to the equations of motion
\begin{equation}
	\mathrm{A} = \frac{\partial \mathbf{f}(\mathbf{q},\mathbf{\dot{q}})}{\partial [\mathbf{q},\mathbf{\dot{q}}]}
\end{equation}
where $\mathbf{q} = [r \quad \theta \quad \phi_1 \quad \phi_2]$.

\begin{figure}[htb]
\begin{center}
	\includegraphics[width=0.6\textwidth]{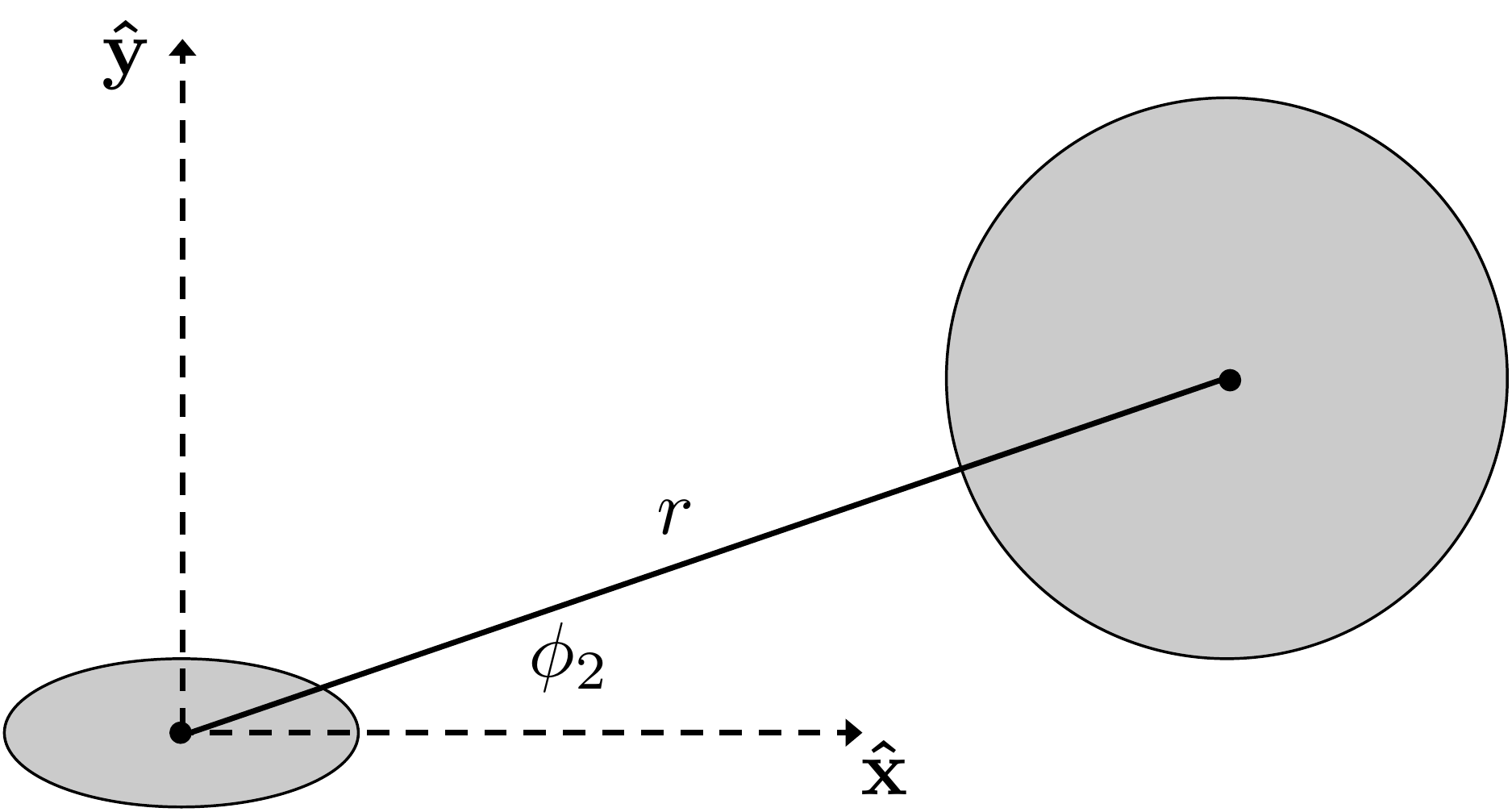}
	\caption{Definition of the secondary fixed relative frame.}
	\label{fig:secondary_frame}
\end{center}
\end{figure}

First, however, we note that we can reduce the order of the system by recalling from Section \ref{sec:integrals} that $\phi_1$ is ignorable, and that through the conservation of angular momentum we can remove $\theta$ through the use of Eq. \eqref{eq:thetadot}. Therefore the equations of motion for the 2 degree-of-freedom system are,
\begin{equation}\label{eq:2dof_rddot}
	\ddot{r} = \frac{(K - \overline{I}_{2_z}\dot{\phi}_2)^2 r}{I_z^2} - \frac{1}{\nu}\frac{\partial V}{\partial r}
\end{equation}
\begin{equation}\label{eq:2dof_phi2ddot}
	\ddot{\phi}_2 = -\left( 1 + \frac{\nu r^2}{\overline{I}_{2_z}}\right)\frac{1}{\nu r^2}\frac{\partial V}{\partial \phi_2} + \frac{2\dot{r}(K - \overline{I}_{2_z}\dot{\phi}_2)}{r I_z}
\end{equation}
This system is depicted from an ellipsoid fixed frame in Fig. \eqref{fig:secondary_frame}.

This frame rotates with the orbit, so that if there is no libration of the secondary, the primary location will be fixed. When the secondary is librating, the primary will appear to move relative to the secondary due to the libration, as well as changing the separation distance as energy is traded between the orbit and the secondary libration.

The dynamic matrix is now a 4x4 matrix with the form,
\begin{equation}
	\mathrm{A} = 
	\begin{bmatrix}
		0 & 0 & 1 & 0 \\
		0 & 0 & 0 & 1 \\
		\frac{\partial \ddot{r}}{\partial r} & \frac{\partial \ddot{r}}{\partial \phi_2} & 0 & \frac{\partial \ddot{r}}{\partial \dot{\phi}_2} \\
		\frac{\partial \ddot{\phi}_2}{\partial r} & \frac{\partial \ddot{\phi}_2}{\partial \phi_2} & \frac{\partial \ddot{\phi}_2}{\partial \dot{r}} & \frac{\partial \ddot{\phi}_2}{\partial \dot{\phi}_2}
	\end{bmatrix}
\end{equation}
where the state vector has been organized as, $\mathbf{x} = [r \quad \phi_2 \quad \dot{r} \quad \dot{\phi}_2]$. 
This matrix will be used in the following sections to analyze the stability of the system around any relative equilibrium points that may exist.

\section{Equilibrium Conditions}\label{sec:equilibrium}


In this section, we determine the relative equilibria between the two bodies. This means that these equilibria are places where the relative dynamics given by Eqs. \eqref{eq:2dof_rddot} and \eqref{eq:2dof_phi2ddot} are stationary. However, the primary is free to rotate at a constant rate, and the secondary (ellipsoidal) body will orbit at a constant radius and orbit rate ($\dot{\theta}$). 
As in Section \ref{sec:system}, many of the relationships given here are simplifications of the general relationships derived in \cite{scheeres_planarF2BP_stability}. Further details are available in the Appendix.

\subsection{Relative Equilibrium Point Locations}\label{sec:eqpt_loc}

The relative equilibrium locations are found by solving the following polynomial for $r$,
\begin{equation}\label{eq:equil_pts}
	\begin{split}
	r^6 &- \frac{K^2}{\nu^2}r^5 + \left[ \frac{2 \overline{I}_{2_z}}{\nu} + \frac{3}{2}\left(C_2^\pm + \overline{I}_{1_z} - \overline{I}_s\right) \right] r^4 \\
	&+ \left[ \frac{\overline{I}_{2_z}^2}{\nu^2} + \frac{3 \overline{I}_{2_z} }{\nu}\left(C_2^\pm + \overline{I}_{1_z} - \overline{I}_s\right) \right]r^2 + \frac{3}{2} \frac{\overline{I}_{2_z}^2}{\nu^2} \left(C_2^\pm + \overline{I}_{1_z} - \overline{I}_s\right) = 0
	\end{split}
\end{equation}
where
\begin{equation}\label{eq:c2pm}
	C_2^\pm = 
	\begin{cases}
		-2\overline{I}_{2_x} + \overline{I}_{2_y} + \overline{I}_{2_z} & \text{if } \phi_2 = 0 \text{, } \pi \\
		\overline{I}_{2_x} - 2 \overline{I}_{2_y} + \overline{I}_{2_z} & \text{if } \phi_2 = \pm \pi/2
	\end{cases}
\end{equation}

Note that the relative equilibria found here is identical to finding the equilibria of the second-order system given in Eqs. \eqref{eq:2dof_rddot} and \eqref{eq:2dof_phi2ddot}. The stationary conditions for energy also enforce that $\dot{r} = \ddot{r} = \dot{\phi}_2 = \ddot{\phi}_2 = 0$.

It is interesting to compare this result to the full case from Scheeres \cite{scheeres_planarF2BP_stability} when there is no assumption on the shape of the primary. In that case we have a dependence on $\phi_1$, and the stationary conditions also tell us that $\dot{\phi}_1 = 0$ and $\phi_1 = 0$, $\pm\pi/2$, or $\pi$. Recalling the equation from Scheeres \cite{scheeres_planarF2BP_stability},
\begin{equation}\label{eq:equil_pts_scheeres}
	\begin{split}
	r^6 &- \frac{K_{tot}^2}{m^2 \mu}r^5 + \left[ \frac{2 \left(I_{1_z} + I_{2_z}\right)}{m} + \frac{3}{2}\left(C_1^\pm + C_2^\pm \right) \right] r^4 \\
	&+ \left[ \frac{\left(I_{1_z} + I_{2_z}\right)^2}{m^2} + \frac{3 \left(I_{1_z} + I_{2_z}\right) }{m}\left(C_1^\pm + C_2^\pm\right) \right]r^2 + \frac{3}{2} \frac{\left(I_{1_z} + I_{2_z}\right)^2}{m^2} \left(C_1^\pm + C_2^\pm\right) = 0
	\end{split}
\end{equation}
The differences between Eq. \eqref{eq:equil_pts} and \eqref{eq:equil_pts_scheeres} (aside from the normalization) are that for the case when the primary is not exactly symmetric about its rotation axis, we must consider the entire angular momentum, $K_{tot}$, and therefore we also see $I_{1_z}$ appearing in terms with $I_{2_z}$. Also, as the primary body becomes spherically symmetric, the dependence on the primary orientation disappears and $C_1^\pm \rightarrow \overline{I}_{1_z} - \overline{I}_s$. This causes the system to decrease the total number of equilibrium configurations from 8 to 4, and since all the dependence on the primary has been removed, the primary can also be spinning at an arbitrary speed, $\dot{\phi}_1$, and the system can still be in a relative equilibrium.


The location of the equilibrium points depends on three main parameters: the mass ratio, the angular momentum, and the body shapes. The influences of these three parameters are studied in Section \ref{sec:vary_eq_pts}. 

Throughout the remainder of the paper, any reference to a nominal system refers to the system outlined in Table \ref{tab:kw4_params}, which are based on the binary asteroid system 1999 KW4 \cite{ostro,gene1999kw4}. The new parameter that appears in Table \ref{tab:kw4_params} is
\begin{equation}
	\chi = \frac{R_p}{R_e}
\end{equation}
which is the ratio of the polar radius to the equatorial radius of the oblate primary body. This parameter becomes important for scaling the system.


\begin{table}[htdp]
\caption{The non-dimensionalized parameters for the nominal test system.}
\begin{center}
\begin{tabular}{|c|cc|ccc|cc|ccc|}
	$\nu$ & $\beta$ & $\gamma$ & $R_p$ & $R_e$ & $\chi$ & $\overline{I}_{1_z}$ & $\overline{I}_{s}$ & $\overline{I}_{2_x}$ & $\overline{I}_{2_y}$ & $\overline{I}_{2_z}$ \\
	\hline
	0.9257 & 0.8109 & 0.6112 & 2.3590 & 2.6182 & 0.9010 & 2.4034 & 2.1175 & 0.1973 & 0.2913 & 0.3434\end{tabular}
\end{center}
\label{tab:kw4_params}
\end{table}


The nominal system has an angular momentum of $K = 2.8382$, which results in equilibria at $r^+ = 9.2442$ with $E^+ = -0.0497$, and $r^- = 9.2869$ with $E^- = -0.0496$ where the superscripts indicate which equilibrium point we are referring to; the $^+$ indicates the equilibrium point at $\phi_2 = 0$, and the $^-$ refers to the equilibrium at $\phi_2 = 90\degree$, as with $C_2^\pm$.

\subsection{Equilibrium Point Linear Stability Analysis}

Given that the linearized system near the equilibrium point is simplified significantly, we can find the eigenvalues of the system analytically. The dynamics matrix around the equilibrium points has the form,
\begin{equation}
	\mathrm{A} = 
	\begin{bmatrix}
		0 & 0 & 1 & 0 \\
		0 & 0 & 0 & 1 \\
		P & 0 & 0 & Q \\
		0 & R & S & 0
	\end{bmatrix}
\end{equation}
where the non-zero partials ($\frac{\partial \ddot{r}}{\partial r}$, $\frac{\partial \ddot{r}}{\partial \dot{\phi}_2}$, $\frac{\partial \ddot{\phi}_2}{\partial \phi_2}$, and $\frac{\partial \ddot{\phi}_2}{\partial \dot{r}}$) have been represented by simpler variable expressions ($P$, $Q$, $R$, and $S$). The characteristic equation for this matrix is simply,
\begin{equation}
	\lambda^4 - (P + R + QS)\lambda^2 + PR = 0
\end{equation}
The roots of this equation are,
\begin{equation}
	\lambda^2 = \frac{P + R + QS \pm \sqrt{(P + R + QS)^2 - 4PR}}{2}
\end{equation}
The determination of the eigenvalues can easily be carried out numerically on a case-by-case basis to determine the dynamic behavior around the equilibrium point.

The eigenvectors can also be computed analytically by writing the eigenvalue problem as,
\begin{equation}
	\bigg[ \lambda \mathrm{I} - \mathrm{A} \bigg] \mathbf{v} = 0
\end{equation}
where the eigenvector is $\mathbf{v} = \left[ v_1 \quad v_2 \quad v_3 \quad v_4 \right]^T$. The four equations that result are,
\begin{equation}
	\lambda v_1 - v_3 = 0
\end{equation}
\begin{equation}
	\lambda v_2 - v_4 = 0
\end{equation}
\begin{equation}
	-P v_1 + \lambda v_3 - Q v_4 = 0
\end{equation}
\begin{equation}
	-R v_2 = S v_3 + \lambda v_4 = 0
\end{equation}
The eigenvector can be computed from these relationships to be
\begin{equation}
	\mathbf{v} = 
	\begin{bmatrix}
		1 \\
		\sigma \\
		\lambda \\
		\sigma \lambda
	\end{bmatrix}
\end{equation}
where 
\begin{equation}
	\sigma = \frac{\lambda^2 - P}{Q \lambda}
\end{equation}

In the nominal system, the eigenvalues of the equilibrium point $r^+$ are
\begin{gather}
	\lambda_{1,2} = 0 \pm 0.0302i \\
	\lambda_{3,4} = 0 \pm 0.0362i
\end{gather}
which makes it a center point. However, we will refer to this equilibrium point as stable in the remainder of the study since trajectories can be bound around it, as will be explored. The eigenvalues of the equilibrium point at $r^-$ are
\begin{gather}
	\lambda_{1,2} = \pm 0.0307 \\
	\lambda_{3,4} = 0 \pm 0.0351i
\end{gather}
This equilibrium point has a unstable and stable asymptote associated with the real eigenvalues. 
It is interesting to note that the eigenstructure of this problem is nearly identical to that of the collinear Lagrange points in the restricted three-body problem studied by Conley \cite{conley}. 




\subsection{Equilibrium Point Energy}\label{sec:eq_pt_energy}

It has been shown in Section \ref{sec:eqpt_loc} that the equilibrium points are stationary points for the total energy. In this section, we determine if these stationary points are local maxima, minima, or saddle points for the energy. To find this, we must investigate the second derivatives of the energy with respect to the state,
\begin{equation}
	E_{\mathbf{x} \mathbf{x}} = 
	\begin{bmatrix}
		E_{rr} & E_{r\phi_2} & E_{r\dot{r}} & E_{r\dot{\phi}_2} \\
		E_{\phi_2 r} & E_{\phi_2 \phi_2} & E_{\phi_2 \dot{r}} & E_{\phi_2 \dot{\phi}_2} \\
		E_{\dot{r}r} & E_{\dot{r}\phi_2} & E_{\dot{r}\dot{r}} & E_{\dot{r}\dot{\phi}_2} \\
		E_{\dot{\phi}_2r} & E_{\dot{\phi}_2\phi_2} & E_{\dot{\phi}_2\dot{r}} & E_{\dot{\phi}_2\dot{\phi}_2}
	\end{bmatrix}	
\end{equation}
where the subscripts indicate partial derivatives. 

Through investigation of Eqs. \eqref{eq:dEdr_eq} - \eqref{eq:dEdphi_eq}, it is quickly clear that,
\begin{equation}
	E_{r\dot{r}} = E_{r\dot{\phi}_2} = E_{\phi_2 \dot{r}} = E_{\phi_2 \dot{\phi}_2} = E_{\dot{r}\dot{\phi}_2} = 0
\end{equation}

The only cross second derivative remaining is with respect to $r$ and $\phi_2$. Looking back at the energy in Eq. \eqref{eq:free_energy}, we see that the only part of the energy which contains both of these variables is the potential. Therefore,
\begin{equation}
	\frac{\partial^2 E}{\partial r \partial \phi_2} = \frac{\partial^2 V}{\partial r \partial \phi_2} = 0
\end{equation}
where the equality to zero at the equilibrium point was shown in Eq. \eqref{eq:dVdrdphi_eq}. We have shown that all of the cross partials are zero, so that $E_{\mathbf{x} \mathbf{x}}$ is a diagonal matrix. The diagonal entries of the matrix are the eigenvalues, and the definiteness of the matrix is found by looking at the sign of the eigenvalues. The diagonal entries are,
\begin{equation}\label{eq:Err}
	E_{rr} = \frac{K^2 \nu}{I_z^3} \left( 2 \nu r_{eq}^2 - \overline{I}_{2_z} \right) -\frac{2 \nu}{r_{eq}^3} \bigg\{ 1 + \frac{3}{r_{eq}^2}\bigg[\left(\overline{I}_{1_z} - \overline{I}_s\right) + C_2^\pm  \bigg] \bigg\}
\end{equation}
\begin{equation}
	E_{\phi_2 \phi_2} = V_{\phi_2 \phi_2} = \pm \frac{3 \nu}{r_{eq}^3} \left(\overline{I}_{2_y} - \overline{I}_{2_x}\right)
\end{equation}
\begin{equation}
	E_{\dot{r}\dot{r}} = \nu
\end{equation}
\begin{equation}
	E_{\dot{\phi}_2\dot{\phi}_2} = \frac{\overline{I}_{2_z} \nu r_{eq}^2}{I_z}
\end{equation}

These results are straight forward to interpret. The velocity partials, $E_{\dot{r}\dot{r}}$ and $E_{\dot{\phi}_2\dot{\phi}_2}$ are always greater than zero. The angle partial, $E_{\phi_2 \phi_2}$, is positive for the equilibrium points at $\phi_2 = 0$ or $\pi$, and is negative for the equilibrium points at $\phi_2 = \pm \pi/2$. The position partial, $E_{rr}$, is much more complicated and can be positive or negative depending on the system parameters and location of the equilibrium point. 

Combined, this tells us that an equilibrium point at $\phi_2 = 0$ or $\pi$ can be a local minimum or a saddle point depending on the sign of $E_{rr}$. An equilibrium point at $\phi_2 = \pm \pi/2$ is always a saddle point. The fact that the only energetically stable solutions are found at $\phi_2 = 0$ solutions was first shown by \cite{scheeres2003}.
In the nominal case, $r^+$ is energetically stable, and is a local minimum in energy. The other equilibrium point, $r^-$, is an energetic saddle, being a local minimum in the radial direction, but a local maximum in the $\phi_2$ direction.

\subsection{Osculating Orbit Elements at Equilibrium}\label{sec:eq_oes}

In considering a planar problem the orbital elements of interest are the semi-major axis, eccentricity, true anomaly, and argument of periapse. The behavior of these elements at equilibrium are discussed in this section.
The other orbital elements, namely inclination and argument of the node are effectively meaningless, and won't be discussed here since they are either constant, undefined, and/or always zero. 

We can derive the Keplerian energy and angular momentum as,
\begin{equation}\label{eq:EK_general}
	E^K = \frac{1}{2}v^2 - \frac{1}{r} = \frac{1}{2}r^2 \dot{\theta}^2 + \frac{1}{2} \dot{r}^2 - \frac{1}{r}
\end{equation}
\begin{equation}\label{eq:HK_general}
	H^K = |\mathbf{H}^K| = |\mathbf{r}\times\mathbf{v}| = r^2 \dot{\theta}
\end{equation}
where $\mathbf{v}$ is the inertial velocity vector. At equilibrium,
\begin{equation}\label{eq:thetadot_eq}
	\dot{\theta} = \frac{K}{I_z} = \sqrt{\frac{1}{r^3}\left[ 1 + \frac{3\left(\overline{I}_{1_z} - \overline{I}_s + C_2^\pm\right)}{2 r^2}\right]}
\end{equation}
which was derived from Eqs. \eqref{eq:thetadot} and by setting \eqref{eq:partEpartr} equal to zero as at equilibrium. The Keplerian energy and angular momentum at equilibrium are then determined as,
\begin{equation}
	E^K = -\frac{1}{2r}\left[ 1 - \frac{3\left(\overline{I}_{1_z} - \overline{I}_s + C_2^\pm\right)}{2 r^2}\right]
\end{equation}
\begin{equation}\label{eq:HK_eq}
	H^K = \sqrt{r\left[ 1 + \frac{3\left(\overline{I}_{1_z} - \overline{I}_s + C_2^\pm\right)}{2 r^2}\right]}
\end{equation}

Then the osculating semi-major axis and eccentricity can be computed in terms of the energy and angular momentum as,
\begin{equation}\label{eq:sma_kep}
	\begin{split}
		a &= \frac{-1}{2 E^K} \\
		&= r \left[ 1 - \frac{3\left(\overline{I}_{1_z} - \overline{I}_s + C_2^\pm\right)}{2 r^2}\right]^{-1}
	\end{split}
\end{equation}
\begin{equation}\label{eq:ecc_kep}
	\begin{split}
		e^2 &= 1 +  2 E^K \left(H^{K}\right)^2 \\
		&= 1 - \left[ 1 - \frac{3\left(\overline{I}_{1_z} - \overline{I}_s + C_2^\pm\right)}{2 r^2}\right] \left[ 1 + \frac{3\left(\overline{I}_{1_z} - \overline{I}_s + C_2^\pm\right)}{2 r^2}\right] \\
		&= \left[ \frac{3\left(\overline{I}_{1_z} - \overline{I}_s + C_2^\pm\right)}{2 r^2}\right]^2
	\end{split}
\end{equation}
so that
\begin{equation}\label{eq:ecc_def_eq}
	e= \frac{3\left(\overline{I}_{1_z} - \overline{I}_s + C_2^\pm\right)}{2 r^2}
\end{equation}

Using Eqs. \eqref{eq:sma_kep} and \eqref{eq:ecc_kep}, we find that at equilibrium we have
\begin{equation}
	a = \frac{r}{1- e}
\end{equation}
Due to the relationship in Eq. \eqref{eq:ecc_kep}, we can see that $e > 0$ at the $\phi=0$ equilibrium points because $\overline{I}_{1_z} - \overline{I}_s \ge 0$ and $C_2^+ > 0$. At the $\phi=90$ equilibrium points, $e$ can be positive, negative, or zero because $C_2^-$ can be negative. This tells us that at equilbrium the system is always locked at periapse or apoapse, depending on the sign of $e$.


However, at a relative equilibria, the orbit rate is given by Eq. \eqref{eq:thetadot_eq}, and is constant. On a Keplerian orbit, this would imply that the orbit is circular, and this orbit rate is identical to the mean motion. In this case the eccentricity is in general non-zero and constant, and the semi-major axis is constant, along with the radius. Combining these results indicates that in fact the true/mean anomaly are constant (equal to 0 or 180$\degree$ as discussed above) and the argument of perigee is precessing at the orbit rate to enforce this condition. 

This can be shown by investigating the evolution of the eccentricity vector, which points to periapse. The eccentricity vector is defined by,
\begin{equation}\label{eq:eccvec_kep}
	\mathbf{e} = \mathbf{v} \times \mathbf{H}^K - \hat{\mathbf{r}}
\end{equation}
and in the secondary fixed frame at equilibrium the eccentricity vector becomes through use of Eqs. \eqref{eq:thetadot_eq} and \eqref{eq:HK_eq}
\begin{equation}
	\mathbf{e} = (r^3\dot{\theta}^2 - 1)\hat{\mathbf{r}} = e \hat{\mathbf{r}}
\end{equation}
where $e$ was defined in Eq. \eqref{eq:ecc_def_eq}. 

The rate of change of the eccentricity vector can be determined by using the transport theorem and resolving in the secondary fixed frame. At the $\phi_2 = 0$ equilibrium point it becomes,
\begin{equation}
	\dot{\mathbf{e}} = e \dot{\theta} \hat{\mathbf{y}}
\end{equation}
and at the $\phi_2 = 90\degree$ equilibrium point it becomes,
\begin{equation}
	\dot{\mathbf{e}} = -e \dot{\theta} \hat{\mathbf{x}}
\end{equation}
Given that in this rotating frame, the inertial rate of change of the unit vectors are
\begin{equation}
	\dot{\hat{\mathbf{x}}} = \dot{\theta} \hat{\mathbf{y}}
\end{equation}
\begin{equation}
	\dot{\hat{\mathbf{y}}} = -\dot{\theta} \hat{\mathbf{x}}
\end{equation}
we can see that the angle between an equilibrium point radius vector and the eccentricity vector is constant; therefore the anomalies (true, mean, and eccentric) are fixed at 0 or 180$\degree$. If we assume that the argument of periapse is measured from a fixed direction in the $\hat{\mathbf{x}}-\hat{\mathbf{y}}$ plane, then we can see that the rate of change of the argument of periapse is exactly equal to $\dot{\theta}$.


In summary, a system which is at equilibrium will appear to an outside observer to be moving on a circular orbit. Due to the stability of the equilibrium points, this will likely only actually occur at the $\phi_2 = 0$ point, which is commonly referred to as a synchronous orbit. If such a system is observed and fitted to Keplerian dynamics only, modeling each body as a point mass (or sphere), the result would be a circular orbit with an eccentricity of zero. The computed semi-major axis would then imply an incorrect $\mu$, and thus an incorrect mass of the system. Therefore we reiterate that it is crucial to account for the non-spherical shape of both bodies when fitting orbits to celestial objects.

For reference and later comparison, the nominal system has a semi-major axis of $a^+ = 9.3269$ and an eccentricity of $e^+ = 8.87\times10^{-3}$ at the $\phi_2 = 0$ equilibrium point. The values at the $\phi_2 = 90\degree$ equilibrium point are $a^- = 9.3270$ and  $e^- = 4.31\times10^{-3}$.




\subsection{Equilibirum Point Variation for Various System Parameters}\label{sec:vary_eq_pts}

The basis for a system to have librational motion is in the properties of the equilibirum points. We have shown in the preceding sections that the equilibrium point at  $\phi_2 = 0$ is spectrally stable and an energetic minimum; therefore this is the equilibrium point about with the system will librate. The equilibrium point at  $\phi_2 = 90\degree$ is spectrally unstable and an energetic saddle, which means systems will generally not stay near this equilibrium point. Before moving on to study the actual trajectories of the system, we first study the locations and properties of the equilibrium points for different systems. The parameters that define these systems are the mass ratio, the angular momentum, and the shapes of the bodies. The effect of varying these parameters are discussed in this section.




First, we investigate how the variation in mass fraction affects the equilibrium points. In order to isolate the effects from varying the mass fraction from the effects of the body shapes, we vary the sizes of the bodies along with the mass fraction to keep the moments of inertia constant as follows. Assuming equal density, the mass fraction is equivalent to the volume fraction,
\begin{equation}
	\nu =  \frac{V_1}{V_1 + V_2} = \frac{R_e^2 R_p}{R_e^2 R_p + \beta \gamma}
\end{equation}
This can be solved for the oblate equatorial radius by using $\chi$ to get,
\begin{equation}
	R_e^3 = \frac{\beta \gamma}{\chi (1 - \nu)}
\end{equation}
Using this equatorial radius, the moments of inertia of the oblate body become,
\begin{equation}
	\overline{I}_{1_z} = \frac{2}{5} R_e^2
\end{equation}
\begin{equation}
	\overline{I}_s = \frac{1}{5} \left( R_e^2 + R_p^2 \right) = \frac{1}{5} R_e^2 \left( 1 + \chi^2 \right)
\end{equation}

This paper is mainly concerned with systems where the ellipsoidal body is the smaller body, which would correspond to $\nu \ge 0.5$. However, for the sake of completeness, we show in this section the locations of the equilibrium points for all values of $\nu$. When the ellipsoidal body is larger, our perspective changes and we think of this in terms of studying the location of orbits of an oblate satellite, instead of the libration of an ellipsoidal satellite. 

\begin{figure}[htb]
\begin{center}
	\includegraphics[width=0.8\textwidth]{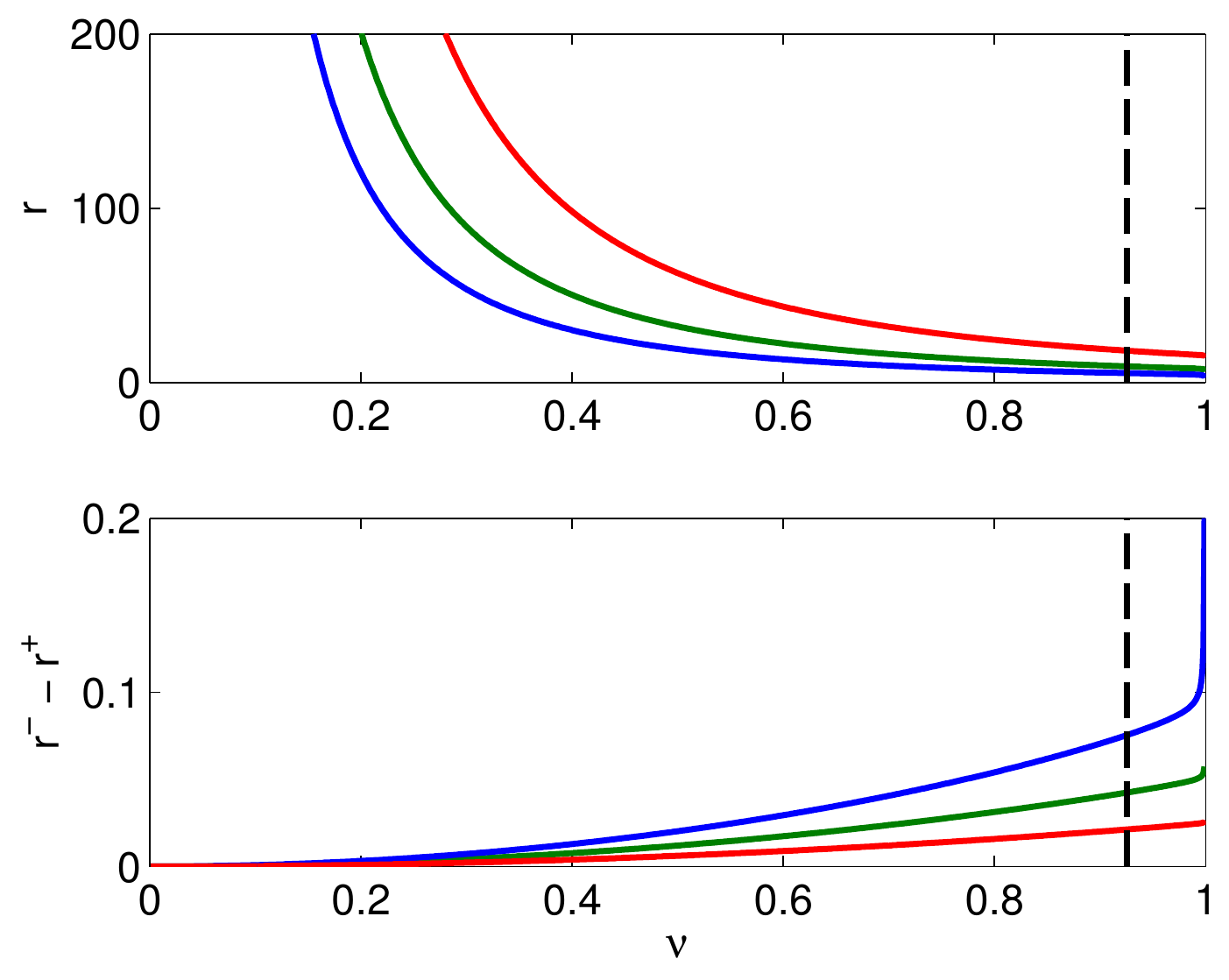}
	\caption{The equilibrium location for varying values of the mass fraction are shown. Three lines are plotted at different values of angular momentum: $K = 2.8382$, the nominal value in green, $K = 3.9625$ in red, and $K = 2.1963$ in blue. The dotted vertical line indicates the nominal value of $\nu = 0.9257$. The lower figure shows the radial difference between the location of the two equilibrium points.}
	\label{fig:mass_frac_eqs}
\end{center}
\end{figure}

The locations of the equilibrium points are shown in Fig. \ref{fig:mass_frac_eqs}. The general behavior is that as the mass fraction becomes smaller, the equilibrium points move to larger radii. The second plot shows the difference in radial distance between the two equilibrium points since they are too close to appear as separate lines in the scale of the first figure. The opposite trend is seen here in that the two equilibrium points are further apart radially as $\nu \to 1$. At the nominal value of angular momentum, the two equilibrium points are 0.04 apart as was given at the end of Section \ref{sec:eqpt_loc}.

The astute reader will notice that this does not seem to match the results from Bellerose \cite{bellerose}; this is because that work included a factor of $\nu$ in their computation of angular momentum we do not include (e.g. $K_{Bellerose} = \nu K$). The value of angular momentum used in this paper is absolute for any system. Also note that Bellerose discusses cases for which multiple equilibria appear; we find these cases as well for small values of $K$. However, since we are interested in studying the outer pair of equilibrium points that have the librational structure, we don't study the inner equilibrium points that may appear. 

\begin{figure}[htb]
\begin{center}
	\includegraphics[width=0.6\textwidth]{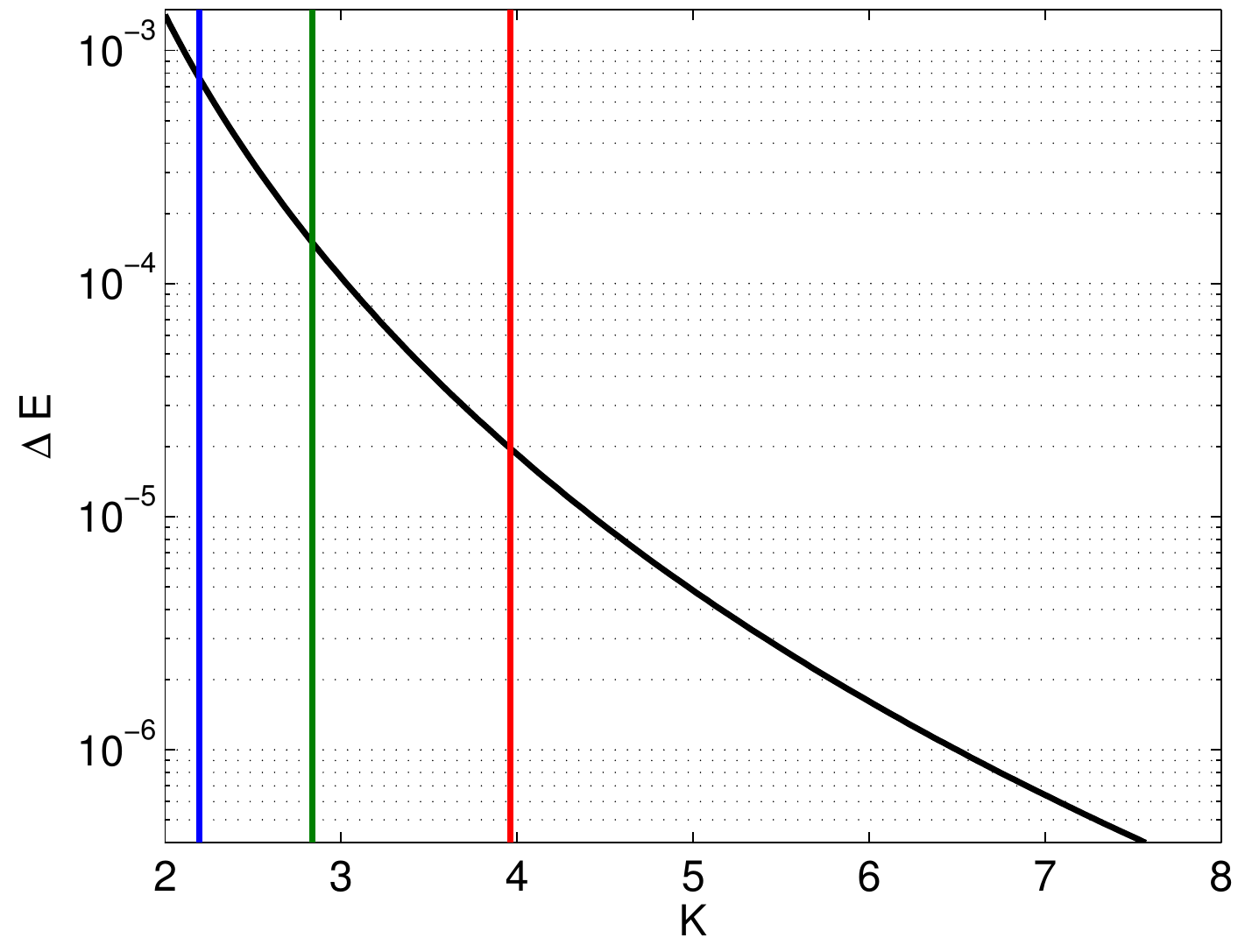}
	\caption{The difference in energy between the stable and unstable equilibrium points, $\Delta E = E_{un} - E_{stab}$, is plotted in black for the nominal system. The three vertical lines indicate the three different values of angular momentum, classified by the same color scheme as used in Fig. \ref{fig:mass_frac_eqs}.}
	\label{fig:angmom_dele}
\end{center}
\end{figure}

The main result of varying the angular momentum as illustrated in Fig. \ref{fig:mass_frac_eqs} is that for higher values of $K$, the equilibrium points move to larger radii. This also has the effect of making the absolute difference in energy levels between the equilibrium points smaller, as is seen in Fig. \ref{fig:angmom_dele}.


The shape of the bodies can have a number of effects on the equilibria. In terms of the oblateness of body 1, the more oblate the body, the larger the difference $\overline{I}_{1_z} - \overline{I}_s$ becomes. This generally means that the location of the equilibrium points moves outward with a more oblate body. The effect on the $\phi_2 = 0$ and $\phi_2 = 90\degree$ equilibria are roughly the same.

The effects of the secondary shape can be more varied due to the fact that the main asymmetry in this problem is due to the ellipsoid shape/moments of inertia. The effects are encompassed by the differences in $C_2^+$ and $C_2^-$, which can be changed due to variations of any of the three moments of inertia. Recall that changing $\overline{I}_{2_z}$ also changes the value of the system moment of inertia, $I_z$. 

It is interesting to recall that the semi-major axis and eccentricity depend largely on the quantity $(\overline{I}_{1_z} - \overline{I}_s + C_2^\pm)$. While the equilibrium point at $\phi_2 = 0$ is always at periapsis, the $\phi_2 = 90\degree$ equilibrium can be at periapse, apoapse, or on a circular orbit depending completely on the moments of inertia of the bodies. Also, due to the same quantity, the $\phi_2 = 90\degree$ equilibrium point can become a local maximum in the radial direction (see Eq. \eqref{eq:Err}), although the point will still be an energetic saddle in general.

In all cases studied here, the results of the stability of the nominal system equilibria holds for the outer set of equilibria points. Namely, the outer equilibrium point at $\phi_2 = 0$ is spectrally stable and an energetic minima, while the outer equilibrium point at $\phi_2 = 90\degree$ is spectrally unstable and an energetic saddle.

\clearpage

\section{Sufficient Condition for Bounded Motion}\label{sec:suff_bounded}

When the system is not at equilibrium, the trajectory will evolve in $r-\phi_2$ space according to the equations of motion given in Eqns. \eqref{eq:2dof_rddot} - \eqref{eq:2dof_phi2ddot}. As this system is non-integrable, we are particularly interested in finding conditions which classify or restrict the trajectories that will occur. In this section, we present a sufficiency condition for bounded motion, which is taken to mean that the libration angle will be less than 90$\degree$ for all time. We then investigate how the sufficiency condition changes for different system parameters. Finally, we look at some properties of the system trajectories that meet the bounded sufficiency conditions.

\subsection{Bounded Motion}\label{sec:zerovel}

The ZVCs can be used to investigate if the current system can become unbounded by reaching a phi = 90, which leads to the sufficiency condition for bounded motion:

\begin{theorem} \textbf{Sufficient Condition for Bounded Motion}\\
\label{thm:suff_cond_bounded}
Given a system with angular momentum $K$ so that a stable equilibrium point exists at $\phi_2 = 0$, the librational motion is bounded ($\phi_2 <90\degree$ always) if $E < E^-$. 
\\
\textbf{Proof:}
The relationship for the free energy of the system, Eq. \eqref{eq:free_energy}, can be rearranged to become
\begin{equation}\label{eq:free_energy_rearranged}
	E - \frac{1}{2}\frac{K^2}{I_z(r)} - V(r,\phi_2) =  \frac{1}{2}\nu \dot{r}^2 + \frac{1}{2}\frac{\overline{I}_{2_z}\nu r^2\dot{\phi}_2^2}{I_z(r)}
\end{equation}
The right hand side of Eq. \eqref{eq:free_energy_rearranged} is always positive, therefore we can state the relationship for a zero-velocity curve (ZVC),
\begin{equation}\label{eq:zero_vel}
	E - \frac{1}{2}\frac{K^2}{I_z(r)} - V(r,\phi_2) \ge 0
\end{equation}
or equivalently,
\begin{equation}\label{eq:E_zero_vel}
	E \ge \frac{1}{2}\frac{K^2}{I_z(r)} + V(r,\phi_2)
\end{equation}

The stable equilibrium is the minimum  energy location, and when $E = E^+$, the ZVC defines only the stable equilibrium. As $E$ is increased, the ZVC will encompass larger areas in $r-\phi_2$ space. If $E$ is increased to $E^-$, the ZVC will touch the unstable equilibria, and because $E^-$ is the minimum energy radius with $\phi_2 = 90\degree$ (see Section \ref{sec:eq_pt_energy}), this is the minimum energy at which the ZVC allows the system to reach $\phi_2 = 90\degree$.

That this is a sufficient, but not necessary, condition is shown in Section \ref{sec:openZVCs}.
\\
$\Box$
\end{theorem}

A given system will have some area in the $r-\phi_2$ space which it can reside in based upon the free angular momentum and energy. Given a value for the free angular momentum of the system, the entire phase space can be mapped with varying energy levels determined by Eq. \eqref{eq:E_zero_vel}. The easiest way to visualize this is by looking at the system from a secondary fixed frame, as shown in Fig. \ref{fig:secondary_frame}. For a given value of free energy for the system, there will be some area to which the primary is constrained to reside. Note that due to the fact that this relationship is an inequality, the primary can be anywhere inside the free energy level, not only on the surface. Therefore this relationship clearly doesn't solve the equations of motion to tell us what the state is at any given time, but it does tell us absolutely that the state is always inside the area bounded by that free energy. 

\begin{figure}[htb]
\begin{center}
	\includegraphics[width=0.6\textwidth]{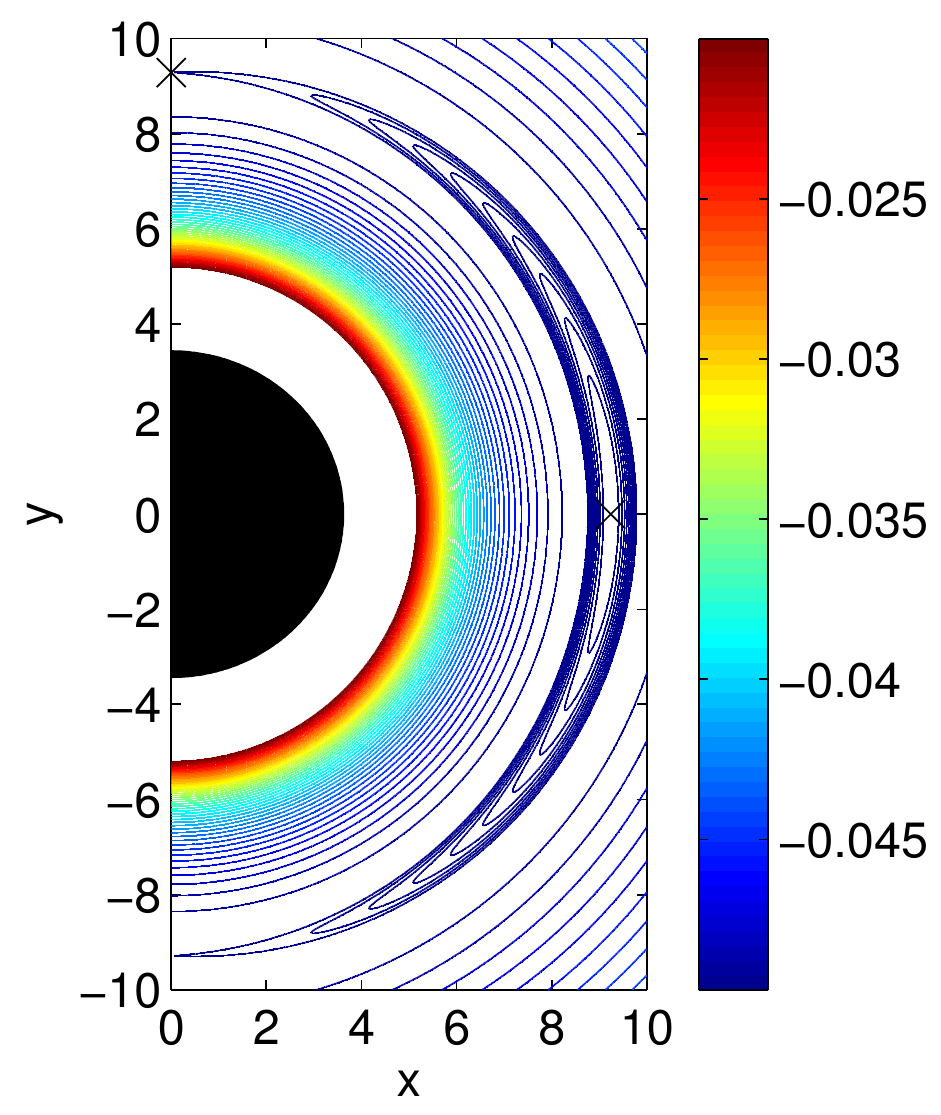}
	\caption{Phase space for the nominal test system described in Table \ref{tab:kw4_params}. The closed ZVCs are at increments of energy of 10\% of $\delta E$, so that the first close curve has energy $E = E^+ + 0.1 \delta E$. This first curve limits the libration angle to a maximum of $\phi_2 = 18.4\degree$. }
	\label{fig:kw4_nom_zvcs}
\end{center}
\end{figure}

\begin{figure}[htb]
\begin{center}
	\includegraphics[width=1\textwidth]{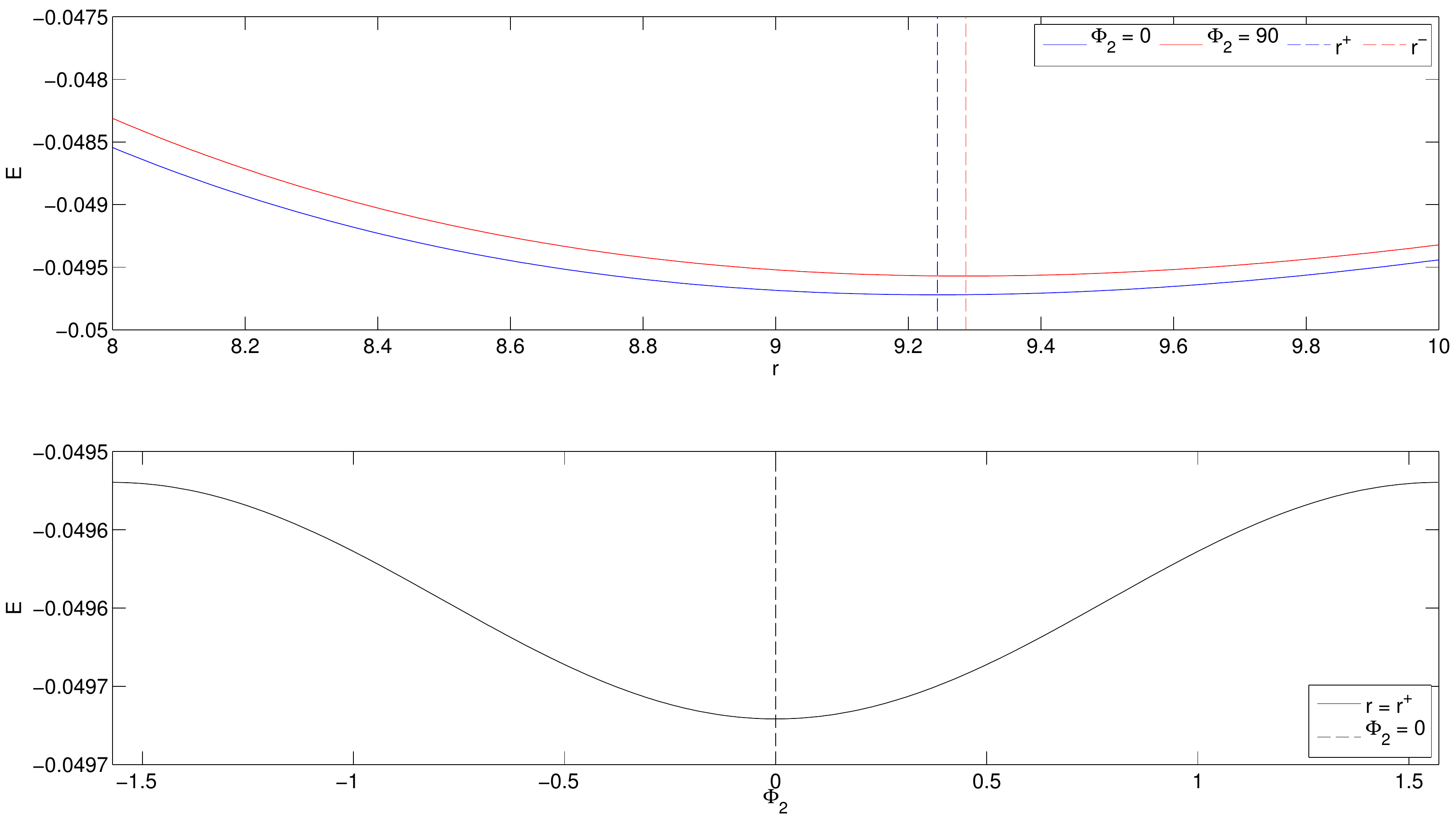}
	\caption{Variation of the energy near the equilibrium points for the nominal system. The upper plot shows the variation with respect to the radial distance at $\phi_2 = 0$ and $\phi_2 = 90\degree$. The location of the equilibria are marked by the vertical dotted lines. The lower plot shows the variation in the energy at the stable equilibrium point radius as $\phi_2$ is varied. The points at $\pm \pi/2$ on this plot are actually equivalent to where the blue vertical dotted intersects the red line in the upper plot, not precisely at the unstable equilibrium radius. However, the same trend of being a local maximum with respect to variation in $\phi_2$ holds at the unstable equilibrium point radius.}
	\label{fig:energy_cross}
\end{center}
\end{figure}

Theorem \ref{thm:suff_cond_bounded} is verified graphically for the nominal system in Figures \ref{fig:kw4_nom_zvcs} and \ref{fig:energy_cross}. The nominal system zero-velocity curves are shown in Fig. \ref{fig:kw4_nom_zvcs}. This is the typical structure for the zero-velocity curves seen in most situations where the libration between the two outer equilibrium points is being examined. In these types of plots, the stable minimum energy equilibrium point is along the x-axis ($\phi_2 = 0$), while the unstable equilibrium point is on the y-axis ($\phi_2 = 90\degree$). As the energy is increased from the minimum at the x-axis equilibrium point, the zero-velocity curves allow for larger libration angles until the unstable equilibrium energy is reached. At energies above the unstable equilibrium point energy the ZVCs open (becoming two separate ZVCs) and the secondary is free to circulate in this area. The black shape around the origin is the projection of the ellipsoid plus the equatorial radius of the oblate body; if the energy is high enough so that this region is inside the ZVC bounds then an impact between the bodies is possible. The color bar lists the values of energy ($E$) corresponding to each ZVC. Recall from Section \ref{sec:eqpt_loc} that the stable equilibrium point has energy $E^+ = -0.0497$ and the unstable equilibrium has an energy of $E^- = -0.496$, where the difference between the two is $\delta E = 1.500\times10^{-4}$. 

In order to make clear the behavior of the energy in the vicinity of the equilibrium points, we plot what is effectively a cross section of Fig. \ref{fig:kw4_nom_zvcs} in Fig. \ref{fig:energy_cross}. This clearly shows the variation of the energy in the radial and circumferential directions. It is clear that both of these equilibrium points are minima in the $r$ direction, however only the x-axis equilibrium point is also a minima in the $\phi_2$ direction. 

It should be noted that the same behavior is seen around the stable equilibrium point at $\phi_2 = 180\degree$, and Theorem \ref{thm:suff_cond_bounded} can be applied there as well. In this paper, we generally only look at the $\phi_2 = 0$ equilibrium point for clarity.

\begin{figure}[htb]
\begin{center}
	\includegraphics[width=0.6\textwidth]{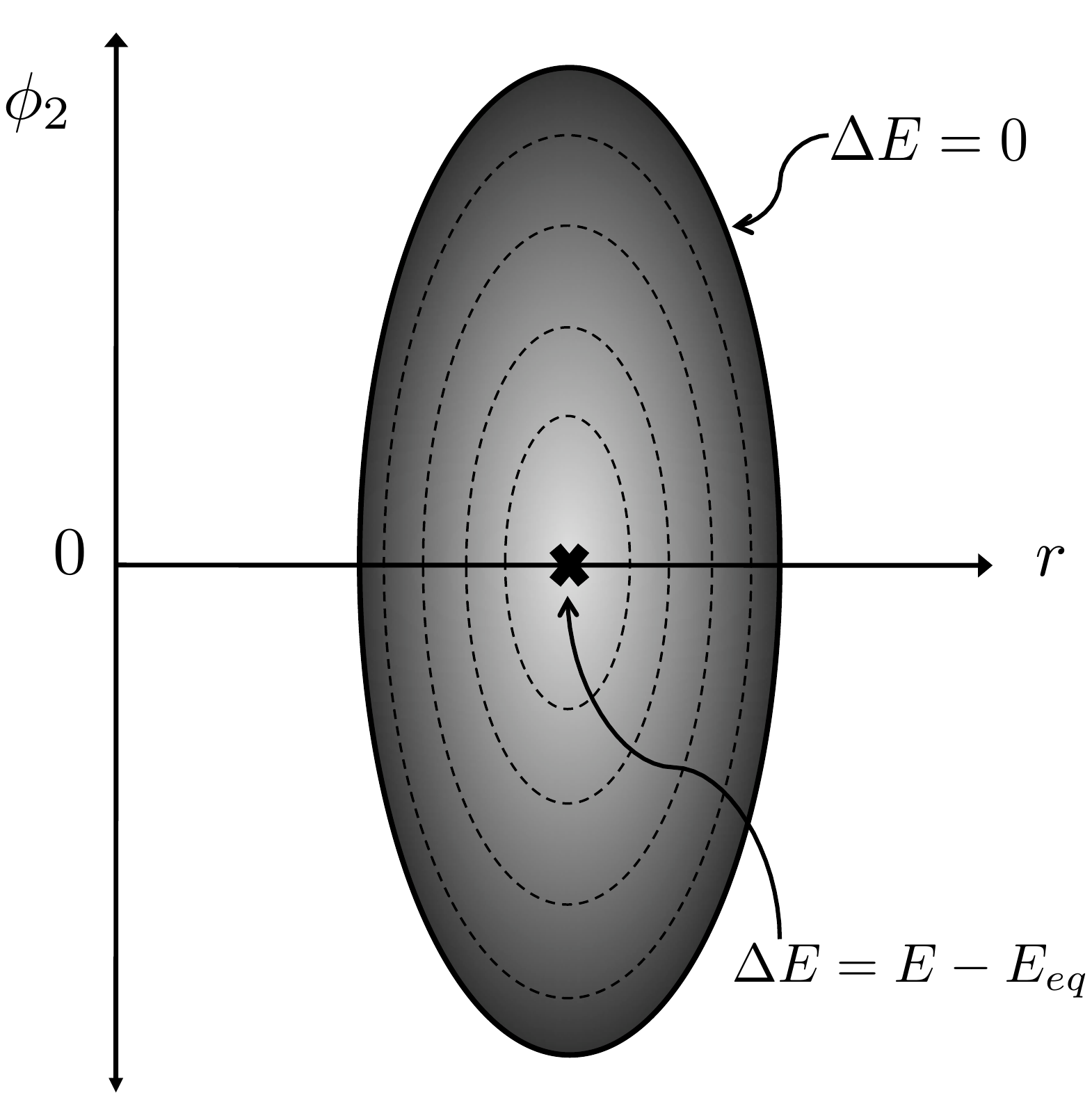}
	\caption{The zero-velocity curve for the given energy, $E$, has zero excess energy, as shown. Moving in toward the equilibrium point increases $\Delta E$ until reaching the equilibrium point (marked by the x), where the excess energy is at a maximum. Intervening zero-velocity curves are drawn with dotted lines.}
	\label{fig:deltaecartoon}
\end{center}
\end{figure}

Consider a zero velocity curve with the free energy, $E$, and the free angular momentum, $K$. When the inequality in Eq. \eqref{eq:zero_vel} is precisely equal to zero, we know that $\dot{r}=0$ and $\dot{\phi}_2 = 0$; all of the free kinetic energy in the system has been transferred to the potential energy. At any point inside the zero velocity curve(s) defined by $E$, the inequality will be greater than zero as there is an excess of energy defined from the ZVC definition, 
\begin{equation}
	\Delta E (r,\phi_2)= E - \frac{1}{2}\frac{K_0^2}{I_z} - V(r,\phi_2) =  \frac{1}{2}m\dot{r}^2 + \frac{1}{2}\frac{I_{2_z}mr^2\dot{\phi}_2^2}{I_z}
\end{equation}
This situation is depicted in the cartoon shown in Figure \ref{fig:deltaecartoon}. Returning to Eq. \eqref{eq:free_energy_rearranged}, it is clear that in this case, the left hand side is greater than zero, and therefore either $|\dot{r}|>0$, $|\dot{\phi}_2| > 0$, or some combination of the two.

Note that the excess energy, $\Delta E$ is not constant on a given trajectory as it varies with the position variables. The maximum value that $\Delta E$ can reach is at the $\phi_2 = 0$ equilibrium point since this is the minimum potential energy location. At any given point on a trajectory, the excess energy can be used to determine the range of possible values of the velocities based on the kinetic energy in the radial and spin components. 

The split in the kinetic energy will be determined by the factor $\kappa$, which dilineates what percentage of the excess energy goes into the radial velocity. The scale factor is therefore bounded such that
\begin{equation}
	0 \le \kappa \le 1
\end{equation}
This means that the radial velocity and ellipsoid rotation rate are defined as,
\begin{equation}\label{eq:rdot_delE}
	\dot{r}^2 = \frac{2 \kappa \Delta E}{\nu}
\end{equation}
\begin{equation}\label{eq:phi2dot_delE}
	\dot{\phi}_2^2 = \frac{2(1-\kappa)\Delta E I_z}{\overline{I}_{2_z}\nu r^2}
\end{equation}

\begin{figure}[htb]
\begin{center}
	\includegraphics[width=1\textwidth]{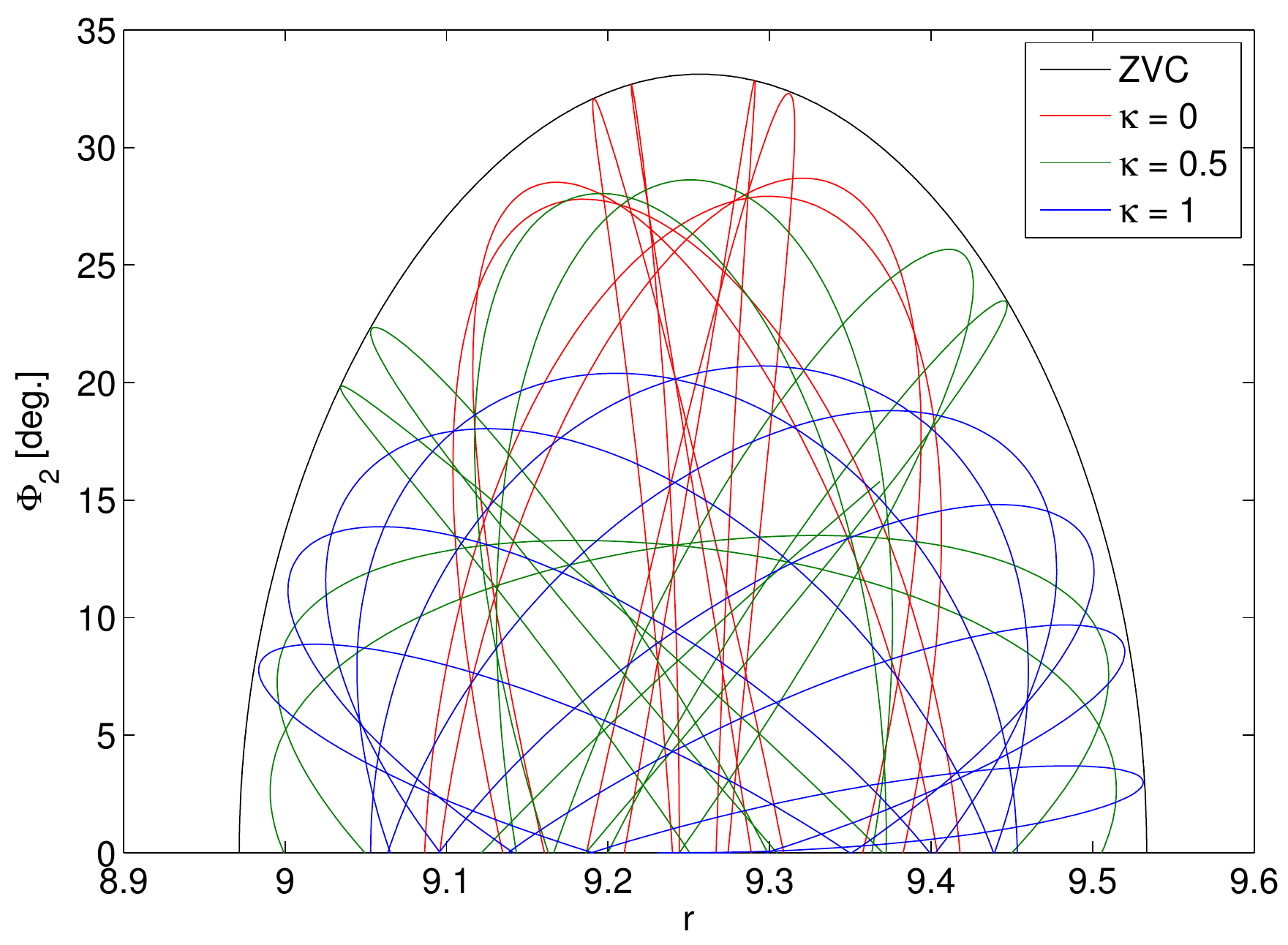}
	\caption{Three trajectories for the nominal system with an energy of $E = E^+ + 0.3 \delta E$, which means the zero-velocity curve is closed, as shown in black. All three trajectories are initiated at $r^+$ with $\kappa = 0$, $\kappa = 0.5$,  and $\kappa = 1$.}
	\label{fig:traj_closed_zvc}
\end{center}
\end{figure}

Fig. \ref{fig:traj_closed_zvc} shows the trajectories of three cases for the nominal system with a closed ZVC, each beginning with a different amount of excess energy in the radial and angular forms of the kinetic energy. These results demonstrate that the ZVC is indeed accurate, and that the trajectories stay bound within this area of $r$-$\phi_2$ space. Furthermore, depending on the initial velocities, the area within the ZVC that is filled by each trajectory is limited further; the trajectory with all the initial velocity in the radial direction (plotted in blue) explores the phase space the furthest in the radial direction, but not as far in the angular direction. The opposite is true for the case with all of the initial velocity in the angular direction, which is plotted in red. 

The ZVCs are a crucial property of the system which limit the phase space in which the system can reside. Since the conditions in Theorem \ref{thm:suff_cond_bounded} are so important, we investigate how the ZVCs vary when the system parameters are changed.

\clearpage

\subsection{Parameter Influence on Zero-Velocity Curves}\label{sec:kw4_zvcs}

Eq. \eqref{eq:zero_vel} shows us that the zero-velocity curves are defined first and foremost by the angular momentum and energy in the system. However, the exact form of the zero-velocity curves depends on all the system parameters such as the moments of inertia of the bodies, and the mass fraction of the system. In this section we explore how the nominal zero-velocity curves shown in Fig. \ref{fig:kw4_nom_zvcs} change when the system parameters are varied.

\subsubsection{Variation Due to Mass Ratio}\label{sec:vary_nu_zvc}

In this section, we look at the ZVCs for three different mass ratio systems pulled from Section \ref{sec:vary_eq_pts}. By comparing these different mass ratio systems to our nominal system in Fig. \ref{fig:kw4_nom_zvcs}, we get a good idea of the effect of changing the mass ratio on the ZVC structure. 

\begin{figure}[htb]
\begin{center}
	\includegraphics[width=0.6\textwidth]{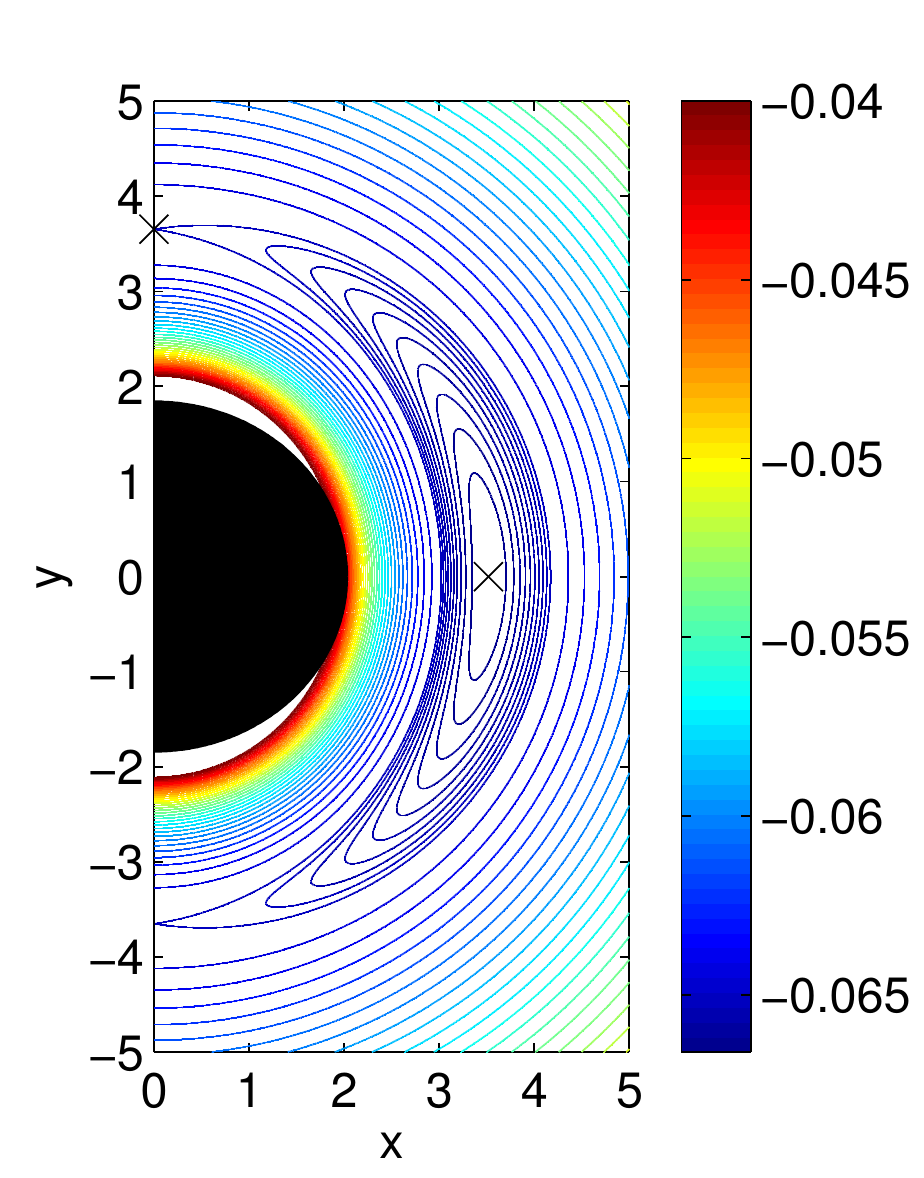}
	\caption{Zero-velocity curves for the scaled system with $\nu=0.5$ and $K = 1.0007$. The first 10 zero-velocity curves surrounding the stable equilibrium point are plotted in increments of energy of $0.1\delta E$.}
	\label{fig:kw4_diffnu}
\end{center}
\end{figure}

The first case is the nominal system scaled to a mass ratio of $\nu = 0.5$, shown in Fig. \ref{fig:kw4_diffnu}. Even with the scaling of the bodies to equal masses, the structure of the ZVCs is very similar. The values of the energy and the location of the equilibrium points have changed due to a change in angular momentum, but the interesting point is that the behavior of the system between and around the equilibrium points found in our nominal case of a relatively small ellipsoid orbiting a larger oblate body appears to hold all the way to equal masses.

\begin{figure}[htb]
\begin{center}
	\includegraphics[width=0.6\textwidth]{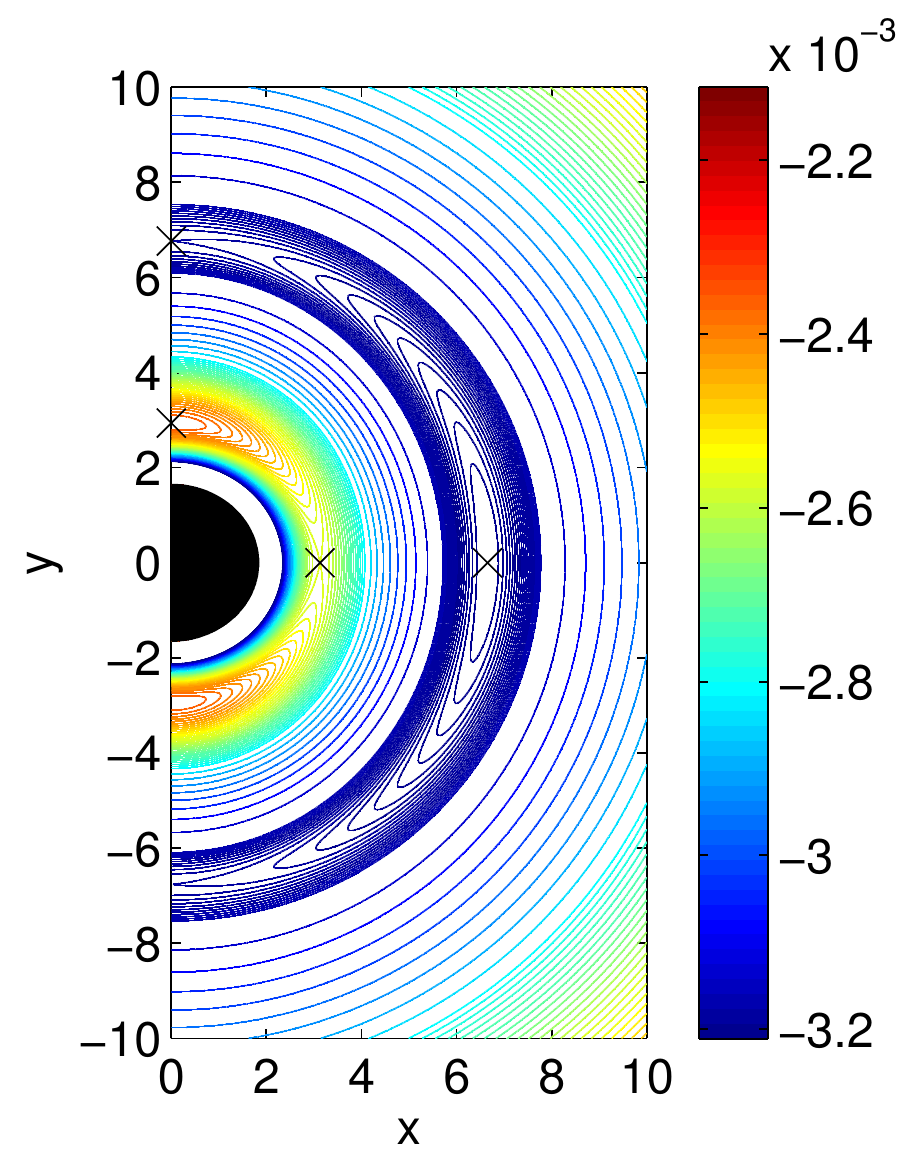}
	\caption{The zero-velocity curves for the system with $\nu = 0.05$ and $K = 0.1479$. The minimum energy point is the outer equilibrium point on the x-axis, however low energy trajectories can exist inside the inner pair of equilibria, although they would likely quickly impact.}
	\label{fig:mass_frac_p05}
\end{center}
\end{figure}

The second case we look at is with a mass ratio of $\nu = 0.05$ shown in Fig. \ref{fig:mass_frac_p05}. This system was chosen because it illustrates a case discussed by Bellerose \cite{bellerose} where there are two equilibrium points on the x-axis, and the outer point is spectrally stable and an energetic minima, while the inner point is spectrally unstable and an energetic saddle. It turns out for these systems that there are also two equilibria on the y-axis. As expected from the systems studied so far, the outer y-axis equilibria is spectrally unstable and an energetic saddle. It is interesting to find that the inner y-axis equilibria is spectrally stable while being an energetic maxima. 

This is interesting, as it implies that as $\nu \to 0$, so that the system being studied is that of a small sphere/oblate body orbiting a large ellipsoid, there will be stable orbits along the y-axis around three times the ellipsoid semi-major axis. The orbits that have the small body near the same radius on the x-axis will be unstable.


\begin{figure}[htb]
\begin{center}
	\includegraphics[width=0.6\textwidth]{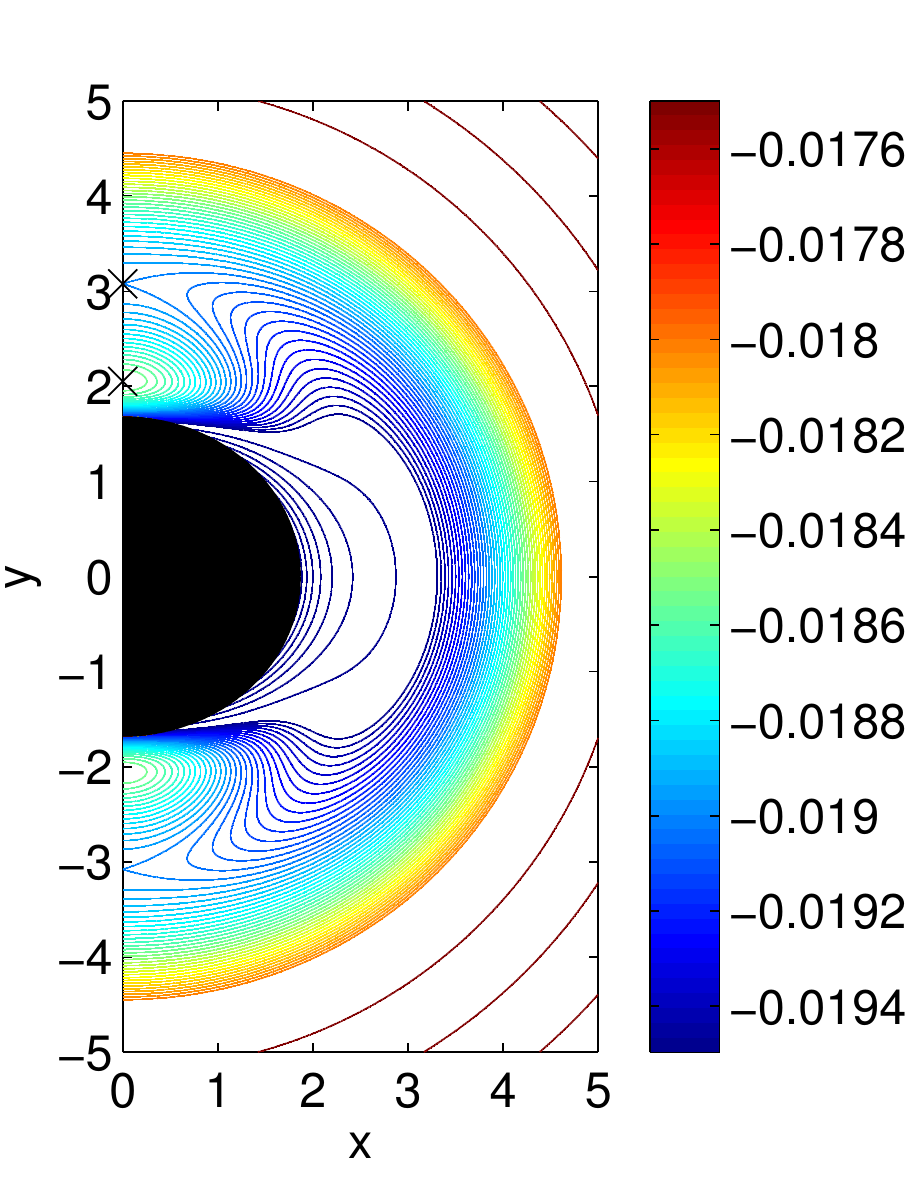}
	\caption{The zero-velocity curves for the system with $\nu = 0.15$ and $K = 0.3211$. At this configuration, the x-axis equilibria have disappeared and the path to impact along the x-axis is open for all energies.}
	\label{fig:mass_frac_p15}
\end{center}
\end{figure}

The final case we show is that with a mass ratio of $\nu = 0.15$, shown in Fig. \ref{fig:mass_frac_p15}. The y-axis equilibrium points keep the same general behavior as those in Fig. \ref{fig:mass_frac_p05}, however in this case the x-axis equilibria have vanished. This appears to make low energy impacts possible between the two bodies starting from basically anywhere in the system; in other words the energy decreases all the way to the surface along the x-axis, and since a given trajectory can explore the region between ZVCs for its energy, they can reach the surface along the x-axis.

In the cases in Figs. \ref{fig:mass_frac_p05} and \ref{fig:mass_frac_p15}, it is important to note that there may be some non-trivial errors in the values obtained for the closer equilibria due to the fact that we are using a second order expansion for the potential of the bodies. If accurate results are needed for these cases, we suggest calculating the outcomes with the exact potential expression used by Bellerose \cite{bellerose}. However, the qualitative results match these previous investigations with the exact potential. Furthermore, with the main interest of this paper being the outer set of equilibrium points, these points are far enough from the bodies that the differences between our second order approximation and the exact potential should be small enough to make little difference. 

\clearpage

\subsubsection{Variation Due to Angular Momentum}

\begin{figure}[htb]
\begin{center}
	\includegraphics[width=0.48\textwidth]{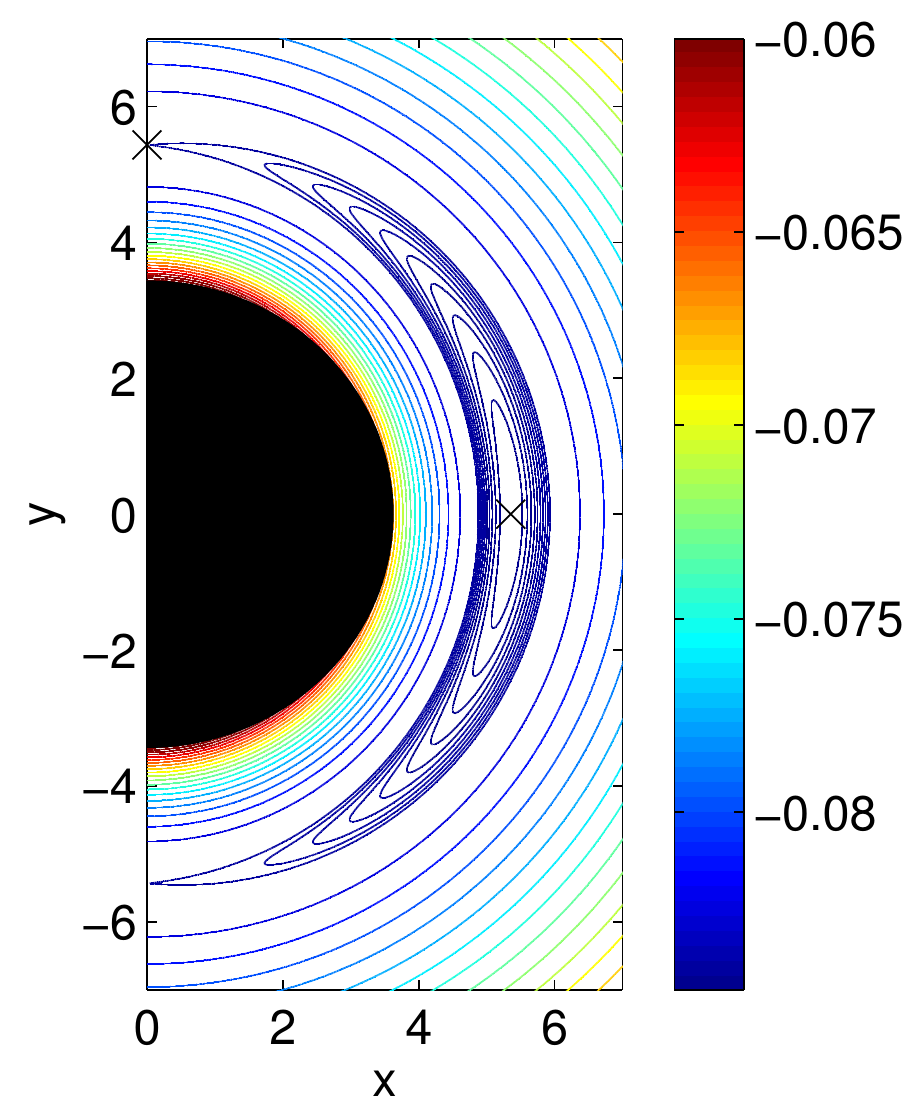}
	\includegraphics[width=0.48\textwidth]{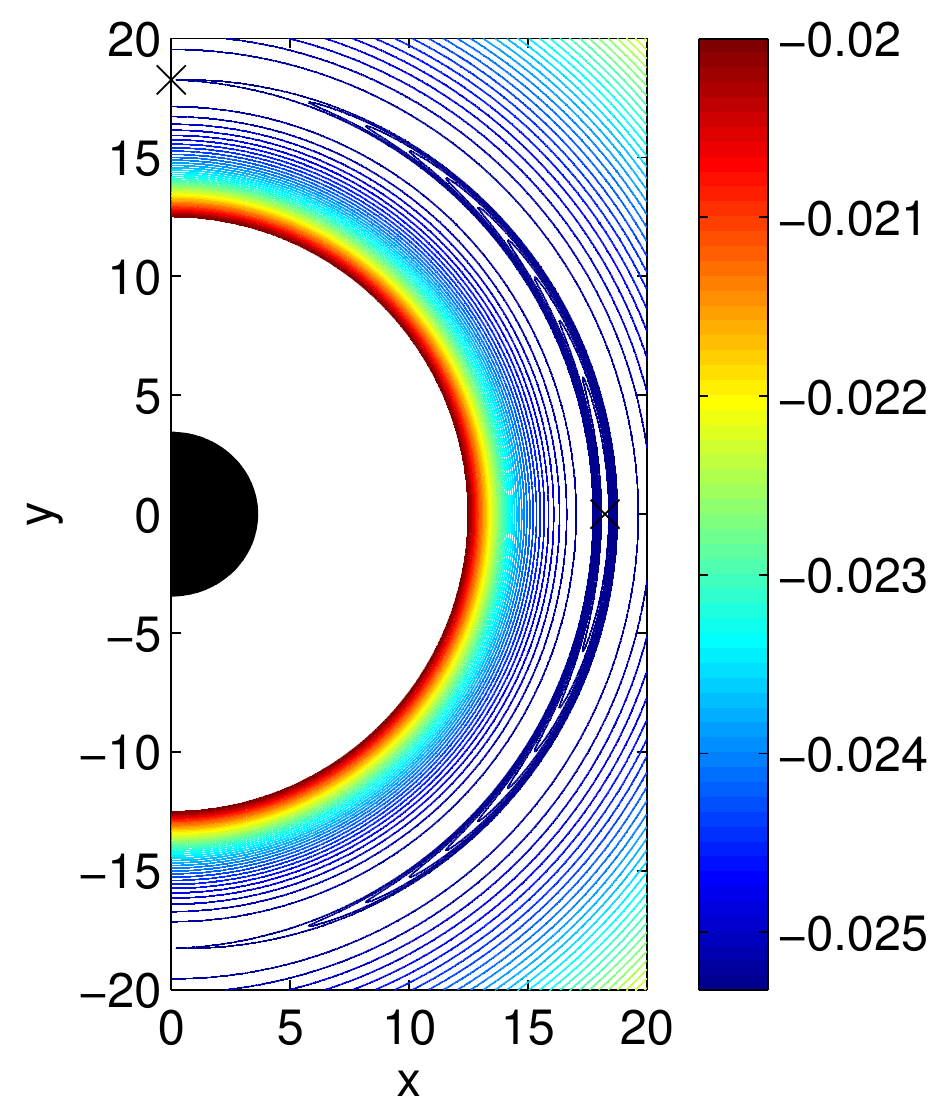}
	\caption{The ZVC structure for two cases of the nominal system with different angular momentum values. The case on the left has a lower angular momentum than the nominal at $K = 2.1963$, while the right figure has a higher value of $K = 3.9625$. Recall that the nominal angular momentum is $K = 2.8382$. The main effect of changing the angular momentum is to move the equilibria out for higher values, and in for lower values.}
	\label{fig:zerovel_1p5}
\end{center}
\end{figure}

As was alluded to in Section \ref{sec:vary_eq_pts}, the main effect of changing the angular momentum is to move the equilibria in or out. This effect is also seen on the ZVC structure in Fig. \ref{fig:zerovel_1p5}; the ZVCs have the same structure as the nominal case, they are basically shifted in or out. It was demonstrated in Section \ref{sec:vary_nu_zvc} that a combination of low $\nu$ and $K$ can cause a second set of equilibrium points to appear, however this situation doesn't happen for larger values of $\nu$ due to the fact that these solutions to Eq. \eqref{eq:equil_pts} are less than the combined radii of the bodies, and usually less than the radius of the secondary.

\clearpage

\subsubsection{Variation Due to Body Shapes}

\begin{figure}[htb]
\begin{center}
	\includegraphics[width=1\textwidth]{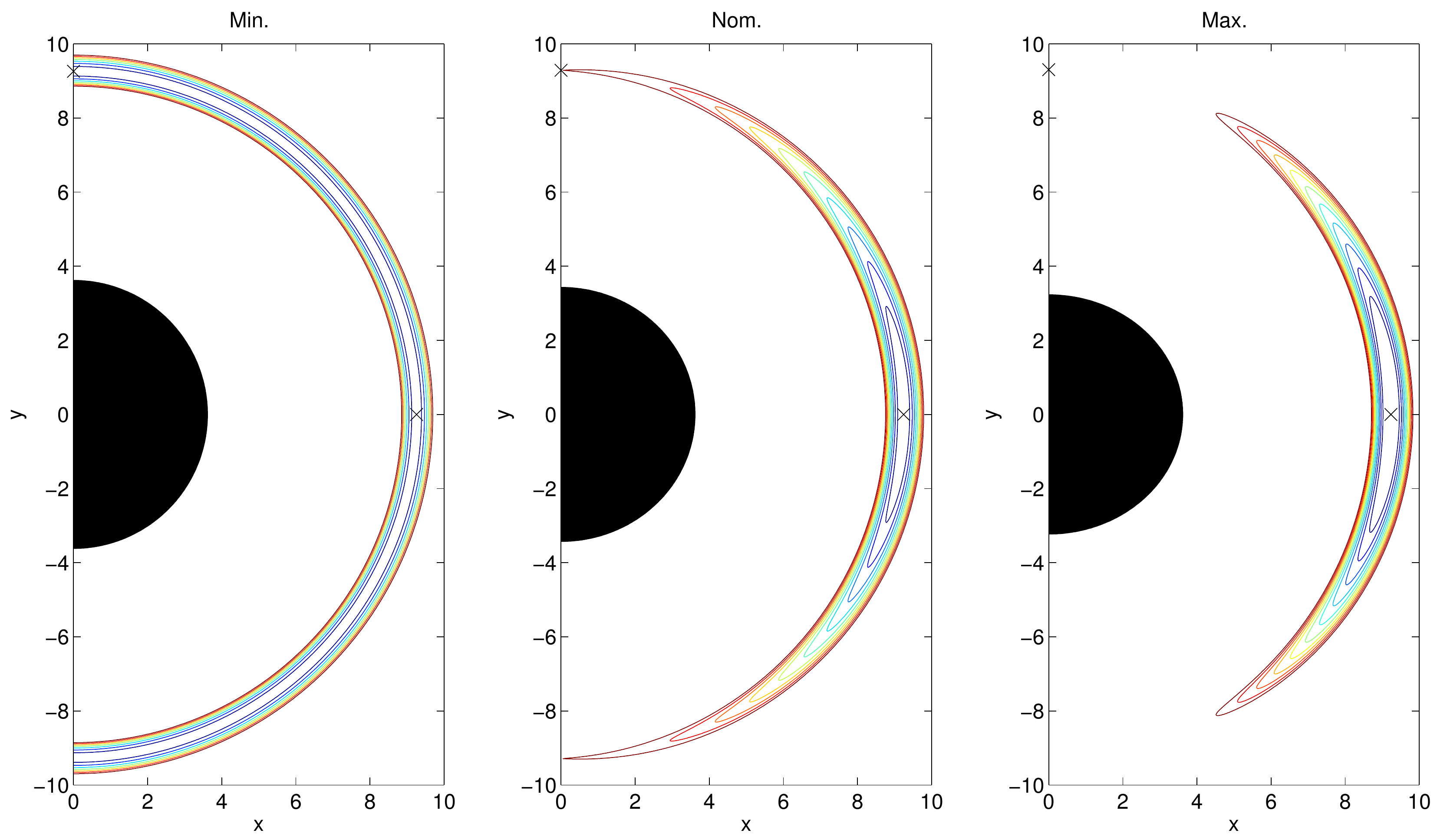}
	\caption{Zero-velocity curves for the nominal system with three different values of the ellipsoid middle moment of inertia. From left to right, the minimum value is $\overline{I}_{2_y} = $, the nominal value is $\overline{I}_{2_y} = $, and the maximum value is $\overline{I}_{2_y} = $. Each of the three figures has ten ZVCs plotted at the same energy levels, clearly illustrating that it is easier to cause an ellipsoid with a smaller difference $(\overline{I}_{2_y} - \overline{I}_{2_x})$, the minimum case, to begin circulating.}
	\label{fig:kw4_diffBA}
\end{center}
\end{figure}

The degree to which the secondary is triaxial plays a large role in the determination of the zero-velocity curves. This is investigated by tweaking the nominal case with two different values of the intermediate moment of inertia, $\overline{I}_{2_y}$, first so that it is just greater than the $\overline{I}_{2_x}$, and second so that it is just less than $\overline{I}_{2_z}$. The main contribution of modifying the ellipsoid moments of inertia, however, is that the difference $(\overline{I}_{2_y} - \overline{I}_{2_x})$ controls the dependence of the potential and therefore the ZVC on the libration angle. The ZVCs for these two cases are compared to the nominal case in Fig. \ref{fig:kw4_diffBA}. The ZVCs of the same color are at the same energy across the three figures, making it easy to see that as the intermediate moment of inertia is made smaller, the body can circulate much easier.

Note that, as discussed in Section \ref{sec:vary_eq_pts}, changing the oblateness of the other body will move the location of the equilibrium points in or out. However, there is very little difference beyond this radial shift on the ZVCs, so this effect isn't illustrated. 

\clearpage

\subsection{Time-Variation of Osculating Orbital Elements}

The orbital elements of a non-equilibrium trajectory are no longer useful as metrics for the system because they change constantly throughout the course of the trajectory. The groundwork for calculating the orbital elements has already been laid in Sections \ref{sec:eq_oes} and \ref{sec:zerovel}. At any given point in $r-\phi_2$ space, given the energy and angular momentum, the possible values of $\dot{r}$ and $\dot{\phi}_2$ can be calculated using Eqs. \eqref{eq:phi2dot_delE} and \eqref{eq:rdot_delE}. The Keplerian energy and angular momentum can then be calculated from Eqs. \eqref{eq:HK_general} and \eqref{eq:EK_general}, which in turn allows us to calculate the semi-major axis and eccentricity with Eqs. \eqref{eq:sma_kep} and \eqref{eq:ecc_kep}. Note that for these calculations, the sign of $\dot{r}$ makes no difference, however the sign of $\dot{\phi}_2$ directly effects the value of $\dot{\theta}$, which then changes the value of the Keplerian energy and angular momentum. 

The eccentricity vector can again be calculated through Eq. \eqref{eq:eccvec_kep}. Note that in this case, $\dot{r}$ is no longer zero in general, so that 
\begin{equation}
	\mathbf{e} = (r^3 \dot{\theta}^2 - 1) \hat{\mathbf{r}} - r^2 \dot{r} \dot{\theta} \hat{\theta}
\end{equation}
where $\hat{\theta} = \hat{\mathbf{H}}^K \times \hat{\mathbf{r}}$. This implies that whenever $\dot{r} \ne 0$, the eccentricity vector no longer lies along the $\hat{\mathbf{r}}$ direction, and therefore the system is no longer at an apse.

\begin{figure}[htb]
\begin{center}
	\includegraphics[width=1\textwidth]{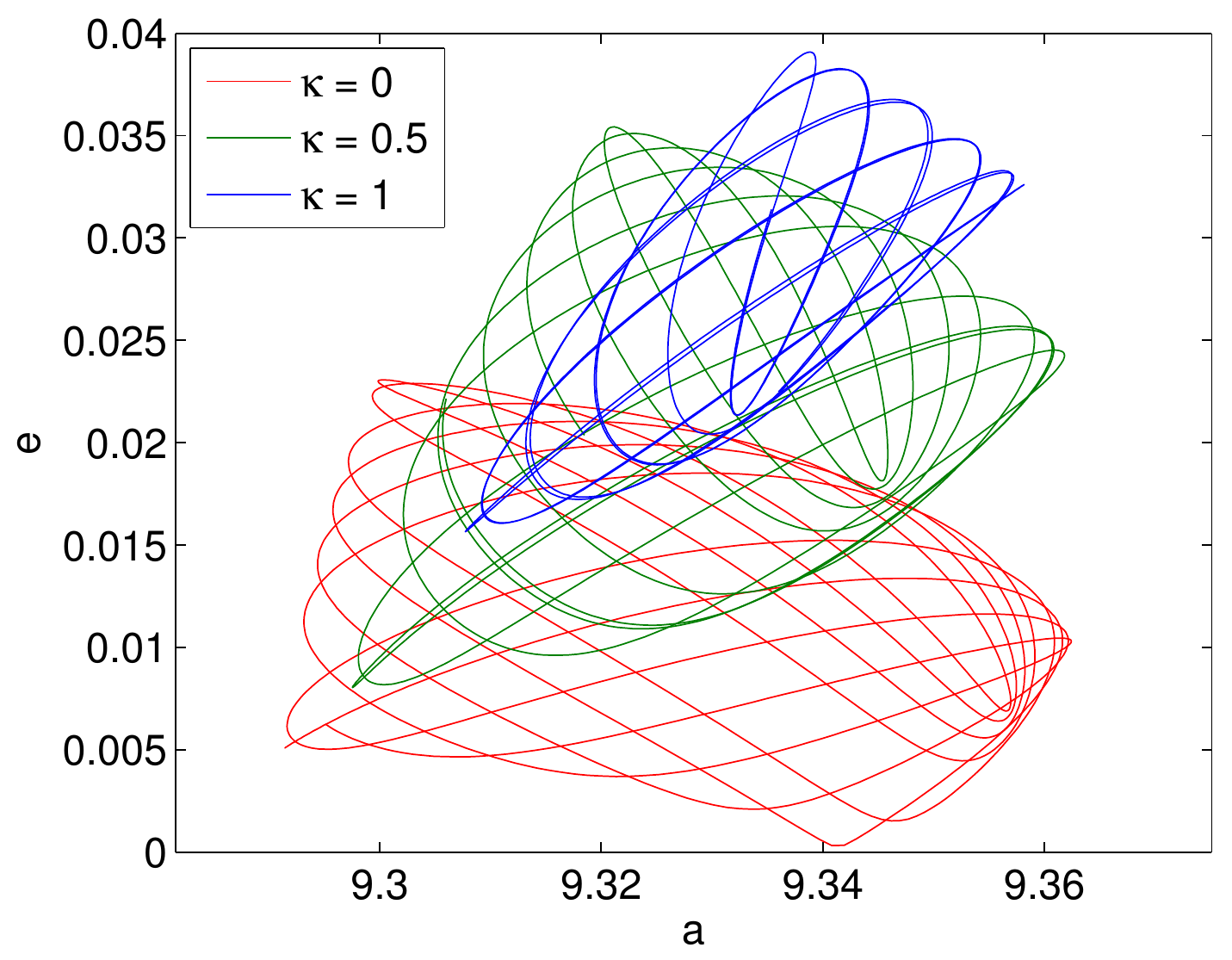}
	\caption{Eccentricity versus semi-major axis for the three trajectories shown in Fig. \ref{fig:traj_closed_zvc}.}
	\label{fig:oes_closed_zvc}
\end{center}
\end{figure}

To illustrate the behavior the semi-major axis and eccentricity on a non-equilibrium trajectory, the corresponding values of the semi-major axis and eccentricity for each of the trajectories in Fig. \ref{fig:traj_closed_zvc} are shown in Fig. \ref{fig:oes_closed_zvc}. As with the radial-angular phase space, each trajectory occupies a distinct region of $a-e$ space despite the fact that they each have the same energy.

\begin{figure}[htb]
\begin{center}
	\includegraphics[width=1\textwidth]{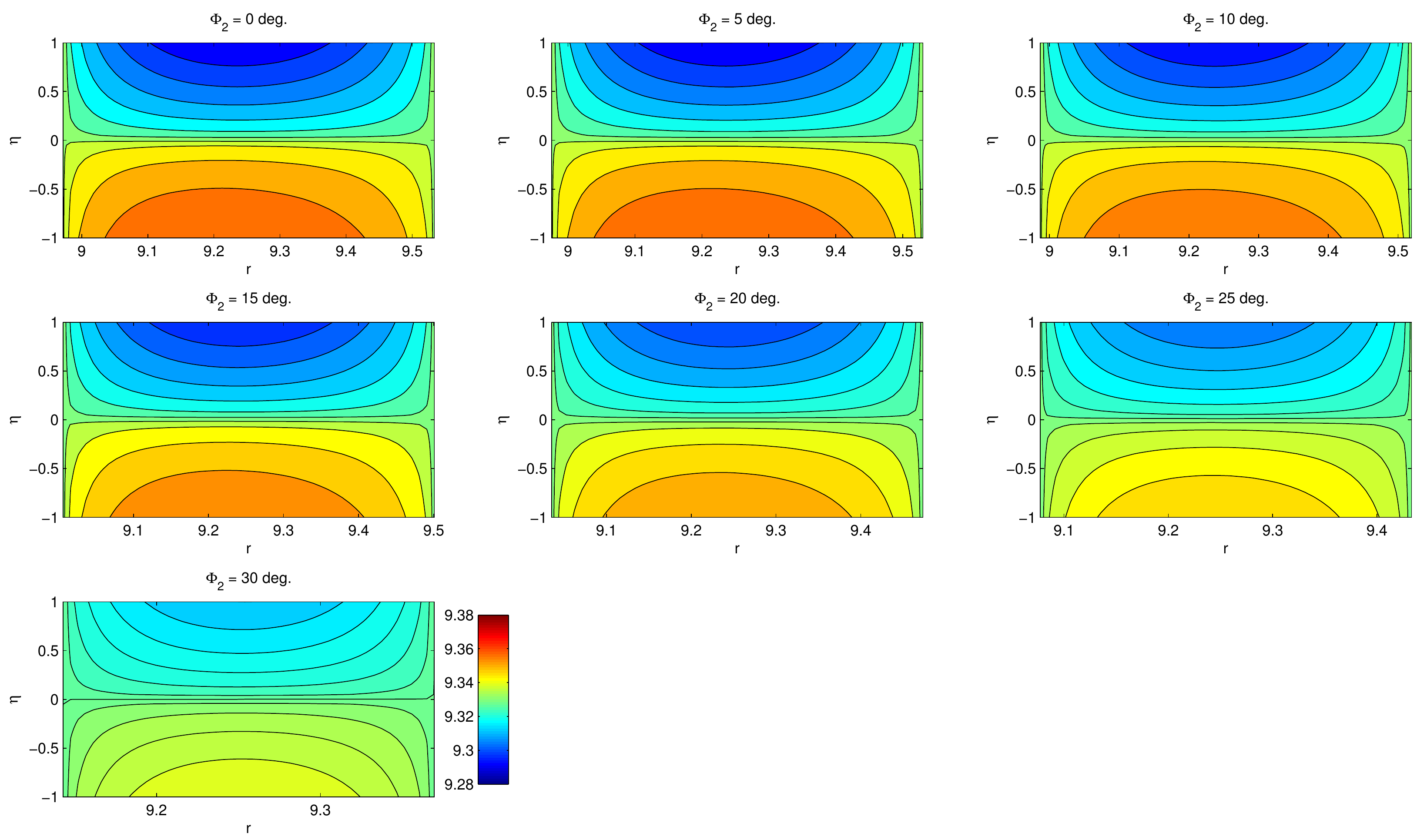}
	\caption{Semi-major axis values for various libration angles with the same energy as Fig. \ref{fig:traj_closed_zvc}.}
	\label{fig:kw4_nom_a}
\end{center}
\end{figure}

\begin{figure}[htb]
\begin{center}
	\includegraphics[width=1\textwidth]{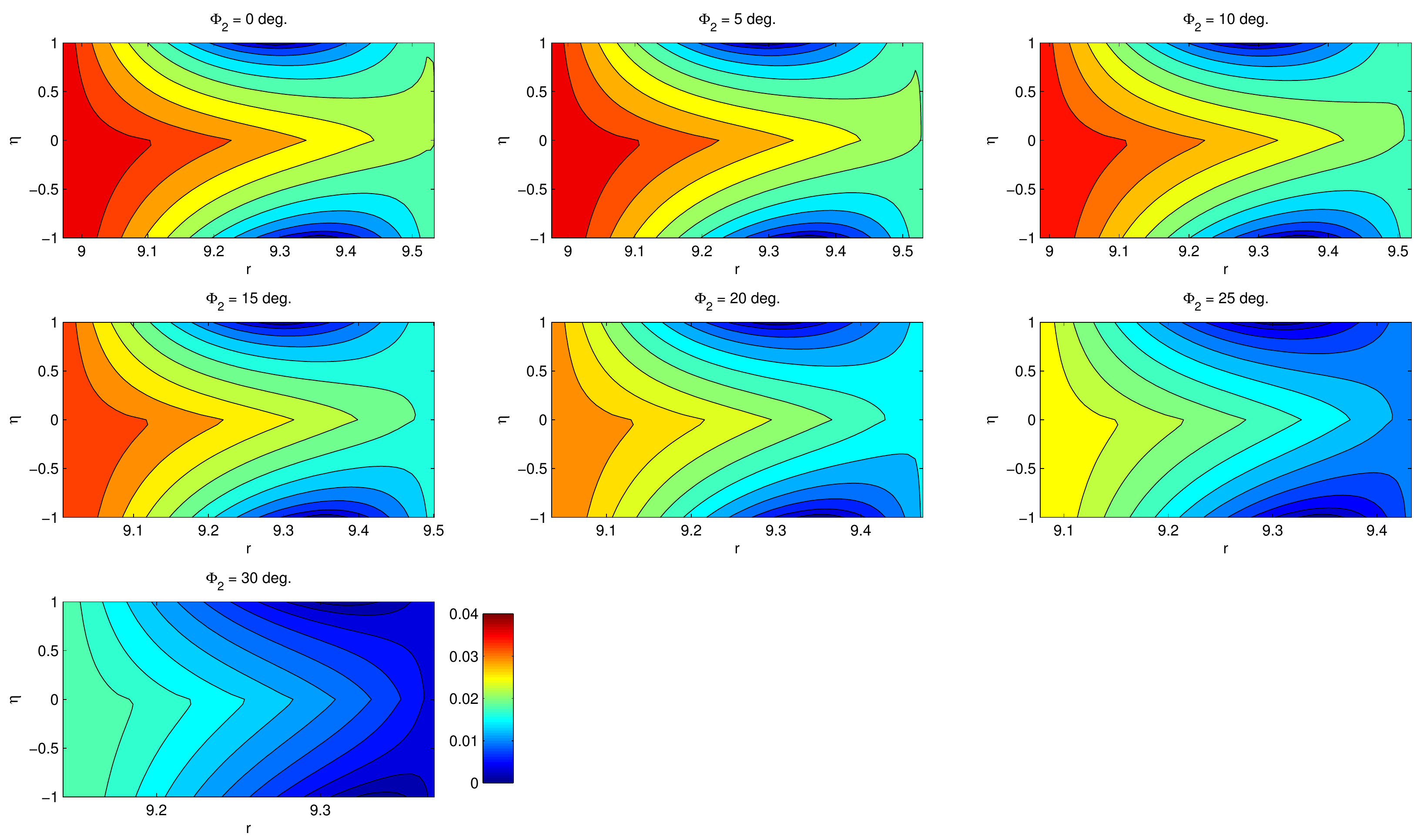}
	\caption{Eccentricity values for various libration angles with the same energy as Fig. \ref{fig:traj_closed_zvc}.}
	\label{fig:kw4_nom_e}
\end{center}
\end{figure}

To further understand the possible values of semi-major axis and eccentricity that can be reached along the trajectories, the values in the entire region inside a give ZVC can be mapped. Figs. \ref{fig:kw4_nom_a} - \ref{fig:kw4_nom_e} show these results. Each subplot can be imagined as a cross section of a ZVC plot (e.g. Fig. \ref{fig:deltaecartoon}) at a different value of $\phi_2$, where the out of plane axis tells us how the excess energy is apportioned between the radial and angular velocities through the use of the parameter $\eta$, which is defined as
\begin{equation}
	\eta = 
	\begin{cases}
		1-\kappa \quad \dot{\phi}_2 \ge 0 \\
		\kappa - 1 \quad \dot{\phi}_2 < 0
	\end{cases}
\end{equation}
so that negative values of $\eta$ correspond to negative values of $\dot{\phi}_2$. The radial bounds of each subplot correspond to the inner and outer edges of the ZVC for this energy level, which is why at larger values of $\phi_2$, the radial range is smaller.

From these figures, we see that the semi-major axis reaches its maximum and minimum values near the stable equilibrium point, with all of the energy in negative or positive angular velocity, respectively. The eccentricity, on the other hand, reaches its maximum value near the inner boundary of the ZVC at $\phi_2 = 0$ with most of the energy in the radial velocity, but some in a negative angular velocity. The minimum eccentricity, which is zero, is seen just outside of the equilibrium point distance with most of the energy in either positive or negative angular velocity. Note that the colorbars near the last subplot are valid for all values of $\phi_2$. Clearly a wide variety of combinations of semi-major axis and eccentricity can be seen over the course of trajectory.

\clearpage

\section{Transient Dynamics}\label{sec:transient}

In this section, we explore trajectories for systems with open ZVCs (those with $E > E^-$). Specifically, we explore the existence of trajectories that behave as if they are bounded, but do not meet the sufficiency conditions outlined in Section \ref{sec:suff_bounded}. These results prove that Theorem \ref{thm:suff_cond_bounded} is in fact a sufficient, but not necessary, condition for bounded motion. We also investigate the special case of bounded motion for periodic trajectories, and the families of periodic trajectories that exist as the energy is varied.

\subsection{Existence of Bounded Motion with Open Zero-Velocity Curves}\label{sec:openZVCs}

\begin{figure}[htb]
\begin{center}
	\includegraphics[width=.45\textwidth]{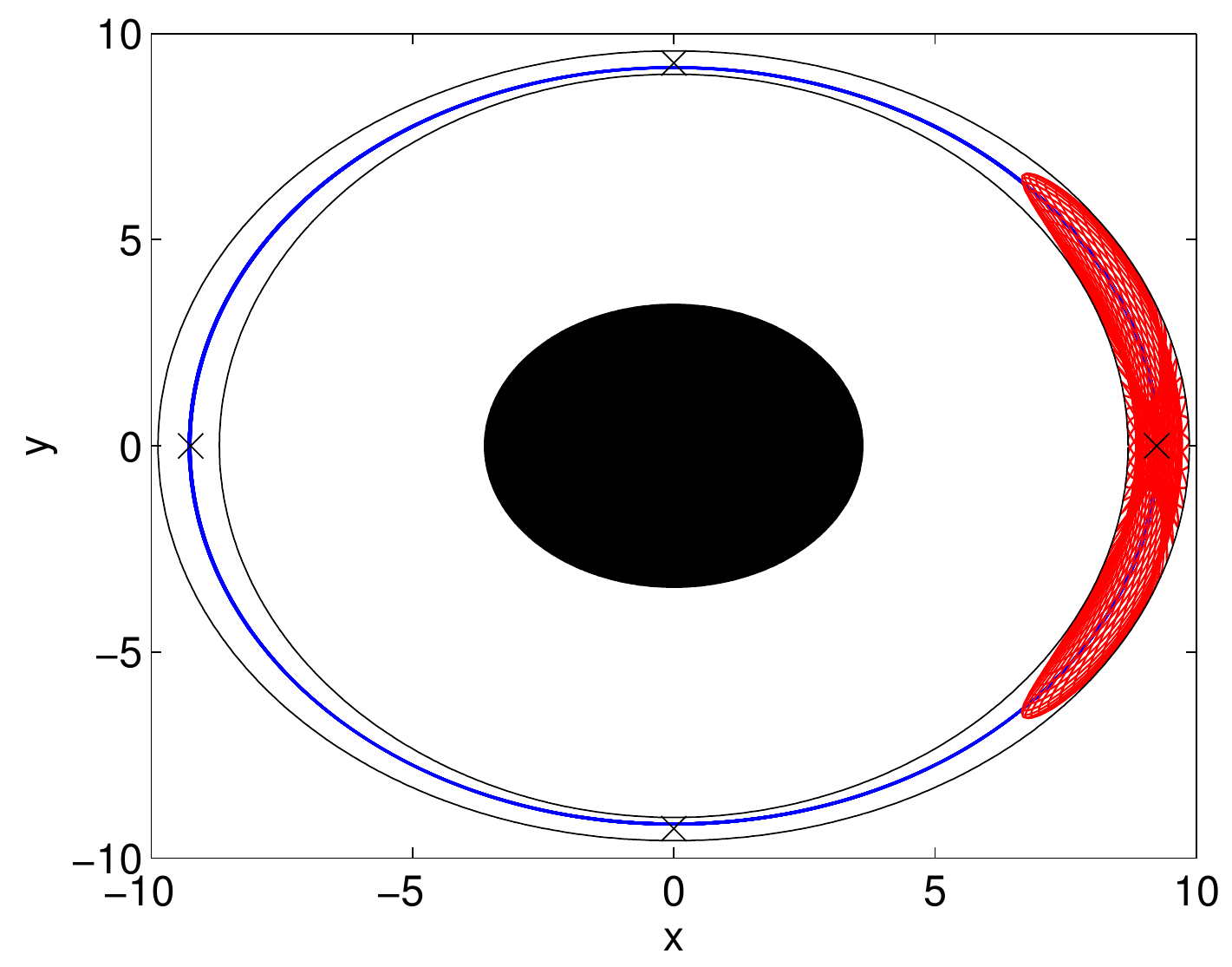}
	\includegraphics[width=.45\textwidth]{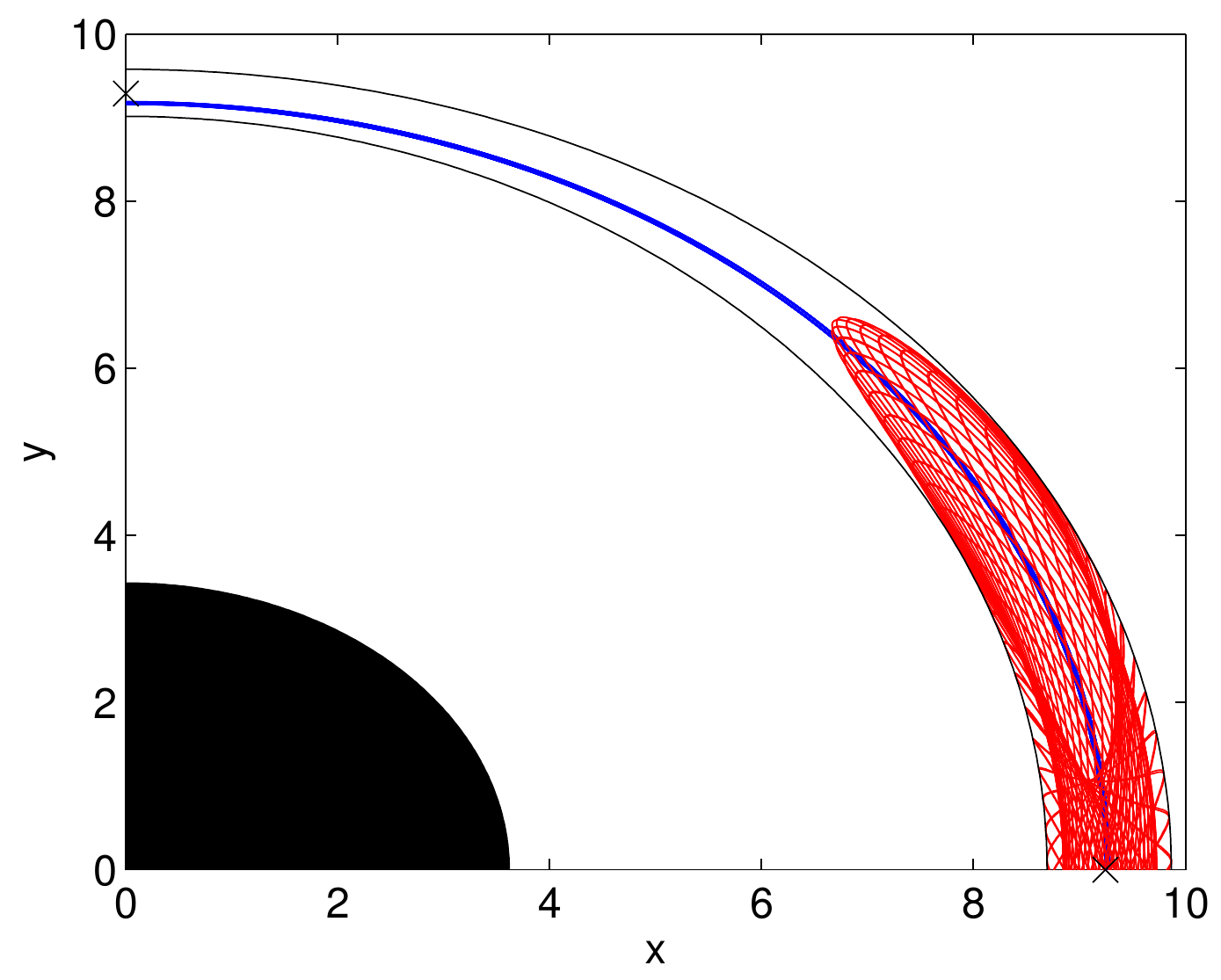}
	\caption{Two trajectories of the nominal system with an energy of $E = E^- + 0.1\delta E$, which is high enough to have an open zero-velocity curve. Although there is no energy barrier to either trajectory circulating, one does and one does not.}
	\label{fig:traj_open_zvc}
\end{center}
\end{figure}

The existence of bounded trajectories with open ZVCs is shown through numerical examples. Two such trajectories are shown in Fig. \ref{fig:traj_open_zvc}. Although the ZVC is open, not every trajectory circulates. This is an extension of what was seen above with the closed ZVC trajectories; depending on the initial conditions of the trajectory at $\phi_2 = 0$, the trajectory will only explore a portion of the $r - \phi_2$ phase space inside the ZVC. The existence of this bounded trajectory proves that Theorem \ref{thm:suff_cond_bounded} is a sufficient, but not necessary condition, as this example does not meet the requirement of the theorem. Furthermore, using the ZVC corresponding to the energy of the unstable equilibrium point is a \emph{conservative} estimate of the energy at which a system will have a circulating, as opposed to librating, secondary body. 

\subsection{Periodic Orbits}\label{sec:periodic}

Periodic orbits can be found in this system when a given orbit repeats itself exactly after a certain period. These orbits are found when the state transition matrix  or monodromy matrix have unity eigenvalues. A detailed methodology for finding periodic orbits in this manner is given in \cite{bellerose}. In this system, due to the symmetry about $\phi_2 = 0$, most of the periodic orbits will have a crossing of this line of symmetry with $\dot{r}=0$, although this is not strictly required for a periodic orbit. However, it is true that every trajectory that crosses $\phi_2 = 0$ twice with $\dot{r} = 0$ is periodic. The period may be very long, but due to the symmetry of the problem it will eventually repeat. 

\begin{figure}[htb]
\begin{center}
	\includegraphics[width=1\textwidth]{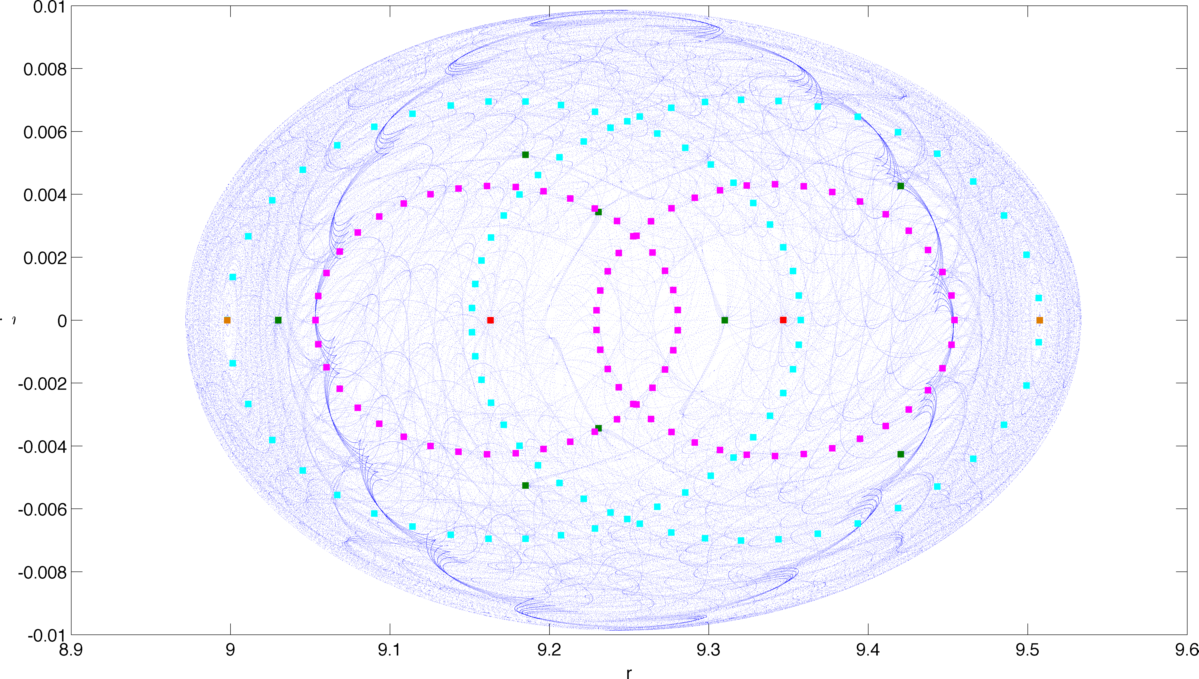} 
	\caption{Poincar\'{e} plot of $\dot{r}$ and $r$ for the nominal system with $E = E^+ + 0.3\delta E$, the same as in Fig. \ref{fig:traj_closed_zvc}. Five different periodic orbits are highlighted. The small blue dots correspond to many other integrated trajectories that are not periodic, and so they pierce the Poincar\'{e} map at different locations each time.}
	\label{fig:periodic_poincare}
\end{center}
\end{figure}

Therefore, our search consisted of starting at different radii with $\dot{r}=0$, and looking at subsequent crossings of $\phi_2 = 0$. The results of this search are combined in a Poincar\`{e} map at of the $r-\dot{r}$ space at $\phi_2 = 0$. One illustrative case is shown in Fig. \ref{fig:periodic_poincare}, where the crossings of several periodic orbits are highlighted.

\begin{figure}[htb]
\begin{center}
	\includegraphics[width=1\textwidth]{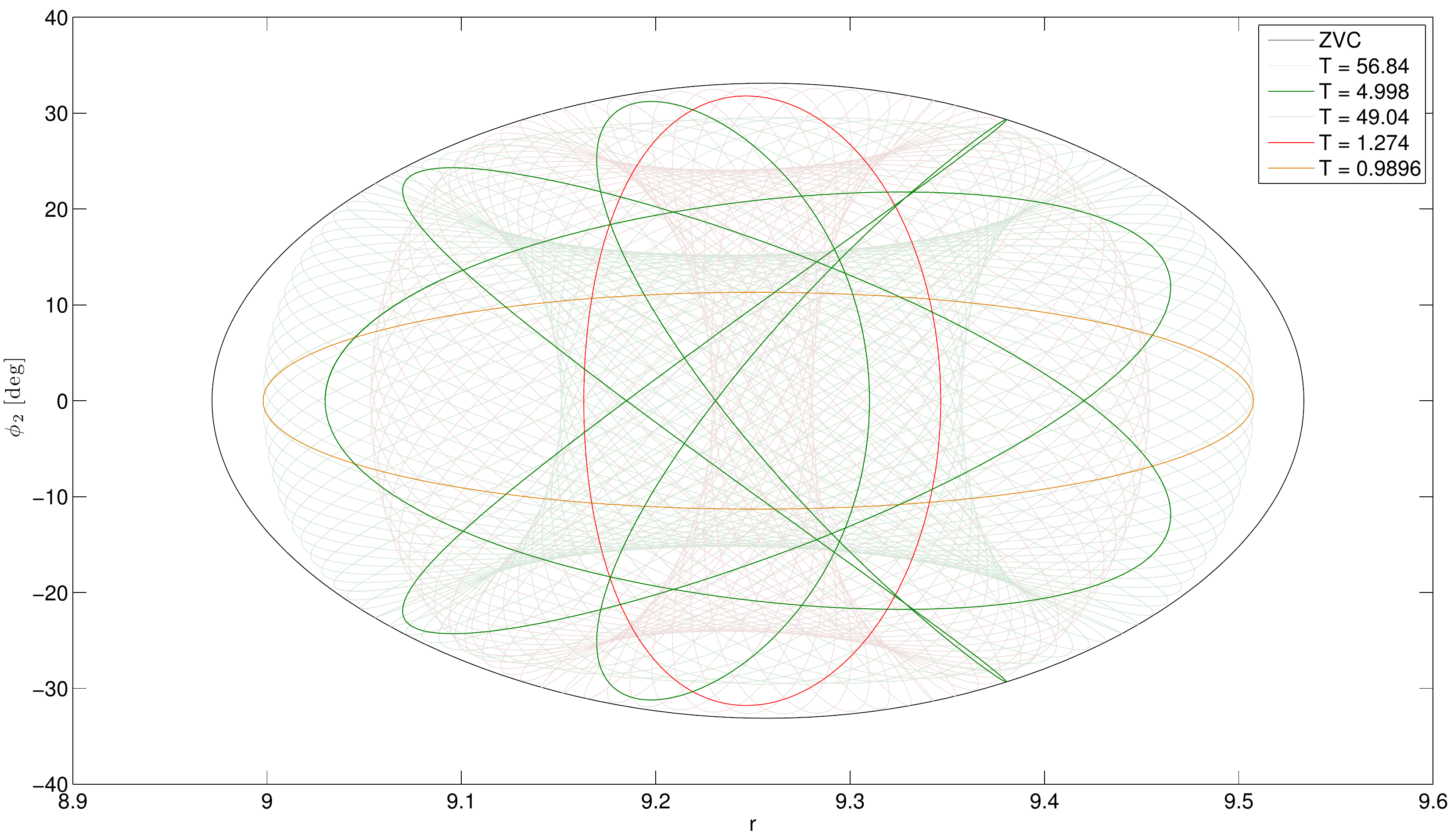}
	\caption{The trajectories for five distinct periodic orbits in the $r$-$\phi_2$ phase plane which were highlighted in Fig. \ref{fig:periodic_poincare}. The trajectories are labeled by the ratio of their period with the period of the equilibrium orbit. Note that the $T = 56.84$ trajectory was plotted in cyan in Fig. \ref{fig:periodic_poincare}, and the $T = 49.04$ was plotted in magenta.}
	\label{fig:periodic_trajs}
\end{center}
\end{figure}

The actual trajectories of several periodic orbits in the $r-\phi_2$ space are shown in Fig. \ref{fig:periodic_trajs}, corresponding to the highlighted cases from Fig. \ref{fig:periodic_poincare}. As alluded to previously, there are actually an infinite number of periodic trajectories in this system, with most of them having very large period ratios such as the $T = 49.04$ and $T = 56.84$ orbits shown. Of particular interest are the two periodic orbits with near $T = 1$. In the example shown in Fig. \ref{fig:periodic_trajs}, these are the $T = 0.9896$ case, which orbits counter-clockwise, and the $T = 1.274$ case which orbits in a clockwise direction. These two orbits roughly span the space inside the closed ZVC in this case, as they are associated with the eigenmodes of the equilibrium point. 

\begin{figure}[htb]
\begin{center}
	\includegraphics[width=.48\textwidth]{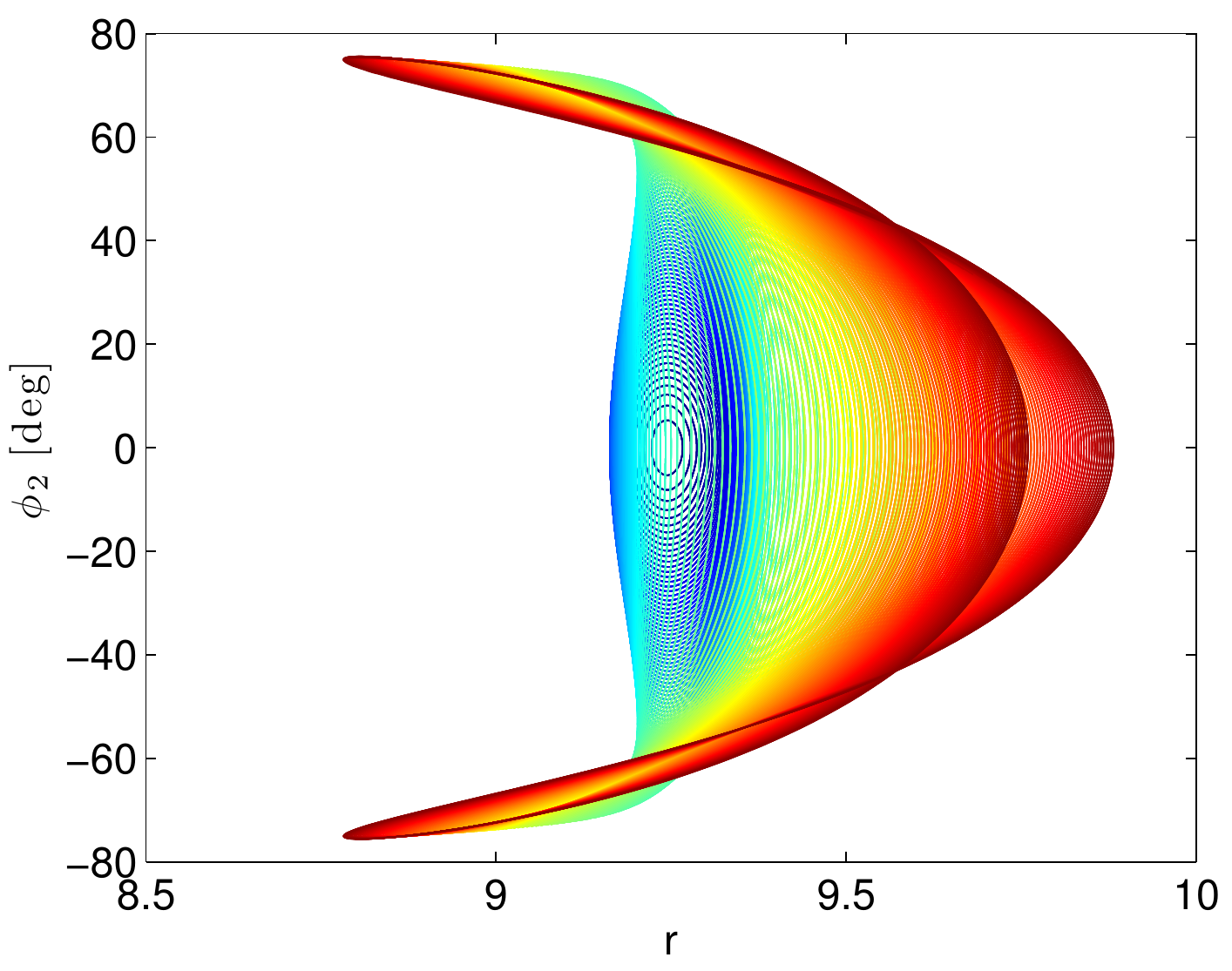}
	\includegraphics[width=.48\textwidth]{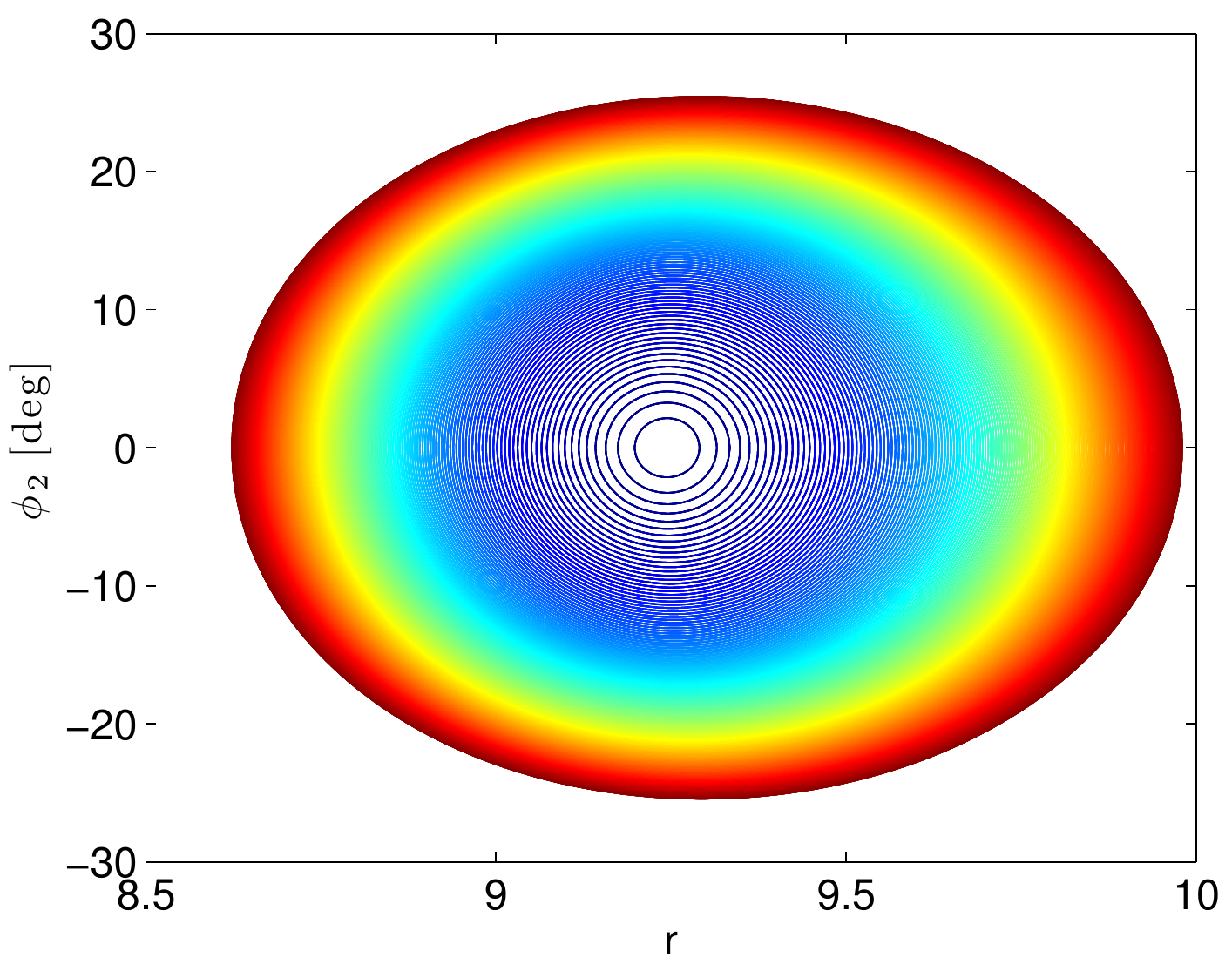}
	\caption{The periodic orbit trajectories are shown, with the clockwise orbits ($T = 1.274$ in Fig. \ref{fig:periodic_trajs}) in the left plot and the counter-clockwise ($T=0.9896$ in Fig. \ref{fig:periodic_trajs}) orbits in the right plot. For this test, the energy was varied from $E = E^+$ to $E = E^- + \delta E$. Note that the lower energy cases are shown in dark blue, varying through green and yellow, up to the highest energy cases in dark red.}
	\label{fig:periodic_trajs_vary_E}
\end{center}
\end{figure}

\begin{figure}[htb]
\begin{center}
	\includegraphics[width=.6\textwidth]{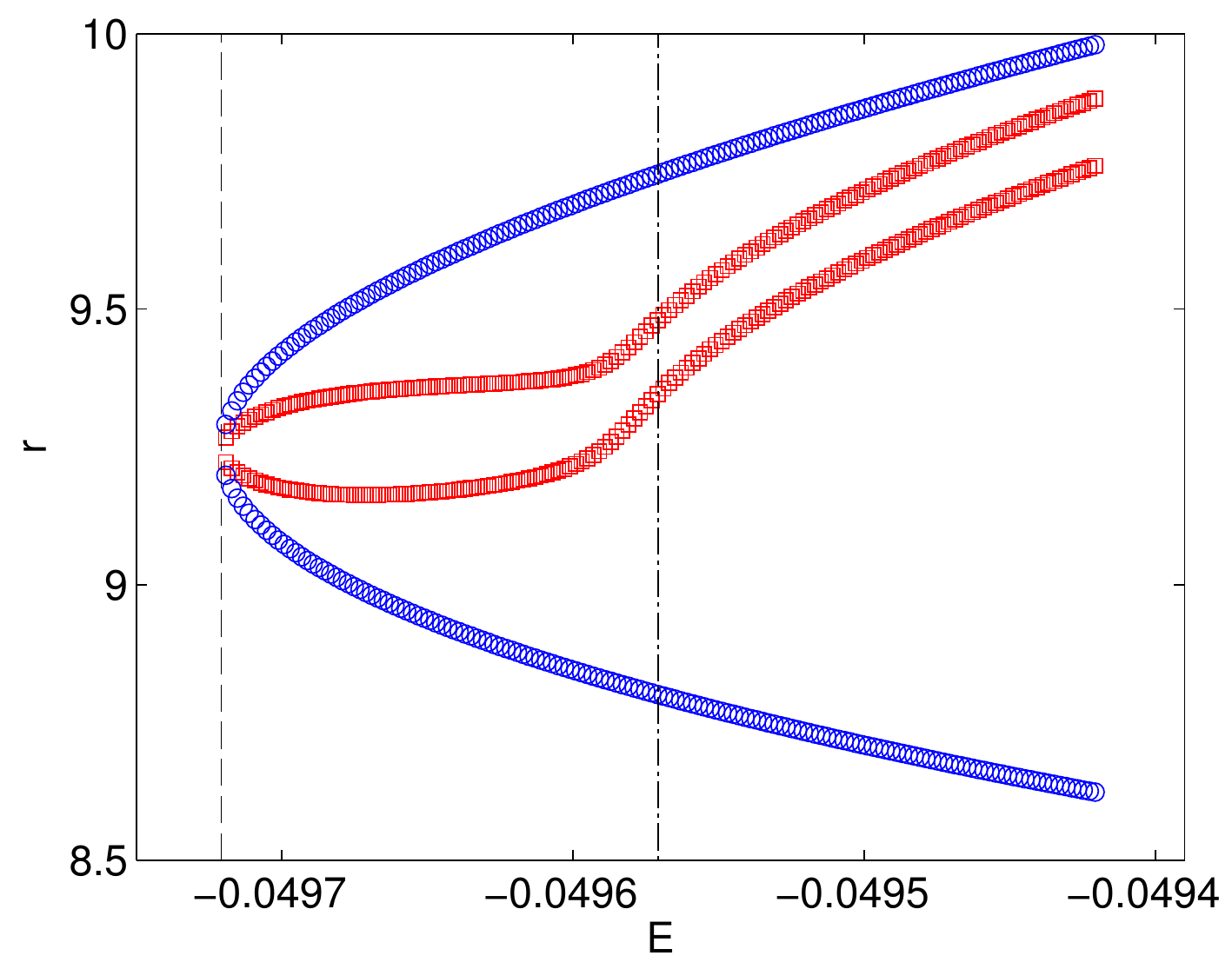}
	\caption{Radial locations of the $\phi_2 = 0$ crossings with $\dot{r} = 0$ at varying energies. The clockwise orbits are shown in red, and the counter-clockwise are shown in blue. For reference, the stable equilibrium point energy is shown as a vertical dashed line, where the periodic orbit families disappear. The unstable equilibrium point energy is also shown as a dash-dot vertical line.}
	\label{fig:periodic_r_vary_E}
\end{center}
\end{figure}

The evolution of the eigenmode periodic orbits with changing values of energy are explored in Figs. \ref{fig:periodic_trajs_vary_E} and \ref{fig:periodic_r_vary_E}. The first plot shows many different trajectories as the energy is increased. Particularly interesting behavior is seen with the clockwise trajectories. When the energy is approximately 75\% of the way to the unstable equilibrium point energy, the shape of these orbits changes from the oval seen in Fig. \ref{fig:periodic_trajs} to a bow shape, and eventually (at the highest energy levels shown) the trajectory crosses itself near $\phi_2 = \pm 50\degree$. The maximum libration angles reached by these periodic orbits is also very high, approaching 80$\degree$ at the energy levels shown. By comparison, the evolution of the counter-clockwise orbits is much more regular. The shape stays basically the same throughout, simply growing in size as the energy grows. Note that the maximum amplitudes reached in these cases is much smaller than the clockwise orbits, getting to a maximum of roughly $\phi_2 = \pm 25\degree$.

A different view of the trajectories is shown in Fig. \ref{fig:periodic_r_vary_E}. This plot shows the radial locations of the $\phi_2 = 0$ crossings with $\dot{r} = 0$ at each energy level. This clearly shows the more regular evolution of the counter-clockwise orbits. It is especially interesting that these periodic orbits exist well beyond the energy of the unstable equilibrium point, as this gives clear examples of stable, libration bounded non-equilibrium orbits with open ZVCs.

\clearpage

\section{Sufficient Conditions for Unbounded Motion}\label{sec:suff_unbounded}

In Section \ref{sec:suff_bounded} we presented a sufficient condition for bounded librational motion. In Section \ref{sec:transient}, we proved this condition was only sufficient due to the existence of bounded trajectories when the conditions from Theorem \ref{thm:suff_cond_bounded} are not met. Significantly there are families of periodic orbits that exist even when the ZVCs of the system are open. In this section, we analyze the converse problem of determining when unbounded librational motion occurs. When the system has a mass fraction near unity, conditions for unbounded motion can be derived analytically. For non-unity mass fractions, strict conditions are unavailable. However, we present analysis for these systems that illustrates the structure of the problem and initial conditions that will lead to circulation.


\subsection{Analytical Limits for Systems with $\nu \simeq 1$}\label{sec:limits_nu_1}

In this section, we show sufficient conditions for circulation for systems with mass ratios close to unity. This restriction limits the effect of the coupling on the orbit of the small ellipsoidal body, which allows us to assume that the orbit itself remains unchanged due to the libration of the secondary. However, due to this assumption the results can only be claimed as sufficient conditions for circulation, as shown in the following theorem.

\begin{theorem} \textbf{Sufficient Condition for Circulation}\\
\label{thm:suff_cond_circ}
Given a system with $\nu \simeq 1$, so that the orbit can be considered unperturbed by the ellipsoidal secondary, with some initial secondary spin rate $\dot{\phi}_{2,0}$ at a libration angle of $\phi = 0\degree$ at some location on the orbit ($r_0$), the secondary body is guaranteed to circulate if the spin rate satisfies the relationship
\begin{equation}\label{eq:theorem1}
	\dot{\phi}_{2,0}^2 \ge \frac{I_{z,0}}{\overline{I}_{2_z} r_0^5} \left[ \overline{I}_{2_y} - \overline{I}_{2_x} + \left( \overline{I}_{2_z} - \overline{I}_{2_x} + \overline{I}_{1_z} - \overline{I}_s \right) \left(1 - \frac{r_0^3}{r_a^3} \right) \right]
\end{equation}
where $r_a$ is the apoapse radius of the orbit.
\\
\textbf{Proof:}
The free energy of the system is written using Eqs. \eqref{eq:potential_oblate}, \eqref{eq:kinetic_total} and \eqref{eq:kinetic} as,
 \begin{equation}\label{eq:etot_breakdown}
	E = V + T - T_1 = E_{orb} + E_\phi + E_{coup}
\end{equation}
where $E_{orb}$ is the orbit energy that is independent of the librational state, $E_\phi$ is the librational energy (identical to a pendulum), and $E_{coup}$ is the ``coupling" energy. The first two energy terms are given by relationships
\begin{equation}\label{eq:ekep_standard}
	E_{orb} = -\frac{\nu}{r} + \frac{\nu}{2}\dot{r}^2 + \frac{\nu K^2}{2 I_z}
\end{equation}
\begin{equation}\label{eq:ephi}
	E_\phi = \frac{1}{2}\overline{I}_{2_z} \dot{\phi}_2^2 + \frac{3 \nu}{2 r^3}\left(\overline{I}_{2_y} - \overline{I}_{2_x} \right) \sin^2 \phi_2
\end{equation}
In the case of a point mass, the orbit energy collapses to the Keplerian energy of the system.
Recalling the trigonometric relationship
\begin{equation}
	\frac{1}{2} \cos 2 \phi_2 = \frac{1}{2} - \sin^2 \phi_2
\end{equation}
allows us to determine the coupling energy from Eq. \eqref{eq:etot_breakdown}, using Eqs. \eqref{eq:free_energy}, \eqref{eq:ekep_standard} and \eqref{eq:ephi} as
\begin{equation}\label{eq:ecoup}
	E_{coup} = -\frac{\nu}{2 r^3}\left( \overline{I}_{1_z} - \overline{I}_s + C_2^+ \right) - \frac{\overline{I}_{2_z}^2 \dot{\phi}_2^2}{2 I_z}  
\end{equation}
where $C_2^+$ was defined in Eq. \eqref{eq:c2pm}. 

If the orbit energy is constant, then the available energy is determined as
\begin{equation}
	\Delta E_{lib} = E - E_{orb} = E_\phi + E_{coup}
\end{equation}

Since this value is a constant at any point on the trajectory, we can relate two points by
\begin{equation}
	\Delta E_{lib}(r_0,\phi_2 = 0, \dot{\phi}_{2,0}) = \Delta E_{lib}(r_m,\phi_{2,m}, \dot{\phi}_{2} = 0)
\end{equation}
This expression relates the available energy for a point on the orbit at radius $r_0$ with some initial libration rate $\dot{\phi}_{2,0}$ and zero libration angle, to another point on the orbit at radius $r_m$ with libration angle $\phi_{2,max}$ and a libration rate of zero. In other words, this tells us the maximum achievable libration angle at radius $r_m$ given a system that had an initial libration rate at $r_0$. Note that this relationship is independent of $\dot{r}$. Writing out the available energy at these two locations using Eqs. \eqref{eq:ephi} and \eqref{eq:ecoup} gives
\begin{equation}\label{eq:delta_elib_equiv}
	\frac{\overline{I}_{2_z} \nu r_0^2 \dot{\phi}_{2,0}^2}{2 I_{z,0}} - \frac{\nu}{2 r_0^3}\left(\overline{I}_{1_z} - \overline{I}_s + C_2^+\right) = \frac{3 \nu}{2 r_m^3}\left(\overline{I}_{2_y} - \overline{I}_{2_x} \right) \sin^2  \phi_{2,m} - \frac{\nu}{2 r_m^3}\left(\overline{I}_{1_z} - \overline{I}_s + C_2^+\right)
\end{equation}

To determine the bounding condition, we look for the case when $\phi_{2,m} = 90\degree$, and solve for $\dot{\phi}_{2,0}$ to get
\begin{equation}\label{eq:phidot_r0_rm}
	\dot{\phi}_{2,0}^2 \ge \frac{I_{z,0}}{\overline{I}_{2_z} r_0^5} \left[ \overline{I}_{2_y} - \overline{I}_{2_x} + \left( \overline{I}_{2_z} - \overline{I}_{2_x} + \overline{I}_{1_z} - \overline{I}_s \right) \left(1 - \frac{r_0^3}{r_m^3} \right) \right]
\end{equation}
where the fact that $C_2^+ = -2 \overline{I}_{2_x} + \overline{I}_{2_y} + \overline{I}_{2_z}$ was used.

For any given starting location, the highest $\dot{\phi}_{2,0}$ will be required if $r_m$ is as large as possible, so $r_m = r_a$. Therefore if the the libration rate is high enough to circulate the secondary at apoapse, then the body will circulate no matter what point on the orbit it reaches its maximum libration angle. Thus the bounding condition to ensure circulation occurs when
\begin{equation*}
	\dot{\phi}_{2,0}^2 \ge \frac{I_{z,0}}{\overline{I}_{2_z} r_0^5} \left[ \overline{I}_{2_y} - \overline{I}_{2_x} + \left( \overline{I}_{2_z} - \overline{I}_{2_x} + \overline{I}_{1_z} - \overline{I}_s \right) \left(1 - \frac{r_0^3}{r_a^3} \right) \right]
\end{equation*}
\\
$\Box$
\end{theorem}

Theorem \ref{thm:suff_cond_circ} assumes that the orbit perturbations caused by the libration of the secondary are ignorable. Therefore the orbit can be described using Keplerian orbital elements, which will be constant. We can express the initial orbit radius in terms of the eccentricity, semi-major axis, and true anomaly as
\begin{equation}
	r_0 = \frac{a(1-e^2)}{1 + e \cos f}
\end{equation}
and then the apoapsis radius is
\begin{equation}
	r_a = a(1+e)
\end{equation}
and finally the periapse radius is
\begin{equation}
	r_p = a(1-e)
\end{equation}

Using these relationships, the condition for circulation can be expressed in terms of the orbital elements as,
\begin{equation}\label{eq:phidot_suff_oes}
	\dot{\phi}_{2,0}^2 \ge \frac{I_{z,0}}{\overline{I}_{2_z}} \left( \frac{1 + e \cos f}{a(1-e^2)} \right)^5 \left[ \left(\overline{I}_{2_y} - \overline{I}_{2_x} \right) + \left(\overline{I}_{2_z} - \overline{I}_{2_x} + \overline{I}_{1_z} - \overline{I}_{s} \right) \left[ 1 - \frac{ \left(1-e\right)^3 }{ \left(1 + e \cos f \right)^3 } \right] \right]
\end{equation}

Theorem \ref{thm:suff_cond_circ} gives the sufficient condition for circulation for a given starting point, however we can define a single condition for circulation that is sufficient for any location on the orbit by defining the highest value that leads to circulation, which is referred to as the uniform condition.

\begin{remark} \textbf{Uniform Condition for Guaranteed Circulation}\\
\label{rmk:1}
The highest bound for circulation is given by maximizing Eq. \eqref{eq:phidot_suff_oes} in terms of $f$. The maximum occurs when $f=0$, or $r_0 = r_p$, and the circulation condition becomes
\begin{equation}\label{eq:remark1}
	\dot{\phi}_{2,0}^2 \ge \frac{I_{z,p}}{\overline{I}_{2_z} a^5 (1-e)^5} \left[ \left(\overline{I}_{2_y} - \overline{I}_{2_x} \right) + \left(\overline{I}_{2_z} - \overline{I}_{2_x} + \overline{I}_{1_z} - \overline{I}_{s} \right) \left[ \frac{ 2e\left(3 + e^2 \right) }{ \left(1 + e\right)^3 } \right] \right]
\end{equation}
\\
$\Box$
\end{remark}

This condition is considered uniform because it combines the extremes for both $r_0$ and $r_m$ in terms of maximizing the required $\dot{\phi}_{2,0}$. The condition given Eq. \eqref{eq:remark1} is the baseline sufficiency test for circulation. At any point on an orbit, if $\dot{\phi}_{2,0}$ meets this condition with $\phi_2 = 0\degree$, it is guaranteed to circulate directly, meaning that the current trajectory will pass through $\pm90\degree$ before returning to $\phi_2 = 0\degree$.

Depending on the actual location on the orbit, the value of $\dot{\phi}_{2,0}$ which is sufficient to cause circulation will generally be lower, and is given by Eq. \eqref{eq:theorem1}. It is important to note here that Theorem \ref{thm:suff_cond_circ} is a sufficiency condition only, but it is possible that a lower value of $\dot{\phi}_{2,0}$ will lead to circulation. This is because Theorem \ref{thm:suff_cond_circ} depended on the most extreme value for $r_m$. However, due to the fact that it is easier to circulate at all $r_m < r_a$, it is true that lower libration rates will lead to circulation in many cases. This depends on the phasing of the librational motion with the orbit to determine at what radius $\phi_{2,max}$ is achieved, and if $\dot{\phi}_{2,0}$ was large enough at the $\phi_2 = 0$ crossing preceeding this $\phi_{2,max}$, circulation can occur. Therefore the lowest possible spin rate that can lead to circulation can be determined in the opposite limiting case where $\dot{\phi}_{2,0}$ is minimized with respect to $r_m$ and $r_0$. In fact, this becomes a sufficient condition for bounded libration.

\begin{corollary} \textbf{Sufficient Condition for Bounded Motion with $\nu \simeq 1$}\\
\label{cor:suff_bounded_nu1}
Given the system considered in Theorem \ref{thm:suff_cond_circ}, the lowest value of $\dot{\phi}_{2,0}$ which can lead to circulation is computed so that any value below this will not lead to circulation
\begin{equation}
	\dot{\phi}_{2,0}^2 < \frac{I_{z,0}}{\overline{I}_{2_z} r_0^5} \left[ \overline{I}_{2_y} - \overline{I}_{2_x} + \left( \overline{I}_{2_z} - \overline{I}_{2_x} + \overline{I}_{1_z} - \overline{I}_s \right) \left(1 - \frac{r_0^3}{r_p^3} \right) \right]
\end{equation}
where $r_0$ is found as the real positive root of,
\begin{equation}\label{eq:r0_roots}
	r_0^3 - \left[ \frac{3 \nu r_p^3 \left(\overline{I}_{1_z} - \overline{I}_s + C_2^+\right)}{2 \overline{I}_{2_z} \left( \overline{I}_{2_z} - \overline{I}_{2_x} + \overline{I}_{1_z} - \overline{I}_s \right)} \right] r_0^2 - \left[ \frac{5 r_p^3 \left(\overline{I}_{1_z} - \overline{I}_s + C_2^+\right)}{2 \left( \overline{I}_{2_z} - \overline{I}_{2_x} + \overline{I}_{1_z} - \overline{I}_s \right)} \right] = 0
\end{equation}
if the resulting $r_0 < r_a$, otherwise $r_0 = r_a$. 
\\
\textbf{Proof:}
Eq. \ref{eq:phidot_r0_rm} is minimized with respect to $r_0$ and $r_m$. The partial with respect to $r_m$ shows that the function monotonically increases with $r_m$, therefore the minima with respect to $r_m$ is at the lower boundary, $r_p$. 

There exists an extrema of Eq. \ref{eq:phidot_r0_rm} with respect to $r_0$, which is located at the root of Eq. \eqref{eq:r0_roots}. Using the Routh criteria it can be proven that there is one and only one real positive root. It can also be shown that this extrema is in fact a minima by investigating the second partial with respect to $r_0$. However, the location of this minima depends on the system parameters, and it is not guaranteed to lie within the range of possible radius values given by $r_p$ and $r_a$. Therefore, if the minima is found to be greater than $r_a$, the constrained minima is at $r_0 = r_a$.
\\
$\Box$
\end{corollary}

These results outline a set of bounds that can be used to determine if a system will circulate or not depending on the libration rate at $\phi_2 = 0\degree$. The main boundaries to be tested to give information for an entire orbit are given by Remark \ref{rmk:1} and Corollary \ref{cor:suff_bounded_nu1}. If either of these conditions are satisfied, we can immediately state if the body \emph{will} or \emph{will not} circulate. 


When the libration rate is between these two boundaries, further inspection must be made. Theorem \ref{thm:suff_cond_circ} can be used to determine if a given initial condition ($r_0$ and $\dot{\phi}_{2,0}$) will circulate directly. However, this relationship does not tell us if the body will \emph{ever} circulate; only if it will circulate directly. The difficulty is in the phasing of the libration and the orbit. If the initial libration rate is below the bound given in Theorem \ref{thm:suff_cond_circ}, then the body will not circulate this on this oscillation. However in general the body will pass back through $\phi_2 =0\degree$ at a different state, and then the condition must be checked again. 

Corollary \ref{cor:suff_bounded_nu1} is an extension of Theorem \ref{thm:suff_cond_bounded} for the specific case when $\nu \simeq 1$. If the total energy in the system is low enough so that Theorem \ref{thm:suff_cond_bounded} is applicable, Corollary \ref{cor:suff_bounded_nu1} will also be satisfied. However, Corollary \ref{cor:suff_bounded_nu1} can be satisfied when Theorem \ref{thm:suff_cond_bounded} is not applicable.

If circulation does not occur, it is useful to be able to determine the libration amplitudes that can be expected. This is addressed in the following corollary.

\begin{corollary} \textbf{Maximum Libration Angle}\\
\label{cor:max_libration_angle}
Given the system considered in Theorem \ref{thm:suff_cond_circ}, if $\dot{\phi}_{2,0}$ is below the limit for circulation, then the maximum libration angle that can be obtained is given by
\begin{equation}\label{eq:phimax_rprm}
	\sin^2 \phi_{2,max} = \frac{1}{3 \left( \overline{I}_{2_y} - \overline{I}_{2_x} \right)}  \left[ \frac{\overline{I}_{2_z} r_0^5 \dot{\phi}_{2,0}^2}{I_{z,0}} + \left(\overline{I}_{1_z} - \overline{I}_s + C_2^+\right) \left(1 - \frac{r_p^3}{r_0^3} \right) \right]
\end{equation}
\\
\textbf{Proof:}
This condition is determined by rearranging Eq. \eqref{eq:delta_elib_equiv} to get
\begin{equation}\label{eq:phimax_r0rm}
	\sin^2 \phi_{2,max} = \frac{1}{3 \left( \overline{I}_{2_y} - \overline{I}_{2_x} \right)}  \left[ \frac{\overline{I}_{2_z} r_0^5 \dot{\phi}_{2,0}^2}{I_{z,0}} + \left(\overline{I}_{1_z} - \overline{I}_s + C_2^+\right) \left(1 - \frac{r_m^3}{r_0^3} \right) \right]
\end{equation}
This relationship is maximized with a minimum $r_m$, giving $r_m = r_p$, the periapse radius.
\\
$\Box$
\end{corollary}

If the maximum libration relationship from Corollary \ref{cor:max_libration_angle} is tested with $\dot{\phi}_{2,0}$ above the circulation limit, the right hand side of Eq. \eqref{eq:phimax_rprm} will be greater than one, resulting in an imaginary value for $\phi_{2,max}$. This indicates the body will circulate. The maximum libration angle that would be reached at any point on the orbit can be determined by using a radius other than the periapse radius in Eq. \eqref{eq:phimax_r0rm} for $r_m$. 

\begin{remark} \textbf{Simplification to Classical Results}\\
\label{rmk:3}
Classical gravity gradient results assume that the orbit is circular ($r_m=r_0=r$) and that the system is dominated by the orbit so that $\nu r^2 \gg \overline{I}_{2_z}$, and therefore $I_z \to \nu r^2$, and $\nu=1$. Substituting these conditions into Eq. \ref{eq:delta_elib_equiv} gives the classical result
\begin{equation}\label{eq:classic_energy_pendulum}
	\dot{\phi}_{2}^2= \frac{3}{r^3}\left(\frac{\overline{I}_{2_y} - \overline{I}_{2_x} }{\overline{I}_{2_z} }\right) \sin^2 \phi_{2,max} 
\end{equation}
which is recognized as the energy integral of the pendulum equation of motion
\begin{equation}\label{eq:classical_eom}
	\ddot{\phi}_{2} = -\frac{3}{2r^3}\left(\frac{\overline{I}_{2_y} - \overline{I}_{2_x} }{\overline{I}_{2_z} }\right) \sin 2\phi_{2}  
\end{equation}
\\
$\Box$
\end{remark}
Derivation of the classical results are given in \cite{beletskii} and \cite{hughes} among many other sources. Note that Eq. \eqref{eq:classical_eom} has a slight difference to the referenced results because we have normalized the units so that $\mu$ has disappeared.



\begin{figure}[htb]
	\centering\includegraphics[width=.6\textwidth]{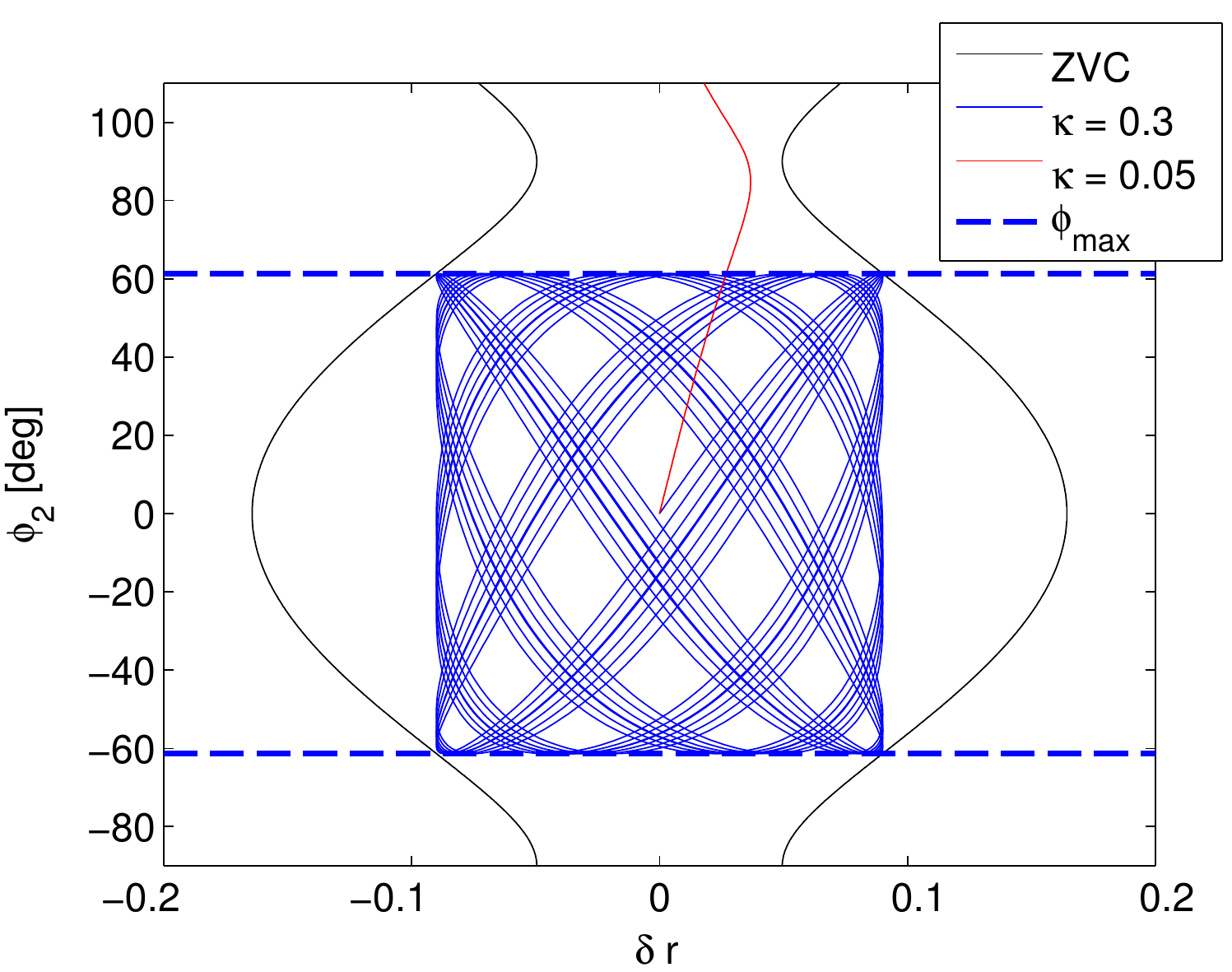}
	\caption{Two different trajectories for the nominal case with energy $10\%$ above the unstable equilibrium energy. The dotted horizontal line indicates the computed $\phi_{2,max} = 61.34\degree$ from Eq. \eqref{eq:phimax_rprm} for the $\kappa = 0.3$ trajectory. Note that on the x-axis, $\delta r = r - r^+$.}
	\label{fig:open_zvc_trajs}
\end{figure}

These results can be illustrated through several numerical systems. First, a basic example is shown in Fig. \ref{fig:open_zvc_trajs} with very low eccentricity ($e = 2.4\times10^{-6}$). The full equations of motion given in Eqs. \eqref{eq:2dof_rddot} - \eqref{eq:2dof_phi2ddot} are integrated for two cases, where one trajectory is initialized with 95\% of the excess energy is given to $\dot{\phi}_2$, while the other has 70\% apportioned to the initial libration rate. Both cases start at $r = r^+$ and $\dot{r} = 0$. Using Theorem \ref{thm:suff_cond_circ}, we can see that the former case is above the limit and therefore will circulate, while the latter case is below. The maximum libration angle is calculated for the $\kappa = 0.3$ case from Eq. \eqref{eq:phimax_r0rm} with $r_m = r^+$, and is found to closely agree with the maximum libration amplitudes seen during the simulation. 

\begin{figure}[htb]
	\centering\includegraphics[width=.6\textwidth]{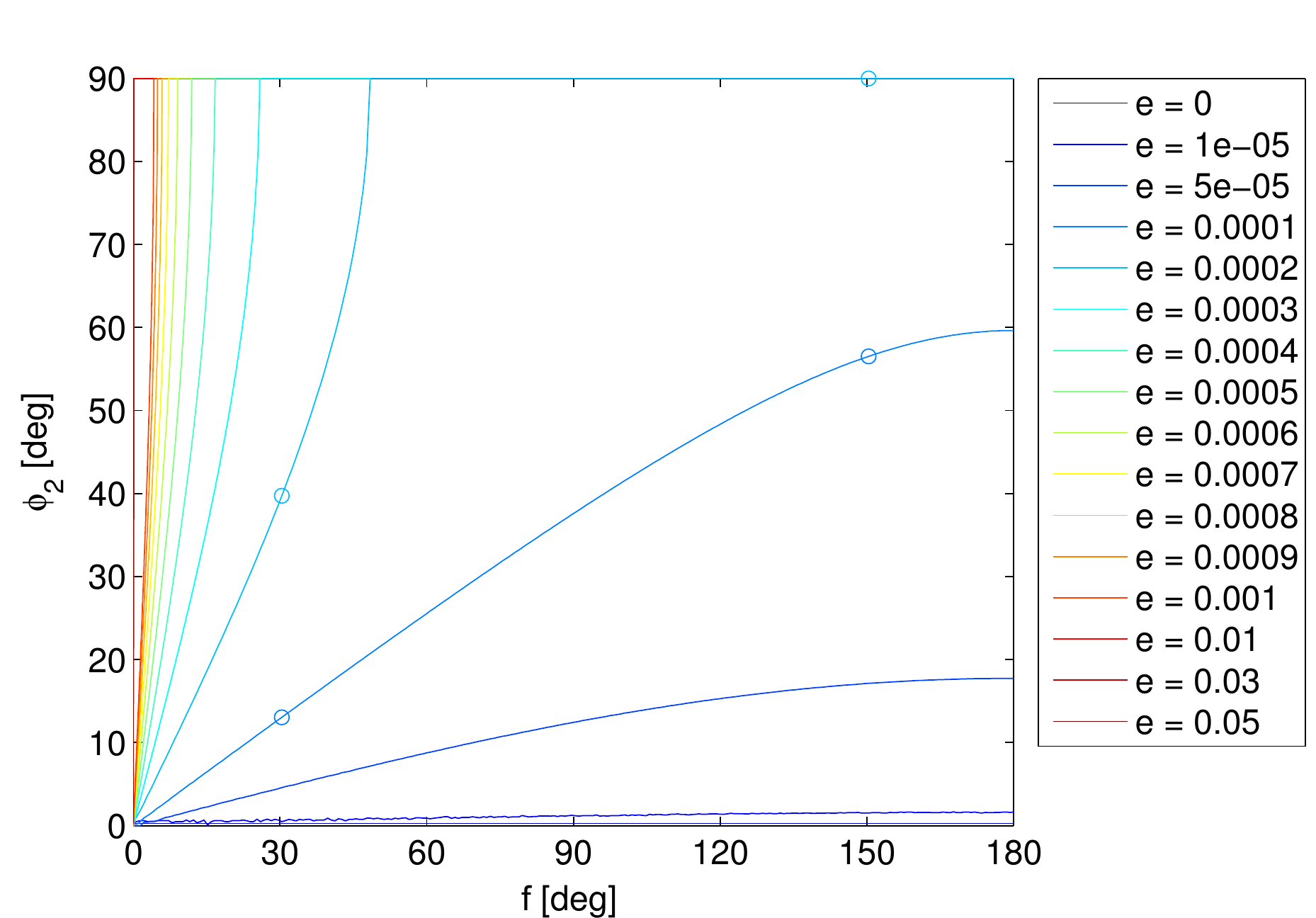}
	\caption{Maximum libration amplitudes at different points on a Keplerian orbit with $\dot{\phi}_2 = 0$. When $e=0$, maximum libration amplitude at apoapsis is only $0.34\degree$ and at $e=1\times10^{-5}$, the amplitude can reach $1.65\degree$ at apoapsis. The unstable eccentricity is between $e=1\times10^{-4}$ and $e=2\times10^{-4}$. For all eccentricities, there are some initial conditions that stay bounded; e.g. for $e = 0.05$, the libration amplitude is bounded for $\nu < 0.0125\degree$.}
	\label{fig:phimax_nu}
\end{figure}

The results developed in this section can also show when the excitation of the libration amplitude from the eccentricity of the orbit can cause the secondary to begin circulating. Application of Corollary \ref{cor:max_libration_angle} for a range of $r_0$ values is shown in Fig. \ref{fig:phimax_nu}. In this case, every case is tested with $\phi_{2,0} = \dot{\phi}_{2,0} = 0$, so the excitation occurs only from the presence of available energy for this orbit due to variation in the radial position. The results are presented in terms of eccentricity and true anomaly, which determine the extents of $r_0$, but are more intuitive for this application. It is shown that for this case, if the eccentricity is less than $1\times10^{-4}$ the libration is bounded for any initial condition. For larger eccentricities, the initial conditions must be nearer to periapsis in order for the libration to remain bounded. It should be noted that each eccentricity line is at a different energy level, but they all have the same angular momentum value for this study.

\begin{figure}[htb]
	\centering\includegraphics[width=.6\textwidth]{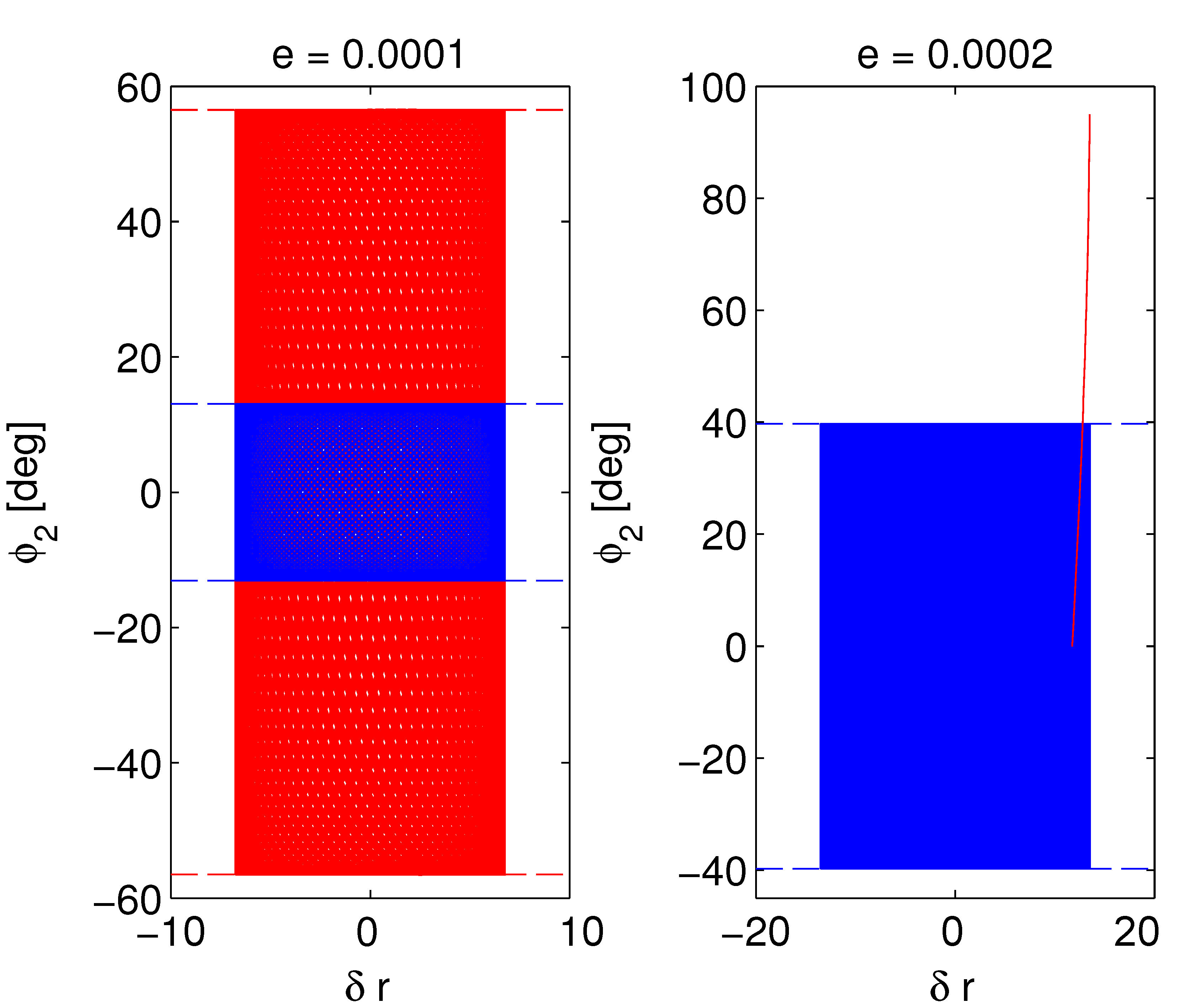}
	\caption{Four trajectories chosen from Fig. \ref{fig:phimax_nu}, integrated for 500 orbits. Integration of the circulating trajectory was stopped upon passing $\phi_2 = 95\degree$. These numerical integrations verify the validity of the relationships shown in Fig. \ref{fig:phimax_nu}.}
	\label{fig:ecc_trajs}
\end{figure}

In order to illustrate the validity of these relationships, four different simulations indicated on Fig. \ref{fig:phimax_nu} with circles were simulated for 500 orbits. These trajectories are plotted in Fig. \ref{fig:ecc_trajs}. It is clear that the bounds are accurately predicting the librational motion, and the unbounded case does circulate directly as predicted by Theorem \ref{thm:suff_cond_circ} for that case. 



It is interesting to investigate the case of Saturn's moon Hyperion, which is known to be in chaotic rotation. Using the moments of inertia for Hyperion from \cite{hyperion} (renormalized by $\alpha = 180$ km for Hyperion), and the known Hyperion orbit and Saturn properties, Eq. \eqref{eq:remark1} is used to find the $\dot{\phi}_2$ that will guarantee circulation. We find that if Hyperion were to have $|\dot{\phi}_2| > 0.00524$ degrees per second at periapse with zero libration angle, it will enter a circulating state. Given the extremely small limit found, it is no surprise that Hyperion is in an unbound chaotic rotation state.

\clearpage

\subsection{Initial Condition Mapping to Circulation for the General Case}

In the general case with $\nu < 1$, the relationships developed in Section \ref{sec:limits_nu_1} can't be used because the orbit energy can not be considered constant due to the coupling from the secondary body's motion. In this section, we investigate which initial conditions at $\phi_2 = 0$ circulate numerically. Our approach is to sample the possible initial conditions at $\phi_2 = 90\degree$ and propagate them backwards in time to see where they originate at $\phi_2 = 0$. 

By fixing one of the states (we choose to fix $\phi_2$), the remaining states are required to lie on the surface of an ellipsoid of constant energy. We use these constant energy surfaces to investigate the states of the simulations at  $\phi_2 = 0$ and $\phi_2 = 90\degree$. These visualizations are essentially 3-D Poincar\`{e} surfaces. The case investigated here is for the nominal system with $E = E^- + 0.1\delta E$.

\begin{figure}[htb]
\begin{center}
	\includegraphics[width=.75\textwidth]{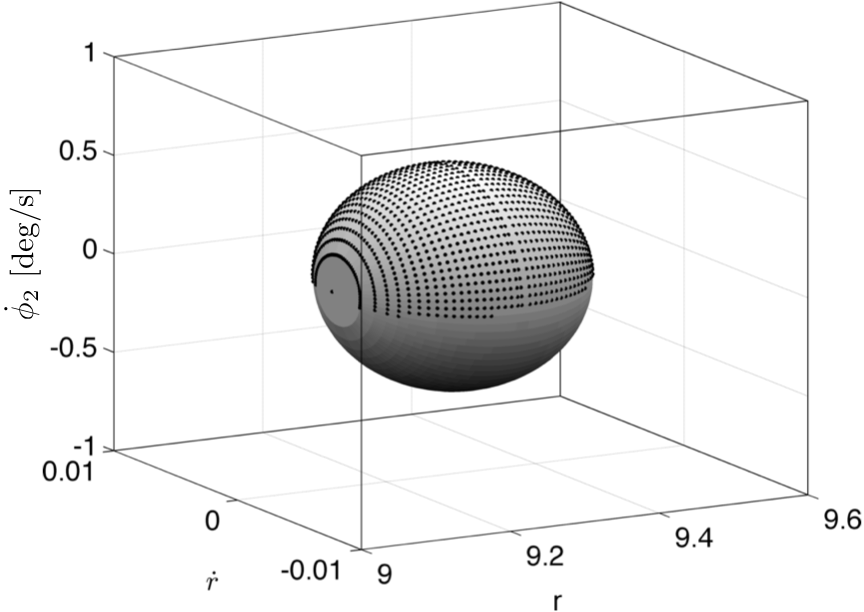}
	\caption{Constant energy surface at $\phi_2 = 90\degree$ for the nominal case with $E = E^- + 0.1\delta E$. The black dots indicate the ICs tested which, when propagated backwards in time, pass through $\phi_2 = 0$.}
	\label{fig:ellipsoid90}
\end{center}
\end{figure}

The constant energy surface at $\phi_2 = 90\degree$ is illustrated in Fig. \ref{fig:ellipsoid90}.  Initial conditions which sampled the entire surface were tested, with those that propagated back to $\phi_2 = 0$ shown as black dots on the ellipsoid. There were 3362 initial conditions tested, exactly half propagated to $\phi_2 = 0$, the other half propagated toward $\phi_2 = 180\degree$ and were ignored for this analysis. As seen in Fig. \ref{fig:ellipsoid90}, the split is basically associated with the sign of $\dot{\phi}_2$; those with a positive libration rate when crossing $\phi_2 = 90\degree$ generally came from $\phi_2 = 0$. However, it is interesting to note that there 22 of the 1681 cases did have negative libration rates, seen as those will small values of $r$ on the ellipsoid.

\begin{figure}[htb]
\begin{center}
	\includegraphics[width=.49\textwidth]{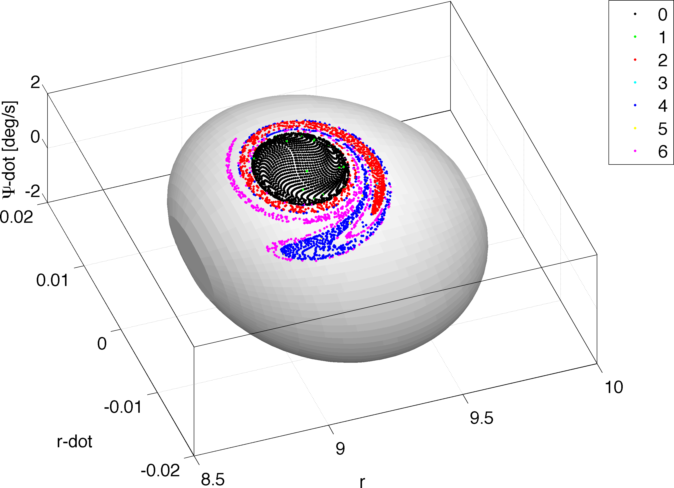}
	\includegraphics[width=.49\textwidth]{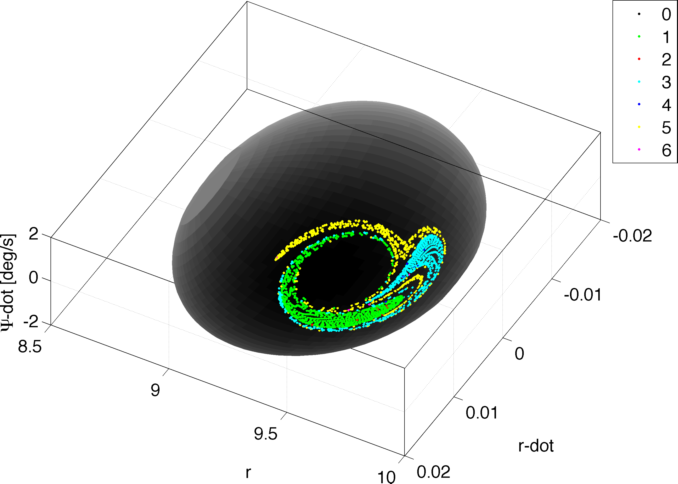}
	\caption{The north pole (left, positive values of $\dot{\phi}_2$) and south pole (right) of the constant energy surface at $\phi_2 = 0$ for the nominal case with $E = E^- + 0.1\delta E$. The initial conditions from Fig. \ref{fig:ellipsoid90} are integrated back in time, with each of the first seven crossings of the $\phi_2 = 0$ surface plotted in different colors as indicated.}
	\label{fig:ellipsoid0}
\end{center}
\end{figure}

The evolution of the backwards propagated trajectories are shown as they cross $\phi_2 = 0$ in Fig. \ref{fig:ellipsoid0}. The initial point where the initial conditions reach $\phi_2 = 0$ are labeled as the zeroth crossing, and plotted also as black dots. We then plot each of the next six crossings in different colors, alternating between the north and south poles of the ellipsoid as the trajectories cross in opposite directions. Note there is definite structure to the region of each crossing on the ellipsoid. 

\begin{figure}[htb]
\begin{center}
	\includegraphics[width=.49\textwidth]{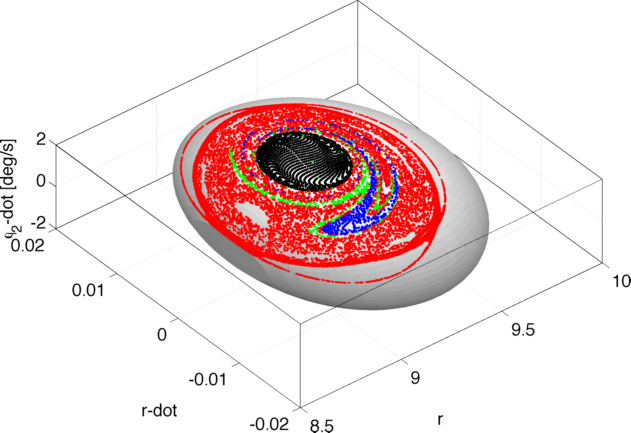}
	\includegraphics[width=.49\textwidth]{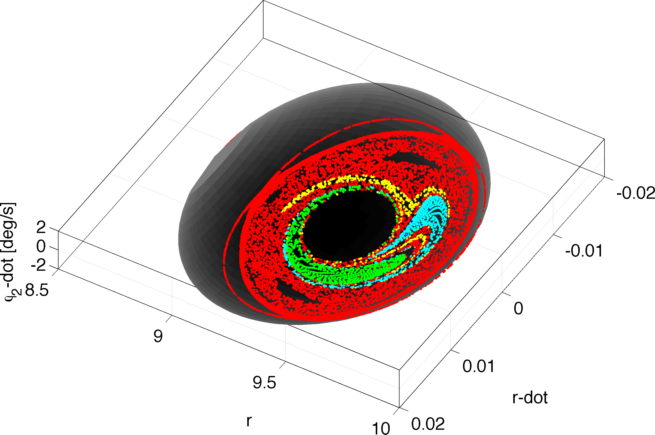}
	\caption{The north pole (left) and south pole (right) from Fig. \ref{fig:ellipsoid0}, with all crossing after the seventh added as red dots.}
	\label{fig:ellipsoid0_all}
\end{center}
\end{figure}

Fig. \ref{fig:ellipsoid0_all} shows the same thing as in Fig. \ref{fig:ellipsoid0}, except that all crossings after the sixth are plotted with red dots. This shows that although the coverage continues to spread and lose some of the distinct structure of the earlier crossings, there are still interesting features. First, there is a large area surrounding the equator where there are no trajectories that are connected to circulation. This tells us that trajectories that reside in this region will stay bounded. We point out that the points around the north pole are biased as a group toward the smaller radius values, while the south pole cases are biased toward the larger radius values. Second, there are some significant areas on the ellipsoids amongst the propagated trajectories that are not filled, such as on the north pole view around $\dot{r}=0$ and small values of $r$. These gaps correspond to the periodic trajectories discussed in Section \ref{sec:periodic}. The third note about this plot is that currently the region immediately surrounding the south pole is unpopulated. These states would correspond to trajectories that would evolve through $\phi_2 = -90\degree$, as opposed to through $\phi_2 = 90\degree$ as are being analyzed here.

\begin{figure}[htb]
\begin{center}
	\includegraphics[width=.49\textwidth]{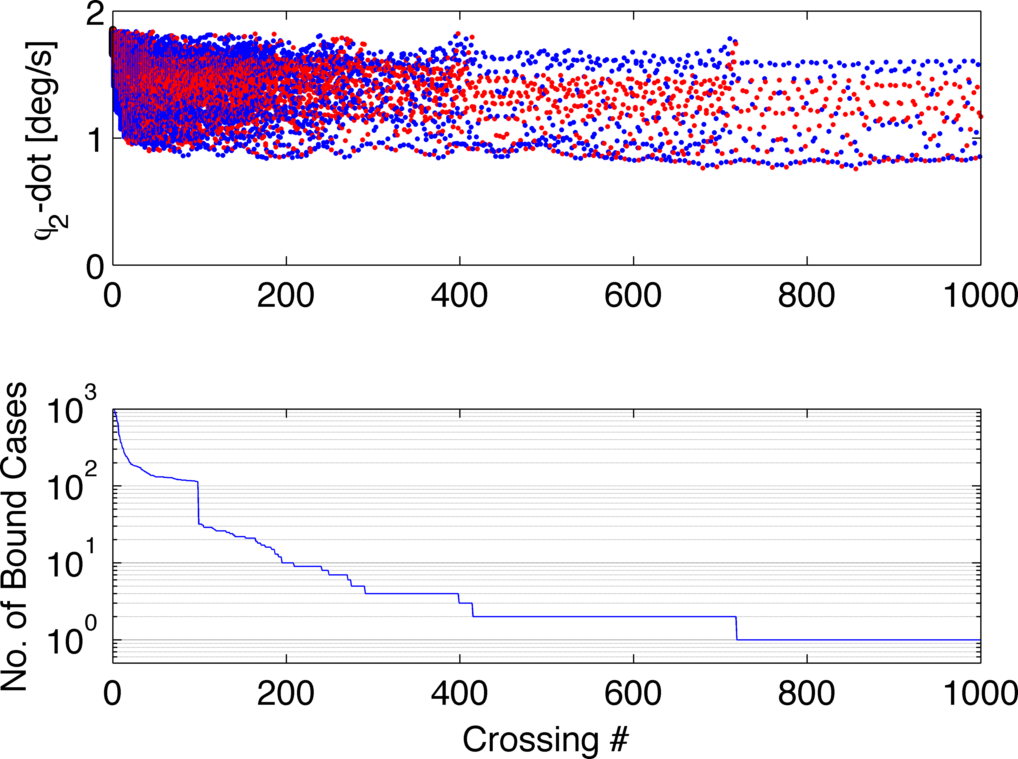}
	\includegraphics[width=.49\textwidth]{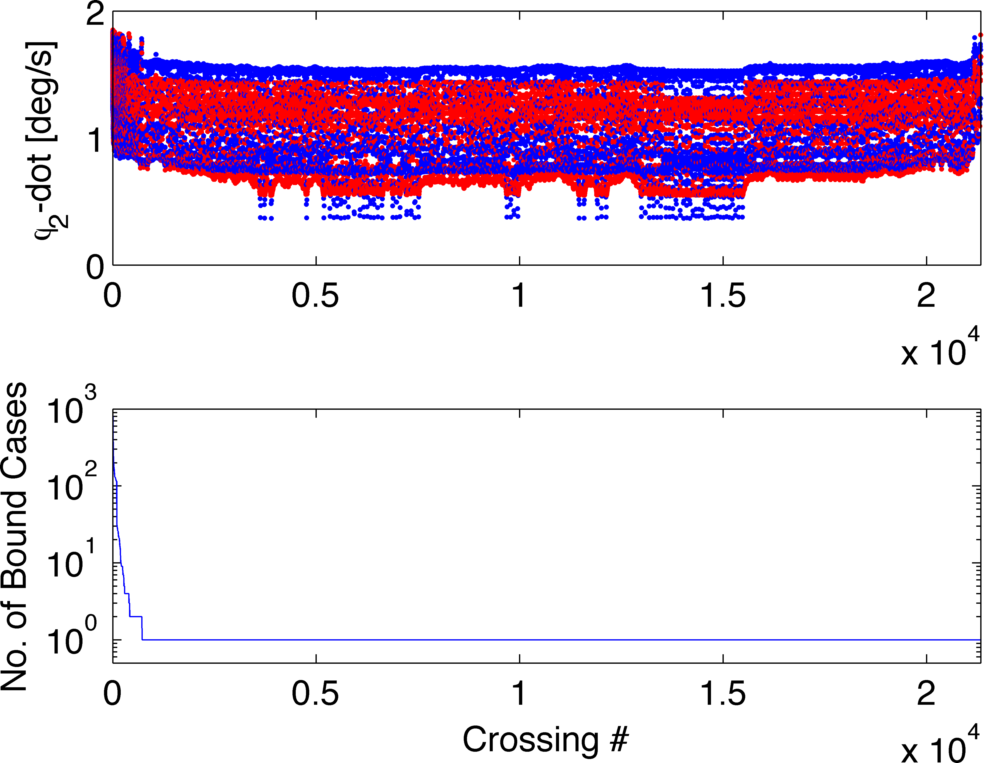}
	\caption{Absolute values of $\dot{\phi}_2$ for each crossing (those with a negative $\dot{\phi}_2$ are plotted in blue), along with the number of trajectories which are bound at that crossing (out of the initial 1681). The left plot is zoomed in on the first 1000 crossings of the right plot; note that after approximately 700 crossings only one sample remains bound. It is clear from the right plot that this one particular long-bound case reaches much lower $\dot{\phi}_2$ values during some crossings than any of the other samples.}
	\label{fig:phi0_crossings}
\end{center}
\end{figure}

\begin{figure}[htb]
\begin{center}
	\includegraphics[width=.49\textwidth]{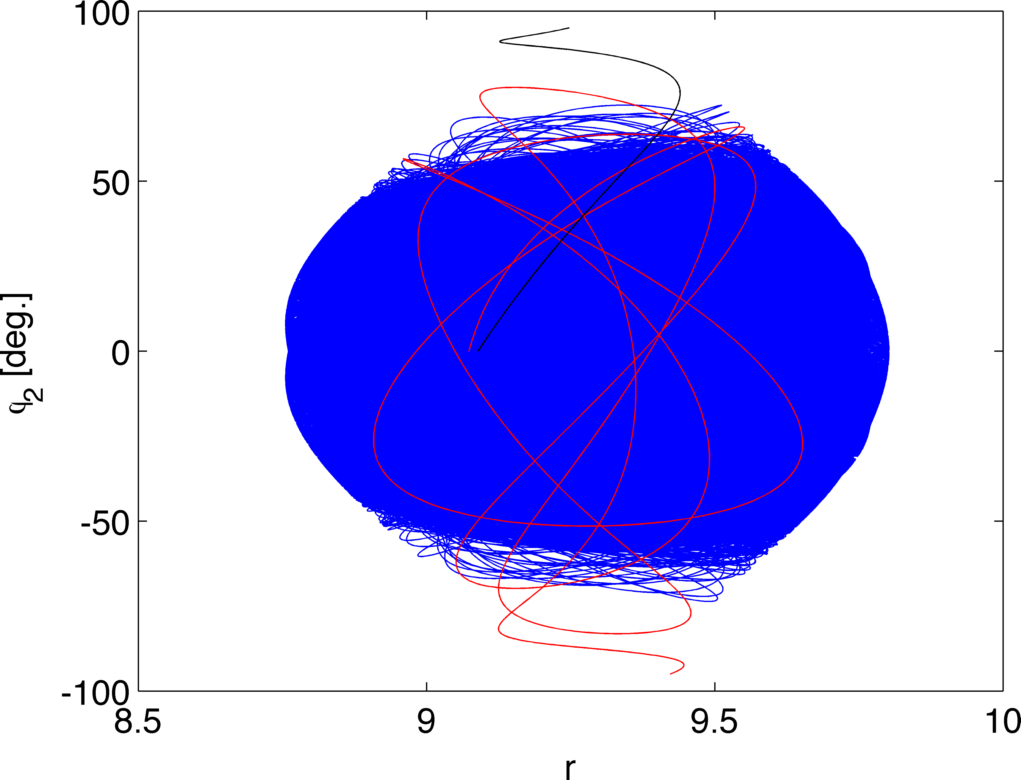}
	\caption{The trajectory of the long lived case from Fig. \ref{fig:phi0_crossings} in $r-\phi_2$ space. The initial trajectory from $\phi_2 = 90\degree \to 0$ is shown in black, and the final portion of the trajectory is shown in red, before it becomes unbound, in this case toward $\phi_2 = -90\degree$.}
	\label{fig:traj_case1564}
\end{center}
\end{figure}

Another view of the evolution of the backwards propagated trajectories from Figs. \ref{fig:ellipsoid90} - \ref{fig:ellipsoid0_all} are shown in Fig. \ref{fig:phi0_crossings}. In this figure, we see the absolute value of $\dot{\phi}_2$ plotted for every trajectory for each crossing. The lower plot shows how many trajectories still exist. The main takeaway from this plot is that most the trajectories that we propagated backwards don't stay in a librational state for very long; of the 1681 initial trajectories, only about 100 cross $\phi_2 = 0$ more than 100 times. Only 2 trajectories last longer than about 400 crossings. Interestingly, one trajectory lasts for a very long time - approximately 22,000 crossings! This tells us that a large portion of the red points seen in Fig. \ref{fig:ellipsoid0_all} can be attributed to one very rare case (this trajectory is shown in Fig. \ref{fig:traj_case1564}). The other thing we find from this plot is that, unlike in the $\nu \simeq 1$ case, there isn't a clear value of $\dot{\phi}_2$ that indicates if circulation will occur or not. This is due to the coupling between the libration and orbital states.

\begin{figure}[htb]
\begin{center}
	\includegraphics[width=.30\textwidth]{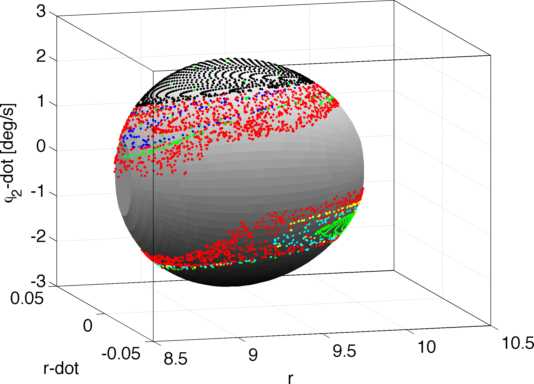}
	\includegraphics[width=.30\textwidth]{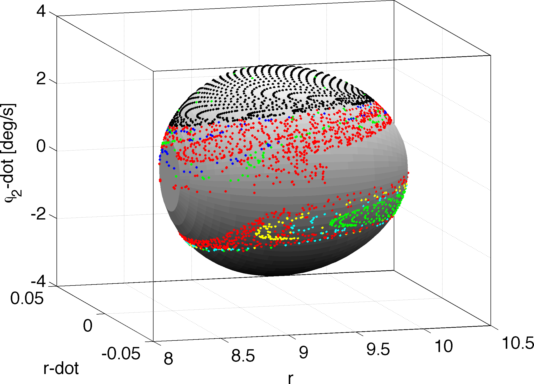}
	\includegraphics[width=.30\textwidth]{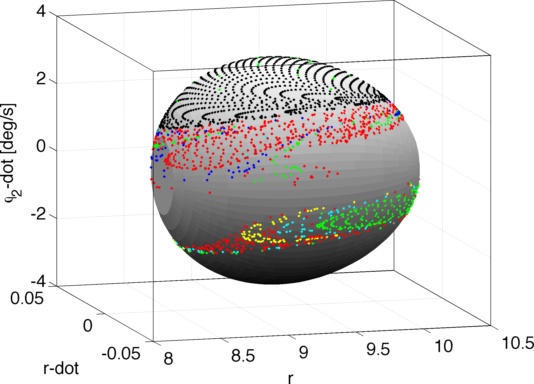}
	\caption{Three constant energy ellipsoids at $\phi_2 = 0$ for cases with $E = E^- + \delta E$, $E = E^- + 2\delta E$, and $E = E^- + 2.5\delta E$ from left to right.}
	\label{fig:ellipsoid0_high_E}
\end{center}
\end{figure}

\begin{figure}[htb]
\begin{center}
	\includegraphics[width=.49\textwidth]{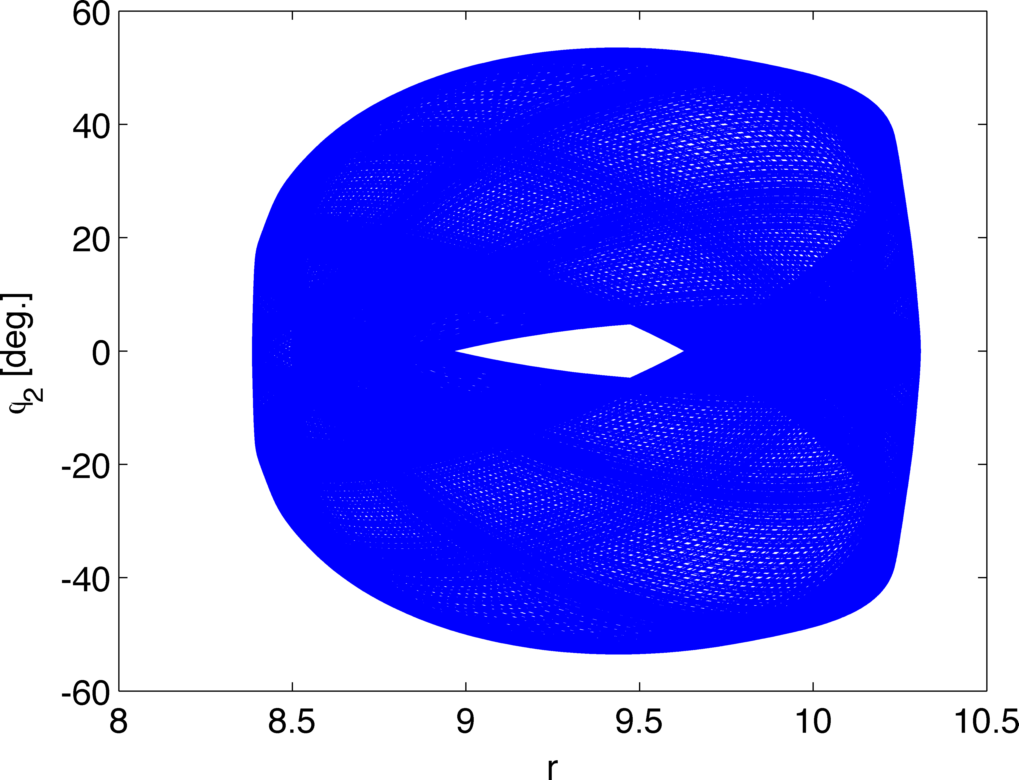}
	\includegraphics[width=.49\textwidth]{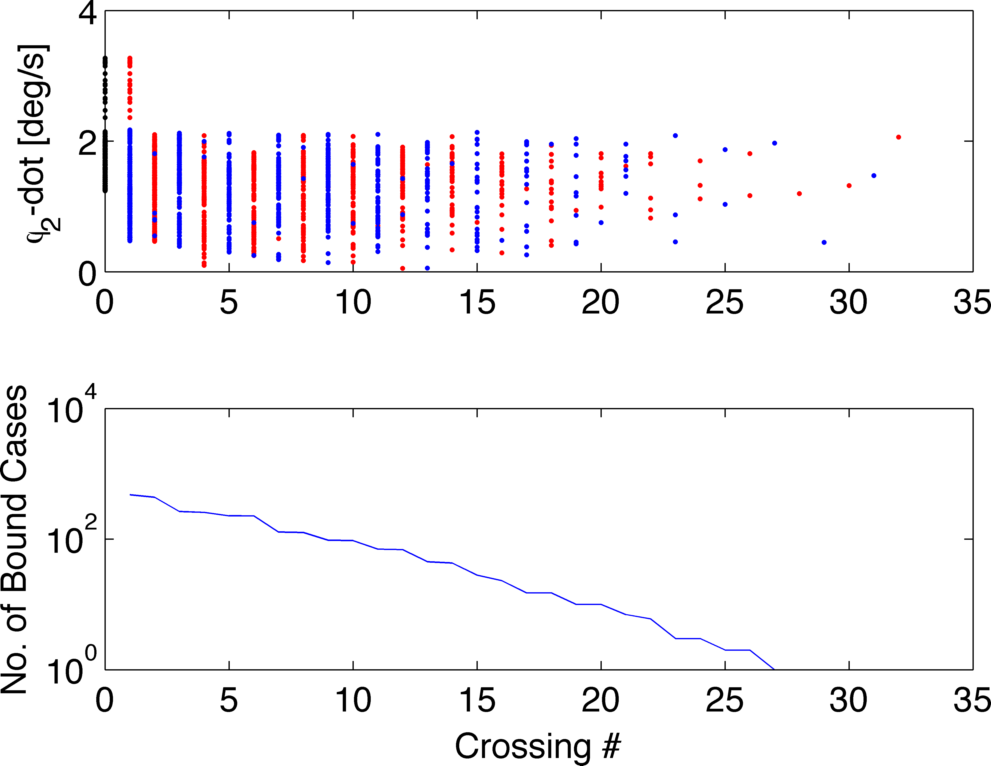}
	\caption{An example of a bounded trajectory and the crossing history for the case when $E = E^- + 2.5\delta E$. The trajectory had initial conditions of $r \simeq 10$, $\phi_2 = 0$, $\dot{r} \simeq 0.02$ and $\dot{\phi}_2 = 0.5$ deg/s and was integrated for 1000 librational periods.}
	\label{fig:highE_stable_trajectory}
\end{center}
\end{figure}

Finally, it is interesting to see how things change when the energy level is increased. Fig. \ref{fig:ellipsoid0_high_E} shows the $\phi_2 = 0$ ellipsoid for three increasingly higher energy levels. These views show that as the energy is increased there are two main impacts. First, the trajectories evolve to much larger areas of ellipsoid, meaning there are smaller sets of initial conditions at $\phi_2 = 0$ that correspond to bounded trajectories. Second, the unbounded trajectories reside for much shorter timespans around  $\phi_2 = 0$. In Fig. \ref{fig:highE_stable_trajectory}, we show the crossing history, which for that case has all trajectories circulating again after less than 30 crossings. We also show a stable trajectory, proving that although the ZVC is significantly opened at this energy level, there are still bounded trajectories.

\section{Conclusion}\label{sec:conclusion}

This paper explored the dynamic system of a triaxial ellipsoid satellite in orbit in the equatorial plane of an oblate body. This system reduces to a 2 degree-of-freedom system in the radial separation and the libration angle of the ellipsoid. The reduced system was then analyzed with the goal of determining limits on the dynamic configurations for which the librational motion is bounded to less than $\pm90\degree$. The relative equilibrium points were found in the ellipsoid-fixed frame, and the stability of these points was discussed. The conservation of energy and angular momentum in the system was exploited to write an expression for zero-velocity curves. These curves are used to determine a sufficiency condition on when the librational motion is bounded. It is shown that bounded trajectories exist beyond these sufficiency conditions, including families of periodic orbits. We addressed the conditions for unbounded motion for all systems. In particular, when the ellipsoid becomes of very small mass compared to the oblate body, analytical relationships were derived that determine the maximum libration angle for any orbit eccentricity.
Future work includes extending this approach to two ellipsoid systems, non-conservative systems, and coupled out-of-plane motion.

\section*{Appendix}\label{sec:appendix}

The mathematical description of the system studied in this paper was determined as a simplified form of the system studied by Scheeres \cite{scheeres_planarF2BP_stability}. The simplifications made to obtain the results used in this paper are derived here in order to make the connection to the previous work explicit. 

\subsection*{Equations of Motion}

The Lagrangian, $L=T-V$ for this system is,
\begin{equation}
	L = \frac{1}{2}\overline{I}_{1_z}\frac{M_1}{M_2}\left(\dot{\theta} + \dot{\phi}_1\right)^2 + \frac{1}{2}\overline{I}_{2_z}\dot{\phi}_2^2 + \frac{1}{2}\nu\dot{r}^2 + \frac{1}{2}\left( \overline{I}_{2_z} + \nu r^2 \right) \dot{\theta}^2 +  \overline{I}_{2_z}\dot{\phi}_2 \dot{\theta} - V(r,\phi_2)
\end{equation}
Using Lagrange's equations with out any external forces,
\begin{equation}
	\frac{\mathrm{d}}{\mathrm{dt}}\left( \frac{\partial L}{\partial \dot{q}_i} \right) = \frac{\partial L}{\partial q_i}
\end{equation}
the equations of motion for this system with the coordinates $r$, $\theta$, $\phi_1$, and $\phi_2$ were found in Scheeres \cite{scheeres_planarF2BP_stability} and are rewritten in normalized units to be,
\begin{equation}
	\ddot{r} = \dot{\theta}^2r - \frac{1}{\nu}\frac{\partial V}{\partial r}
\end{equation}
\begin{equation}
	\ddot{\phi}_1 = -\left( 1 + \frac{\nu r^2}{\overline{I}_{1_z}}\right)\frac{1}{\nu r^2}\frac{\partial V}{\partial \phi_1} - \frac{1}{\nu r^2}\frac{\partial V}{\partial \phi_2} + 2\frac{\dot{r}\dot{\theta}}{r}
\end{equation}
\begin{equation}
	\ddot{\phi}_2 = -\left( 1 + \frac{\nu r^2}{\overline{I}_{2_z}}\right)\frac{1}{\nu r^2}\frac{\partial V}{\partial \phi_2} - \frac{1}{\nu r^2}\frac{\partial V}{\partial \phi_1} + 2\frac{\dot{r}\dot{\theta}}{r}
\end{equation}
\begin{equation}
	\ddot{\theta} = \frac{1}{\nu r^2}\frac{\partial V}{\partial \phi_1} + \frac{1}{\nu r^2}\frac{\partial V}{\partial \phi_2} - 2\frac{\dot{r}\dot{\theta}}{r}
\end{equation}

However, in the case of an oblate body the potential is no longer a function of $\phi_1$ so that
\begin{equation}
	\frac{\partial V}{\partial \phi_1} = 0
\end{equation}
and the equations of motion for the angles become,
\begin{equation}\label{eq:phi1_eom}
	\ddot{\phi}_1 = - \frac{1}{\nu r^2}\frac{\partial V}{\partial \phi_2} + 2\frac{\dot{r}\dot{\theta}}{r}
\end{equation}
\begin{equation}
	\ddot{\phi}_2 = -\left( 1 + \frac{\nu r^2}{\overline{I}_{2_z}}\right)\frac{1}{\nu r^2}\frac{\partial V}{\partial \phi_2} + 2\frac{\dot{r}\dot{\theta}}{r}
\end{equation}
\begin{equation}\label{eq:theta_eom}
	\ddot{\theta} = \frac{1}{\nu r^2}\frac{\partial V}{\partial \phi_2} - 2\frac{\dot{r}\dot{\theta}}{r}
\end{equation}

The partials of the potential for an oblate primary are,
\begin{equation}\label{eq:partVpartr}
	\frac{\partial V}{\partial r} =  \frac{\nu}{r^2} \bigg\{ 1 + \frac{3}{2r^2}\bigg[\left(\overline{I}_{1_z} - \overline{I}_s\right) -\frac{1}{2}\overline{I}_{2_x} -\frac{1}{2} \overline{I}_{2_y} + \overline{I}_{2_z} + \frac{3}{2} \left(\overline{I}_{2_y} - \overline{I}_{2_x}\right) \cos 2 \phi_2  \bigg] \bigg\}
\end{equation}
\begin{equation}\label{eq:partVpartphi2}
	\frac{\partial V}{\partial \phi_2} = \frac{3}{2}\frac{\nu}{r^3} \left(\overline{I}_{2_y}-\overline{I}_{2_x}\right)\sin(2\phi_2)
\end{equation}

\subsection*{Integrals of Motion}

In the current problem with an oblate primary, we have three integrals of motion. The first is the total energy of the system, which is shown to be the Jacobi integral of this system since it is time invariant,
\begin{equation}
	h = \dot{\mathbf{q}}\cdot\frac{\partial L}{\partial \dot{\mathbf{q}}} - L = T + V
\end{equation}

The second integral of motion is the total angular momentum of the system. This is found because the coordinate $\theta$ is ignorable, meaning that $\mathrm{d}/\mathrm{dt}(\partial L / \partial \dot{\theta})=0$, so the integral is written as,
\begin{equation}\label{eq:angmom_total}
	\begin{split}
		K_{tot} &= \frac{\partial L}{\partial \dot{\theta}} \\
		&=  I_z \dot{\theta} + \overline{I}_{2_z}\dot{\phi}_2 + \frac{M_1}{M_2} \overline{I}_{1_z}\dot{\theta}_1
	\end{split}
\end{equation}

The third integral is found by combining Eqs. \eqref{eq:phi1_eom} and \eqref{eq:theta_eom}, so we find that 
\begin{equation}
	\ddot{\theta}_1 = \ddot{\phi}_1 + \ddot{\theta} = 0
\end{equation}
Therefore the inertial angular velocity of the primary, $\dot{\theta}_1$, is an integral of motion. This implies that the terms in the kinetic energy and angular momentum expressions which depend only on $\dot{\theta}_1$ are also conserved. This fact makes intuitive sense as a primary that is symmetric about the spin axis can't have any gravitational torques exerted on it from the secondary since the center of mass and the center of gravity (in the secondary's gravity field) are at the same location in the primary.

\subsection*{Dynamics Matrix Partials}

The partials are, 
\begin{equation}
	\frac{\partial \ddot{r}}{\partial r} = \frac{1}{I_z^2}\left[K^2 - 2K\overline{I}_{2_z}\dot{\phi}_2 + \overline{I}_{2_z}^2\dot{\phi}^2_2  \right] - \frac{1}{I_z^4}\left[\left( 4\overline{I}_{2_z} \nu r + 4 \nu^2 r^3 \right) \left( K^2 - 2K\overline{I}_{2_z}\dot{\phi}_2 + \overline{I}_{2_z}^2\dot{\phi}^2_2 \right) r  \right] - \frac{1}{\nu}\frac{\partial^2 V}{\partial r^2}
\end{equation}

\begin{equation}
	\frac{\partial \ddot{r}}{\partial \phi_2} = - \frac{1}{\nu}\frac{\partial^2 V}{\partial r \partial \phi_2}
\end{equation}

\begin{equation}
	\frac{\partial \ddot{r}}{\partial \dot{\phi}_2} = \frac{2r}{I_z^2} \left[ -K\overline{I}_{2_z} + \overline{I}_{2_z}^2 \dot{\phi}_2 \right]
\end{equation}

\begin{equation}
	\frac{\partial \ddot{\phi}_2}{\partial r} = \frac{2}{\nu r^3}\frac{\partial V}{\partial \phi_2} -\left( 1 + \frac{\nu r^2}{\overline{I}_{2_z}}\right)\frac{1}{\nu r^2}\frac{\partial^2 V}{\partial r \partial \phi_2} - \frac{2 \dot{r}}{r^2 I_z^2} \left[ \left(K - \overline{I}_{2_z}\dot{\phi}_2\right) \left( \overline{I}_{2_z} + 3 \nu r^2 \right) \right]
\end{equation}

\begin{equation}
	\frac{\partial \ddot{\phi}_2}{\partial \phi_2} = -\left( \frac{1}{\nu r^2} + \frac{1}{\overline{I}_{2_z}} \right) \frac{\partial^2 V}{\partial \phi_2^2}
\end{equation}

\begin{equation}
	\frac{\partial \ddot{\phi}_2}{\partial \dot{r}} = \frac{2}{r I_z} \left(K - \overline{I}_{2_z}\dot{\phi}_2\right)
\end{equation}

\begin{equation}
	\frac{\partial \ddot{\phi}_2}{\partial \dot{\phi}_2} = -\frac{2 \overline{I}_{2_z}\dot{r}}{r I_z}
\end{equation}

And the second partials of the potential are given by,
\begin{equation}
	\frac{\partial^2 V}{\partial r^2} = -\frac{2 \nu}{r^3} \bigg\{ 1 + \frac{3}{r^2}\bigg[\left(\overline{I}_{1_z} - \overline{I}_s\right) -\frac{1}{2}\overline{I}_{2_x} -\frac{1}{2} \overline{I}_{2_y} + \overline{I}_{2_z} + \frac{3}{2} \left(\overline{I}_{2_y} - \overline{I}_{2_x}\right) \cos 2 \phi_2  \bigg] \bigg\}
\end{equation}

\begin{equation}
	\frac{\partial^2 V}{\partial \phi_2^2} = \frac{3 \nu}{r^3} \left(\overline{I}_{2_y} - \overline{I}_{2_x}\right) \cos 2 \phi_2
\end{equation}

\begin{equation}
	\frac{\partial^2 V}{\partial r \partial \phi_2} = -\frac{9 \nu}{2 r^4} \left(\overline{I}_{2_y} - \overline{I}_{2_x}\right) \sin 2 \phi_2
\end{equation}

\subsection*{Relative Equilibria Locations}

Following Scheeres \cite{scheeres_planarF2BP_stability}, we find the equilibrium points by searching for places where the variations in energy are stationary at a constant value of angular momentum. In other words, we find when the following conditions hold:
\begin{equation}
	\frac{\partial E}{\partial r} = 0
\end{equation}
\begin{equation}
	\frac{\partial E}{\partial \dot{r}} = 0
\end{equation}
\begin{equation}
	\frac{\partial E}{\partial \phi_2} = 0
\end{equation}
\begin{equation}
	\frac{\partial E}{\partial \dot{\phi}_2} = 0
\end{equation}
at a given value of $K$, as seen in Eq. \eqref{eq:free_energy}.

The following relationships are found,
\begin{equation}\label{eq:dEdr_eq}
	\frac{\partial E}{\partial \dot{r}} = \nu \dot{r}
\end{equation}
\begin{equation}
	\frac{\partial E}{\partial \dot{\phi}_2} = \frac{\overline{I}_{2_z}\nu r^2\dot{\phi}_2}{I_z}
\end{equation}
\begin{equation}\label{eq:dEdphi_eq}
	\frac{\partial E}{\partial \phi_2} = \frac{3}{2}\frac{\nu}{r^3} \left(\overline{I}_{2_y}-\overline{I}_{2_x}\right)\sin(2\phi_2)
\end{equation}
which when combined with the stationarity conditions imply, respectively, that $\dot{r}=0$, $\dot{\phi}_2 = 0$, and $\phi_2 = 0$, $\pm\pi/2$, or $\pi$.  Using these conditions, we can evaluate the partial with respect to $r$ to be,
\begin{equation}\label{eq:partEpartr}
	\frac{\partial E}{\partial r} = -\nu r\frac{K^2}{I_z^2} + \frac{\nu}{r^2}\left[ 1 + \frac{3\left(\overline{I}_{1_z} - \overline{I}_s + C_2^\pm\right)}{2 r^2}\right]
\end{equation}

\subsection*{Relative Equilibria Stability Partials}

Recall from Section \ref{sec:equilibrium}, the equilibrium points must occur at $\dot{r}=0$, $\dot{\phi}_2 = 0$, and $\phi_2 = 0$, $\pm\pi/2$, or $\pi$. Therefore the partials for the dynamic matrix are greatly reduced to become,
\begin{equation}
	\frac{\partial \ddot{r}}{\partial r} = \frac{K^2}{I_z^2} - \frac{1}{I_z^4}\left[\left( 4\overline{I}_{2_z} \nu r_{eq} + 4 \nu^2 r_{eq}^3 \right) K^2 r_{eq}  \right] - \frac{1}{\nu}\left(\frac{\partial^2 V}{\partial r^2}\right)_{eq}
\end{equation}

\begin{equation}
	\frac{\partial \ddot{r}}{\partial \phi_2} = - \frac{1}{\nu}\left(\frac{\partial^2 V}{\partial r \partial \phi_2}\right)_{eq}
\end{equation}

\begin{equation}
	\frac{\partial \ddot{r}}{\partial \dot{\phi}_2} = -\frac{2r_{eq}K}{I_z^2} 
\end{equation}

\begin{equation}
	\frac{\partial \ddot{\phi}_2}{\partial r} = \frac{2}{\nu r_{eq}^3}\left(\frac{\partial V}{\partial \phi_2}\right)_{eq} -\left( 1 + \frac{\nu r_{eq}^2}{\overline{I}_{2_z}}\right)\frac{1}{\nu r_{eq}^2}\left(\frac{\partial^2 V}{\partial r \partial \phi_2} \right)_{eq}
\end{equation}

\begin{equation}
	\frac{\partial \ddot{\phi}_2}{\partial \phi_2} = -\left( \frac{1}{\nu r_{eq}^2} + \frac{1}{\overline{I}_{2_z}} \right) \left(\frac{\partial^2 V}{\partial \phi_2^2}\right)_{eq}
\end{equation}

\begin{equation}
	\frac{\partial \ddot{\phi}_2}{\partial \dot{r}} = \frac{2K}{r_{eq} I_z}
\end{equation}

\begin{equation}
	\frac{\partial \ddot{\phi}_2}{\partial \dot{\phi}_2} = 0
\end{equation}
Where the equilibrium point distance is indicated by $r_{eq}$. Note that from Eq. \eqref{eq:partVpartphi2}, we see that
\begin{equation}
	\left(\frac{\partial V}{\partial \phi_2}\right)_{eq} = 0
\end{equation}

The second partials of the potential evaluated at the equilibrium points are,

\begin{equation}
	\frac{\partial^2 V}{\partial r^2} = -\frac{2 \nu}{r_{eq}^3} \bigg\{ 1 + \frac{3}{r_{eq}^2}\bigg[\left(\overline{I}_{1_z} - \overline{I}_s\right) + C_2^\pm  \bigg] \bigg\}
\end{equation}

\begin{equation}
	\frac{\partial^2 V}{\partial \phi_2^2} =  \pm \frac{3 \nu}{r_{eq}^3} \left(\overline{I}_{2_y} - \overline{I}_{2_x}\right)
\end{equation}

\begin{equation}\label{eq:dVdrdphi_eq}
	\frac{\partial^2 V}{\partial r \partial \phi_2} = 0
\end{equation}
where $C_2^\pm$ was defined in Eq. \eqref{eq:c2pm}. The plus terms correspond to the case where $\phi_2 = 0$ or $\pi$, and the minus terms correspond the the other cases where $\phi_2 = \pi/2$ or $3\pi/2$.

\bibliographystyle{spmpsci}
\bibliography{planar_refs}

\end{document}